\documentclass[10.5pt,twoside,english]{article}
\usepackage[T1]{fontenc}
\usepackage[latin9]{inputenc}
\usepackage{geometry}
\usepackage{babel}
\usepackage{float}
\usepackage{amsthm}
\usepackage{amsmath}
\usepackage{amssymb}
\usepackage{graphicx}
\usepackage{setspace}
\usepackage[skip=5pt]{caption}

\makeatletter

\numberwithin{equation}{section}
\usepackage{epstopdf, epsfig}
\usepackage{bm}
\usepackage[normalem]{ulem}
\usepackage{commath}
\usepackage{units}
\usepackage{xcolor}
\usepackage{hyperref}
\usepackage{soul}
\usepackage{mathrsfs}

\usepackage{mathtools}

\usepackage[comma]{natbib}

\DeclareMathAlphabet{\mathbfsf}{\encodingdefault}{\sfdefault}{bx}{sl}
\newcommand{\tens}[1]{\bm{\mathbfsf{#1}}}

\hypersetup{colorlinks=true, linkcolor={red!50!black}, citecolor={blue!50!black}, urlcolor={blue!80!black}}
\definecolor{dgreen}{rgb}{0.05, 0.5, 0.06}

\newcommand{\mF}{f}
\newcommand{\mG}{g}
\newcommand{\mS}{\mathcal{R}}
\newcommand{\lm}{\lambda}
\newcommand{\thb}{\bar{\mathit{\Theta}}_{\rm H}}

\newcommand{\thbd}{\mathit{\theta}_{\rm h}{}}

\newcommand{\hb}{\overline{h}}

\newcommand{\Sd}{\mathcal{S}}

\newcommand{\Bd}{\mathcal{B}}
\newcommand{\Cd}{\mathcal{C}}
\newcommand{\Qd}{\mathcal{Q}}
\newcommand{\iiii}{i}
\newcommand{\lbeta}{\sigma_n}

\newcommand{\Omfmax}{\Omega_{\rm f, max}}
\newcommand{\Omrmax}{\Omega_{\rm r, max}}
\newcommand{\Omf}{\Omega_{\rm f}}
\newcommand{\Omfone}{\Omega_{\rm f,1}}
\newcommand{\Omr}{\Omega_{\rm r}}
\newcommand{\Omi}{\Omega_{\rm i}}
\newcommand{\OmL}{\Omega_{\rm coll}}
\newcommand{\betaL}{\beta_{\rm coll}}
\newcommand{\lmL}{\lambda_{\rm coll}}
\newcommand{\lmf}{\lambda_{\rm f}}
\newcommand{\betacn}{\beta_{\rm im}}

\newcommand{\Omegacm}{\Omega_{\rm c, max}}
\newcommand{\betacm}{\beta_{\rm c,max}}

\makeatletter
\newcommand{\mypm}{\mathbin{\mathpalette\@mypm\relax}}
\newcommand{\@mypm}[2]{\ooalign{%
\raisebox{.1\height}{$#1+$}\cr
\smash{\raisebox{-.6\height}{$#1-$}}\cr}}
\makeatother

\begin{document}

\title{Oscillatory thermocapillary instability of a film heated by a thick substrate}

\author{W. Batson\thanks{Department of Mathematical Sciences, New Jersey Institute of Technology, Newark, NJ 07102-1982, USA%
}\rm{ }\thanks{Corresponding author, wbatson@gmail.com}\rm{ }, L. Cummings$^{*}$, D. Shirokoff$^{*}$, L. Kondic$^{*}$
}


\maketitle

\begin{abstract}
In this work we consider a new class of oscillatory instabilities that pertain to thermocapillary destabilization of a liquid film heated by a solid substrate.  We assume the substrate thickness and substrate-film thermal conductivity ratio are large so that the effect of substrate thermal diffusion is retained at leading order in the long-wave approximation.  As a result, system dynamics are described by a nonlinear partial differential equation for the film thickness that is nonlocally coupled to the full substrate heat equation.  Perturbing about a steady quiescent state, we find that its stability is described by a non-self adjoint eigenvalue problem.  We show that, under appropriate model parameters, the linearized eigenvalue problem admits complex eigenvalues that physically correspond to oscillatory (in time) instabilities of the thin film height. As the principal results of our work, we provide a complete picture of the susceptibility to oscillatory instabilities for different model parameters. Using this description, we conclude that oscillatory instabilities are more relevant experimentally for films heated by insulating substrates.  Furthermore, we show that oscillatory instability where the fastest-growing (most unstable) wavenumber is complex, arises only for systems with sufficiently large substrate thicknesses.
\end{abstract}

\section{Introduction}
The tendency of thin liquid films to destabilize and form wavy patterns is an important area of research for a wide range of applications.  For some applications, such as coatings and glass manufacturing, one may wish to operate under conditions that avoid these instabilities.  In others, such as multiphase heat/mass transfer technology and nanoscale patterning of liquid metals/polymers, precise control of the emerging wave pattern is of utmost concern.  In either case, the parametric conditions of interest can be determined, most simply, by applying the long-wave approximation to the governing nonlinear equations, see \citet{Oron97} and \citet{Craster09}.  In the long-wave approach, physical effects such as gravity, mean surface tension, thermocapillarity, solutocapillarity, and electromagnetism can be easily be accounted for, and one typically obtains a single nonlinear partial differential equation (PDE) for the spatiotemporal evolution of the local film thickness.  This method assumes the film dynamics are non-inertial and governed by a (first order in time) nonlinear PDE.


The principal phenomenon that a single-equation long-wave model cannot describe is the emergence of instabilities that are \emph{oscillatory} in time, i.e., overstability (see \citet{Nepomnyashchy01}, chapter 5).  Whereas single-equation film models predict \emph{monotonic} perturbations that grow or decay exponentially in time, oscillatory instabilities can only be observed in systems that describe the interaction between processes that occur on distinct time scales.  Thus, oscillatory instabilities are commonly obtained from Orr-Somerfeld type analyses of governing equations of motion that retain inertial effects and diffusive time scales.  Wide-ranging examples that highlight the emergence of oscillatory instabilities in fluid layers include work by \citet{Sternling59}, \citet{Takashima81}, \citet{Anderson96}, and \citet{Rednikov98}.  A common theme to these works is the level of analytical difficulty;  each obtains a linear dispersion relation (describing system stability) that is transcendental and implicit in the perturbation growth rates.  Combined with large parametric spaces and the fact that oscillatory perturbations necessarily reside in the complex plane, concise description of the emergence of oscillatory instability can be a challenging task.   Alternatively, the long wave approximation offers a convenient means to couple free surface deformation to other time-dependent physical processes of interest.

Several authors have investigated oscillatory instabilities of thin liquid films in the context of the long-wave approximation.  In many cases, e.g. \citet{Podolny05} and \citet{Bestehorn10}, such instabilities originate from the coupling between the local thickness and bulk concentration of a film composed of a binary mixture.  In addition to the bulk concentration dynamics, \citet{Morozov14} investigated oscillatory instability with the added effect of absorption/desorption kinetics between interfacial and bulk film surfactant concentration.  In other cases, oscillatory instabilities have been uncovered in multiple stacked layers of films, as described theoretically by coupled sets of film thickness evolution equations (\citet{Nepom07, Beerman07}). Multi-layer film configurations do not, however, guarantee oscillatory modes: for example, such instabilities were not obtained by \citet{Pototsky05} who investigated the dewetting dynamics of isothermal, ultrathin bilayers.  Of particular interest to the present work are oscillatory instabilities reported by \citet{Shklyaev12} in a model of thin-film thermocapillary destabilization from below. While there are similarities between that work and the present, we point out one important difference: in \citet{Shklyaev12}, the instability is driven by imposing a heat flux at the film-substrate interface; instead, in the present work we consider the full time-dependent heat-transfer in the substrate.  We also note that each of these works on oscillatory instabilities of thin liquid films obtains low-order polynomial equations for the perturbation growth rates (in contrast to the transcendental, implicit dispersion we obtain in the present work).

The problem we investigate in this work is the deformational thermocapillary instability.  This classic long-wavelength instability was first introduced by \citet{Scriven64} and later verified experimentally by \citet{VanHook97}.  In short, thermocapillary stresses that destabilize the free surface are generated by heating a film from below (transverse heating).  For sufficiently thin layers, these stresses can surpass capillary stabilization and deform an initially flat film.  Recently, \citet{Dietzel09} connected this mechanism with the formation of nanopillars ($\sim$10 $\mu$m spacing) on  ultrathin ($\sim$100 nm) polymer films.  Continuing work on thermocapillary patterning in (ultra)thin polymer films has been reviewed by \citet{Singer17}.  A patterning application that directly motivates our study is pulsed-laser dewetting of nanometric liquid metal films.  Experiments by \citet{Trice07} demonstrated dewetting pattern wavelengths that were commensurate with the predictions of long-wavelength thermocapillary modes.   Driven by such results, several workers have developed and investigated theoretical models for the pulsed-laser process.  \citet{Atena09} augmented a long-wave theory with pulsed laser irradiation to describe the thermocapillary dewetting of liquid cobalt on silicon oxide substrates.  Notably, they assumed that the substrate was thin, thereby ensuring model dynamics could be described by a (first order in time) single nonlinear PDE for the film thickness.  As a result, oscillatory instabilities do not arise in their model.

Recently, oscillatory modes for pulsed-laser thermocapillary dewetting of liquid metal films were uncovered by \citet{Dong16} and \citet{Seric18}.  These authors made observations primarily via nonlinear simulations of a model that couples the film PDE to the full heat equation for the substrate.  These works have not precisely characterized the emergence of oscillatory instabilities, in particular, because the task is complicated by the parameter space introduced by laser heating.  Thus, in the present work, we investigate the emergence of oscillatory instability for the simpler problem: a film heated by a \emph{thick} solid substrate. To do so, we initially follow the work by \citet{Saeki11, Saeki13} that considered the linear analysis of a coupled film-substrate model, which induces thermocapillary film deformation driven by laser heating.  In the present work we follow their asymptotic assumptions so that the full heat equation of the substrate is retained at leading order in the long-wave expansion of the governing equations.  Effectively, we assume that the substrate-film thermal conductivity and thickness ratios are large.  Although we obtain a dispersion relation that is similar to that of \citet{Saeki13}, it is important to note that they did not observe oscillatory modes.  This may be due to a limited examination of model parameter values in their investigation. 

The manuscript proceeds as follows:  in \S\ref{sec:dim}, we present the dimensional equations of motion and boundary conditions for a deformable liquid layer heated by a thick substrate.  In \S\ref{sec:ndim} we introduce a long-wave asymptotic expansion and derive an evolution equation for the film thickness that is nonlocally coupled to the diffusive (time-dependent) heat conduction problem in the substrate.  In this section we also introduce a unique nondimensionalization that casts the nonlocal model in terms of four dimensionless parameters: (1) $\Bd$, characterizing the mean thermal thickness of the film; 
(2) $\Sd$, characterizing the thermal thickness of the substrate; (3) $\Cd$, characterizing the imposed temperature difference; (4) $\Qd$, depending only on material properties.  In the following section, \S\ref{sec:ls}, we perform a linear analysis of the nonlocal model and demonstrate that its stability is governed by a generalized two-point boundary value problem that is not self-adjoint.  Solution of this problem yields the (implicit) dispersion relation that sets the course of investigation for the remainder of the paper.  In section \S\ref{sec:DispRel} we characterize the root structure of the dispersion relation and introduce the numerical contintuation methods we use to track its roots as functions of the perturbation wavenumbers.  Then, in section \S\ref{sec:charcoal} we classify the two characteristic pathways by which oscillatory instability manifests itself.  Finally, in section \S\ref{sec:emerge}, we provide a complete picture of the emergence of oscillatory instabilities within the considered parameter space.
\section{Dimensional equations}\label{sec:dim}
\begin{figure}
\begin{centering}
\includegraphics[width=30pc]{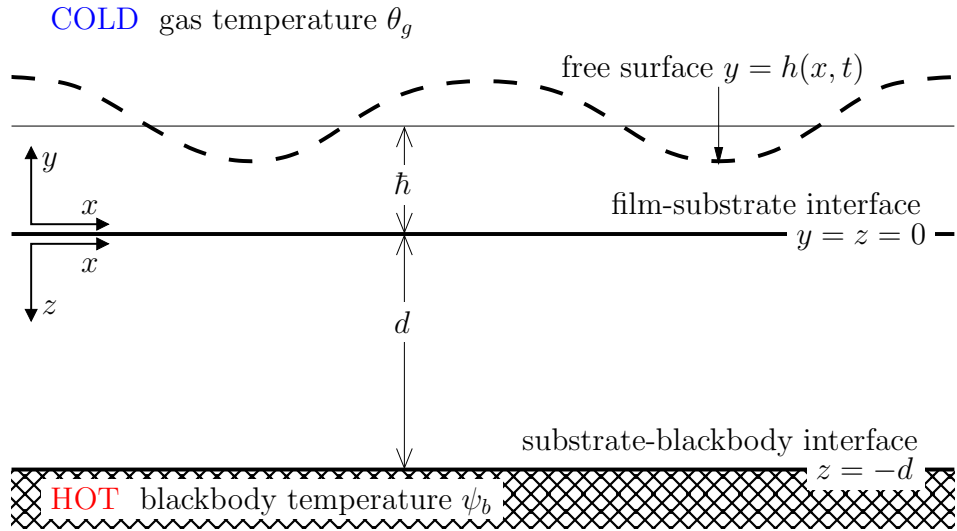}
\caption{Geometric sketch of the problem}
\label{fig:sketch}
\end{centering}
\end{figure}

Here we introduce equations that describe the fluid and temperature dynamics of the laterally infinite, two dimensional film-substrate system depicted schematically in figure \ref{fig:sketch}.  The film is composed of a Newtonian, incompressible liquid with average thickness $\hb$, density $\rho$, dynamic viscosity $\mu$, kinematic viscosity $\nu=\mu/\rho$, thermal conductivity $\kappa_{\rm f}$, and thermal diffusivity $\chi_{\rm f}$. Neglecting gravity, we have
\begin{align}
\rho\left(\partial_t\boldsymbol{v}+\boldsymbol{v}\cdot{}\boldsymbol{\nabla}{\boldsymbol{v}}\right)  &=  -\boldsymbol{\nabla}{p}+\mu\nabla^2 {\boldsymbol{v}}, \label{NSdim} \\
{\boldsymbol{\nabla }} \cdot \boldsymbol{v}  &=  0, \label{cont-dim} \\
\partial_t\mathit{\theta}+\boldsymbol{v}\cdot {\boldsymbol{\nabla}} \mathit{\theta}  &=  \chi_{\rm f} \nabla^2 \mathit{\theta}, 
\label{heateq-dim}
\end{align}
where $\boldsymbol{v}\equiv{}\{u(x,y,t),w(x,y,t)\}$, $p(x,y,t)$, and $\mathit{\theta}(x,y,t)$ are the film velocity, pressure and temperature fields, respectively.   With $\boldsymbol{\nabla}=\{\partial_x,\partial_y\}$, equations (\ref{NSdim}-\ref{heateq-dim}) govern the evolution of these fields in time $t$ on the horizontal domain $x\in(-\infty,\infty)$ and the vertical domain $y\in{[0,h]}$ where $h=h(x,t)$ is the local, instantaneous film thickness.  

The dynamics of this system are decoupled from those of the gas phase by assuming that the ratios between the liquid and gas phase densities, viscosities, and thermal diffusivities are large.  Accordingly, at the free surface, we have the kinematic condition
\begin{gather}
\partial_t{h}+\boldsymbol{v}\cdot\nabla{h}=w\hspace{0.5cm}\textrm{at}\hspace{0.5cm}y=h,
\label{eq:kc}
\end{gather}
which states that the speed of the free surface is equal to the velocity of the fluid.  Using 
\begin{gather}
\thbd=\thbd(x,t)\equiv\mathit{\theta}(x,h,t), 
\label{eq:th}
\end{gather}
to denote the free surface temperature, the normal and tangential stress balances that hold at the free surface are
\begin{align}
p_{\rm g}-p+\tens{T}\cdot\boldsymbol{n}\cdot\boldsymbol{n} = -2{\mathcal H}\sigma(\thbd{})\hspace{0.5cm}&\textrm{at}\hspace{0.5cm}y=h, \label{nstress-dim}\\
\tens{T}\cdot\boldsymbol{n}\cdot\boldsymbol{t} = \boldsymbol{\nabla}\sigma(\thbd{})\cdot\boldsymbol{t}\hspace{0.5cm}&\textrm{at}\hspace{0.5cm}y=h, 
\label{stress-dim}
\end{align}
respectively, with gas pressure $p_{\rm g}$, rate of deformation tensor $\tens{T}=\mu\left[\boldsymbol{\nabla}\boldsymbol{v}+\left(\boldsymbol{\nabla} \boldsymbol{v}\right)^\intercal\right]$ in the liquid phase, surface normal and tangent unit vectors,
\begin{gather}
\boldsymbol{n}=\frac{\boldsymbol{k}-\partial_xh\,\boldsymbol{i}}{\sqrt{1+\left(\partial_xh\right)^2}}\hspace{1cm}\boldsymbol{t}=\frac{\boldsymbol{i}+\partial_xh\,\boldsymbol{k}}{\sqrt{1+\left(\partial_xh\right)^2}},
\label{unitvec}
\end{gather}
and twice mean curvature $2{\mathcal H}=-\boldsymbol{\nabla}\cdot\boldsymbol{n}$.  We consider fluids whose surface tension decreases linearly with temperature according to $\sigma(\thbd{})=\sigma_0-\gamma\left(\thbd{}-\mathit{\theta}_0\right)$ where $\gamma=-d\sigma/d\thbd{}$ is positive and $\mathit{\theta}_0$ is a reference temperature.

Variations in $\theta_{\rm h}$ leading to thermocapillary destabilization are driven by the heat exchanged with the bounding gas phase.  This process is modeled using Newton's Law of Cooling, viz., 
\begin{gather}
{\kappa_{\rm f}}\boldsymbol{\nabla}\mathit{\theta}\cdot{\boldsymbol{n}}+q\left(\thbd{}-{\mathit{\theta}}_g\right)=0\hspace{0.5cm}\textrm{at}\hspace{0.5cm}y=h,
\label{eq:newton}
\end{gather}
where $\theta_{\rm g}$ is the uniform gas temperature and $q$ is the empirical rate of heat transfer between the surface and the gas.  

The film temperature evolves according to (\ref{heateq-dim}), and, at $y=0$, the film is in thermal contact with a rigid substrate of temperature $\psi$, thermal conductivity $\kappa_{\rm s}$, and diffusivity $\chi_{\rm s}$.  No-slip and no-penetration enforce $\boldsymbol{v}=\boldsymbol{0}$, and, continuity of temperature and heat flux require
\begin{align}
\mathit{\theta}=\mathit{\psi}\hspace{0.5cm}&\textrm{at}\hspace{0.5cm}y=z=0, \label{eq:Tcont} \\
\partial_y\mathit{\theta}={\kappa}\,\partial_z{\mathit{\psi}}\hspace{0.5cm}&\textrm{at}\hspace{0.5cm}y=z=0, \label{eq:tsbc} 
\end{align}
where 
\begin{align}
		\kappa=\kappa_{\rm s}/\kappa_{\rm f} \label{eq:kappadef}
\end{align}
is the conductivity ratio. Here, the vertical domain of $\psi$ is assigned to a second vertical coordinate $z\in{}[-d,0]$ in anticipation that two vertical length scales will be introduced in the asymptotic analysis of the thick substrate case. The evolution of $\mathit{\psi}$ throughout the substrate is governed by
\begin{gather}
\partial_t\mathit{\psi}=\chi_{\rm s}\boldsymbol{\nabla}_s^2\mathit{\psi},
\label{eq:thetas}
\end{gather}
where $\boldsymbol{\nabla}_s=\{\partial_x,\partial_z\}$ is defined with respect to $z$ and $\chi_{\rm s}=\kappa_{\rm s}(c_s\rho_s)^{-1}$ is substrate thermal diffusivity.

Opposite the film, we assume the substrate is in perfect thermal contact with a blackbody of uniform temperature $\mathit{\mathit{\psi}}_{\rm b}$  and impose a Dirichlet condition,
\begin{gather}
{\mathit{\psi}}={\mathit{\psi}}_{\rm b}\hspace{0.5cm}\textrm{at}\hspace{0.5cm}z=-d,
\label{eq:fixed2}
\end{gather}  
and define the temperature difference $\Delta\equiv\mathit{\theta_{\rm g}}-\mathit{\psi_{\rm b}}$.  At the cost of introducing a second heat transfer coefficient, a mixed boundary condition accounting for interfacial resistances to heat transfer could also be imposed at $z=-d$.  By assuming instead that the blackbody transfers heat efficiently to the substrate, the parametric burden of the model is lessened.

\section{Dimensionless asymptotic model}
\label{sec:ndim}
In this section we perform a formal asymptotic expansion of the model in Section \ref{sec:dim} that describes the evolution of long wavelength disturbances driven by thermally diffusive substrates. The asymptotic model will be written to depend on four dimensionless quantities,
\begin{gather}\label{eq:dpar}
\Bd=\frac{q \,\hb}{\kappa_{\rm f}}, \hspace{15pt}
\Sd =  \frac{q\, d}{\kappa_{\rm s}}, \hspace{15pt}
\Cd =  \frac{\gamma \,\Delta}{\sigma_0} \frac{\kappa_{\rm s}^2}{\kappa_{\rm f}^2}, \hspace{15pt} 
\Qd =  \frac{q \,\mu\, \chi_{\rm s}}{ \sigma_0 } \frac{\kappa_{\rm s}^2}{\kappa_{\rm f}^3},
\end{gather}
where the Biot numbers $\Bd$ and $\Sd$ measure the thermal thickness of the film and substrate, respectively, $\Cd$ measures the imposed temperature difference, and $\Qd$ measures the combined effects of the film viscosity and substrate thermal diffusivity.  These groups arise in the dimensionless asymptotic model if we choose the characteristic scales
\renewcommand{\arraystretch}{2.15}
\renewcommand{\arraycolsep}{12pt}
\begin{equation}
\left. \begin{aligned}
	\begin{array}{lllll}
		 x' = d, & z' = d,  &  y' = \dfrac{\kappa_{\rm f}}{q},& t' =  \dfrac{d^2}{\chi_{\rm s}},&h' =  \dfrac{\kappa_{\rm f}}{q},  \\
		 u' = \dfrac{x'}{t'} = \dfrac{\chi_{\rm s}}{d},  &  w' =\dfrac{y'}{t'}= \dfrac{\chi_{\rm s} \kappa_{\rm f}}{d^2 q},& p' =  \dfrac{q^2 \,\mu\, \chi_{\rm s} }{\kappa_{\rm f}^2},&	\theta' = \Delta, &   \psi' = \Delta.
	\end{array}
	\label{eq:scales}
\end{aligned} \right\}
\end{equation}
The scales for the film vertical coordinate and its thickness, $y'=h'=\kappa_{\rm f}/q$, are chosen to ensure that $\Bd$ arises as the mean value of the local dimensionless film thickness.  A different vertical scale, $z'=d$, is the natural choice for the substrate, so that the parameter $\Sd$ will enter into the non-dimensional version of the heat flux condition (\ref{eq:tsbc}). Taking $x'=d$ and $t'=d^2/\chi_{\rm s}$ as the lateral length and time scales of the substrate, the lateral film velocity and pressure scales that follow from these choices are as given in (\ref{eq:scales}), which leads to the emergence of the quantities $(\Sd^2\Qd)^{-1}$ and $\Cd\,\Qd^{-1}$ in the dimensionless normal and tangential stress balances, respectively.  Lastly, the film transverse velocity scaling $w'$ is different than $u'$ as a result of the different scaling choices for $x'$ and $y'$, and the (subsequent) nondimensionalization of the continuity equation \eqref{cont-dim}.

To perform the asymptotic expansion, we first define the aspect ratio parameter $\varepsilon\equiv{}y'/x'=(\Sd\kappa)^{-1}$ and require that $\varepsilon\ll{1}$.  Having set $z'=x'$, satisfying $\epsilon\ll{1}$ ensures we consider systems with mean film thicknesses that are small compared to both its lateral variations \emph{and} the substrate thickness.  This definition of $\epsilon$ (i.e., with respect to two system dimensions) contrasts conventional long-wavelength analyses that define $\epsilon$ as the ratio of the film thickness to a characteristic horizontal wavelength.  As a result, in the current problem, $\epsilon$ arises naturally in the model following nondimensionalization with \eqref{eq:scales}.  The formal expansions of the dependent variables take the form
\renewcommand{\arraystretch}{1.2}
\renewcommand{\arraycolsep}{2pt}
\begin{equation}
\left. \begin{aligned}
	\begin{array}{ccrlcllcl}
		u&=& u' & (U_0&+&\varepsilon \,U_1&+\dots+\varepsilon^n \,U_n)&& \\
		w&=& w' & (W_0&+&\varepsilon \,W_1&+\dots+\varepsilon^n \,W_n)&&\\
		p&=& p' & (P_0&+&\varepsilon\, P_1&+\dots+\varepsilon^n \,P_n)&+&p_{\rm g}\\
		h&=& h' & (H_0&+&\varepsilon\, H_1&+\dots+\varepsilon^n \,H_n)&&\\
		\theta&=   & \theta' &(\mathit{\Theta}_0&+&\varepsilon\, \mathit{\Theta}_1&+\dots+\varepsilon^n \,\mathit{\Theta}_n)&+&\theta_{\rm g}\\
		\psi&=  & \psi' & (\mathit{\Psi}_0&+&\varepsilon\, \mathit{\Psi}_1&+\dots+\varepsilon^n \,\mathit{\Psi}_n)&+&\theta_{\rm g}
	\end{array}
	\label{eq:scales2}
\end{aligned} \right\}
\end{equation}
where the variables subscripted with $n=0, 1, ...$ are dimensionless and assumed to be $O(1)$ in magnitude. 

Substituting (\ref{eq:scales2}) into the governing equations (\ref{NSdim}-\ref{eq:fixed2}), we retain the leading order terms, drop the $0$ subscripts on the dependent variables, and obtain dimensionless \emph{long-wavelength}, thick substrate equations and boundary conditions.  Attending first to the film equations of motion, we have, from (\ref{NSdim}), 
\begin{align}
\partial_Y^2U -  \partial_XP = 0\hspace{0.5cm}&\textrm{for}\hspace{0.5cm}Y\in{[0,H]}, \label{NSx-o1} \\
\partial_YP = 0\hspace{0.5cm}&\textrm{for}\hspace{0.5cm}Y\in{[0,H]}, \label{NSz-o1}
\end{align}
a boundary value problem for $U$ and $P$ that is closed by applying the conditions $U=0$ at $Y=0$, and, from (\ref{nstress-dim}) and (\ref{stress-dim}),
\begin{align}
	P = - (\Sd^2\Qd)^{-1}\; \partial_X^2H
	\hspace{0.5cm}&\textrm{at}\hspace{0.5cm}
	{Y}=H, \label{eq:ph} \\
	\partial_YU = -\Cd \Qd^{-1} \; \partial_X \mathit{\Theta}_{\rm H}
	\hspace{0.5cm}&\textrm{at}\hspace{0.5cm}
	{Y}=H, \label{eq:uz}
\end{align}
at the free surface.

This boundary value problem (\ref{NSx-o1})-(\ref{eq:uz}) describes viscous, locally-parallel flows that may be driven by capillary normal stresses or thermocapillary tangential stresses at the free surface.  Because the leading order vertical pressure gradient is equal to zero via (\ref{NSz-o1}), the horizontal pressure gradient appearing in (\ref{NSx-o1}) is independent of $Y$ and is evaluated using the interfacial value specified by (\ref{eq:ph}).  Solution of the boundary value problem for the horizontal velocity $U(Y)$  yields
\begin{gather}
U= (\Qd \,\Sd^2)^{-1} {H}\; \partial_X^3H\left(YH-\frac{1}{2}Y^2\right)-\Cd \,\Qd^{-1} \,Y\partial_X \mathit{\Theta}_{\rm H},
\label{eq:u}
\end{gather}
where $\mathit{\Theta}_{\rm H}(X,T) = \mathit{\Theta}(X,H(X,T),T)$ is the temperature at the free surface. From (\ref{cont-dim}), we use the dimensionless equation for continuity to rewrite the kinematic condition as
\begin{gather}
0=\partial_TH+\partial_{X}\int_{0}^{H} U dY. \label{eq:kc2}
\end{gather}
Evaluating the integral \eqref{eq:kc2} using \eqref{eq:u}, we obtain a nonlinear partial differential equation for the spatiotemporal evolution of $H(X,T)$,
\begin{gather}
\partial_TH+(\Qd \Sd^{2})^{-1} \partial_X\left\{\frac{1}{3}H^3\partial_X^3H-\frac{1}{2}(\Cd \Sd^2)H^2\partial_X\mathit{\Theta}_{\rm H}\right\}=0.
\label{eq:evo}
\end{gather}
This equation represents a standard model for the dynamics of a liquid film subject to capillary stabilization and thermocapillary destabilization.  Via the free surface temperature $\mathit{\Theta}_{\rm H}(X,T)$, (\ref{eq:evo}) is coupled to the long-wave counterparts to equations (\ref{eq:newton})-(\ref{eq:fixed2}), viz.,
\begin{align}
\partial^2_Y \mathit{\Theta}=0\hspace{0.5cm}&\textrm{for}\hspace{0.5cm}Y\in[0,H],\label{eq:Dtf} \\
\partial_T {\mathit{\Psi}}-\partial_X^2\mathit{\Psi}-\partial_{Z}^2\mathit{\Psi}=0 \hspace{0.5cm}&\textrm{for}\hspace{0.5cm}Z\in[-1,0],\label{eq:Dts}
\end{align}
which are subject to
\begin{align}
	\partial_Y\mathit{\Theta}+ \mathit{\Theta}=0
	\hspace{0.5cm}&\textrm{at}\hspace{0.5cm}Y=H, \label{eq:NLC} \\
	\mathit{\Theta}-\mathit{\Psi}=0
	\hspace{0.5cm}&\textrm{at}\hspace{0.5cm}Y=Z=0, \\
	\Sd \, \partial_Y \mathit{\Theta}- \partial_{Z} \mathit{\Psi}=0
	\hspace{0.5cm}&\textrm{at}\hspace{0.5cm}Y=Z=0, \label{eq:sDbc2}\\
	\mathit{\Psi}=1\hspace{0.5cm}&\textrm{at}\hspace{0.5cm}{Z}=-1. \label{eq:sDN2}
\end{align}

To summarize, equations (\ref{eq:evo})-(\ref{eq:sDN2}) represent an asymptotic model that couples, via $\mathit{\Theta}_{\rm H}(X,T)$, a nonlinear partial differential equation for the film thickness $H(X, T)$ to a thermal boundary value problem for temperature profiles $\Theta(X,Y,T)$ and $\Psi(X,Z,T)$ in the film and substrate, respectively.  The model can be recast without $\Theta(X,Y,T)$, given that (\ref{eq:Dtf}) prescribes profiles $\Theta(X,Y,T)$ that are linear in $Y$.  However we find it easier to present the linear analysis that follows by first perturbing the system as written in (\ref{eq:evo})-(\ref{eq:sDN2}).  We also note that the model can be recast to include conventional capillary and Marangoni numbers if (\ref{eq:evo})-(\ref{eq:sDN2}) are instead nondimensionalized with respect to the viscous scales of the film.  However, we find the parameter set $(\Cd, \Qd, \Bd, \Sd)$ is most conducive to a complete presentation of oscillatory instabilities.

\section{Linear analysis}
\label{sec:ls}
In this section we present a linear stability analysis of small perturbations to a steady state solution of (\ref{eq:evo})--(\ref{eq:sDN2}).  The steady state solution, which we will also refer to as the \emph{basic state}, consists of a horizontally uniform (i) flat film of constant height, and (ii) temperature profile that depends linearly on the vertical $Y$ and $Z$-coordinates. Notably, we demonstrate that the resulting linear equations can be cast as a generalized eigenvalue problem that is not self adjoint (in the standard $L^2$ inner product).  The key result of the linear analysis is the determination of the dispersion relation  that characterizes the perturbation growth rate $\Omega$ implicitly in terms of the wavenumber $\beta$.  Approximate and numerical assessment of system stability as governed by the dispersion relation then sets the course of investigation for the remainder of the paper.

To proceed with the linear analysis we introduce a normal-mode perturbation to a steady state solution of (\ref{eq:evo})--(\ref{eq:sDN2}),
\renewcommand{\arraystretch}{1.2}
\begin{equation}
\left. \begin{aligned}
	\begin{array}{@{~}r@{~}c@{~}c@{~}ccl}
		H(X,T) = & \Bd  &+ & \delta\; \hat{H}  \cos{(\beta{X})}\exp(\Omega{T}) &+ & \mathcal{O}(\delta^2) \\
		\mathit{\Theta}_{\rm H}(X,T) = & \thb  & 
		+ & \delta\; \hat{\mathit{\Theta}}_{\rm H} \cos{(\beta{X})}\exp(\Omega{T}) &+ & \mathcal{O}(\delta^2)\\
		\mathit{\Theta}(X,Y,T) = & \bar{\mathit{\Theta}}({Y}) & 
		+ & \delta\; \hat{\mathit{\Theta}}({Y})\cos{(\beta{X})}\exp(\Omega{T}) &+ & \mathcal{O}(\delta^2)\\
		\mathit{\Psi}(X,Z,T) = & \bar{\mathit{\Psi}}({Z}) & 
		+ & \delta\; \hat{\mathit{\Psi}}({Z})\cos{(\beta{X})}\exp(\Omega{T})&+ & \mathcal{O}(\delta^2).
	\end{array}
	\label{eq:Hpert}
\end{aligned} \right\}
\end{equation}
Here $\delta\ll{1}$ is the real amplitude of a horizontally-periodic perturbation of real wavenumber $\beta$ and complex growth rate $\Omega$; while we choose the functions $(\Bd, \thb, \bar{\mathit{\Theta}}({Y}), \bar{\mathit{\Psi}}({Z}))$ to be a steady solution of (\ref{eq:evo})--(\ref{eq:sDN2}). To determine the steady solutions first note that the functions $H(X,T) = \Bd$ and $\mathit{\Theta}_{\rm H}(X,T) =  \thb$ are both constants. The remaining equations (\ref{eq:Dtf})--(\ref{eq:sDN2}) then govern the basic state temperature profiles for $\bar{\mathit{\Theta}}(Y)$ and $\bar{\mathit{\Psi}}(Z)$: 
\begin{align}
\bar{\mathit{\Theta}}'' = 0 \hspace{0.5cm}&\textrm{for}\hspace{0.5cm}Y\in(0,\Bd),\label{eq:tfbzz} \\
\bar{\mathit{\Psi}}'' = 0\hspace{0.5cm}&\textrm{for}\hspace{0.5cm}Z\in(-1,0),\label{eq:tsbzz}
\end{align}
subject to the interface and boundary conditions
\begin{align}\label{eq:ss_ibc1}
\bar{\mathit{\Theta}}'+ \bar{\mathit{\Theta}}=0\hspace{0.5cm}&\textrm{at}\hspace{0.5cm}Y=\Bd,  \\ \label{eq:ss_ibc2}
\bar{\mathit{\Theta}}-\bar{\mathit{\Psi}}=0\hspace{0.5cm}&\textrm{at}\hspace{0.5cm}Y=Z=0, \\ \label{eq:ss_ibc3}
\Sd \, \bar{\mathit{\Theta}}'- \bar{\mathit{\Psi}}'=0\hspace{0.5cm}&\textrm{at}\hspace{0.5cm}Y=Z=0, \\ \label{eq:ss_ibc4}
\bar{\mathit{\Psi}}=1\hspace{0.5cm}&\textrm{at}\hspace{0.5cm}Z=-1.
\end{align}
Solving the linear equations (\ref{eq:tfbzz})--(\ref{eq:ss_ibc4}) yields the complete steady solution:
\begin{align}\label{eq:thetahbar} 
	H = \Bd, \hspace{0.5cm}  \thb  &= (1+ \Bd + \Sd)^{-1},   \hspace{0.5cm} 
	\bar{\mathit{\Theta}}(Y) =
	\frac{1 + \Bd - Y}{1+ \Bd + \Sd},   \hspace{0.5cm} 
	\bar{\mathit{\Psi}}({Z}) =
	\frac{1 + \Bd - \Sd \, Z}{1+ \Bd + \Sd} . 
\end{align}
Note that the value of $\thb$ in (\ref{eq:thetahbar}) is determined from $\bar{\mathit{\Theta}}(Y)$ via $\thb = \bar{\mathit{\Theta}}(\Bd)$, since $\thb$ is defined as the temperature profile $\bar{\mathit{\Theta}}(H)$ at $Y = H = \Bd$. Together equations \eqref{eq:thetahbar} define the film and substrate temperatures of a horizontally-uniform basic state as linear functions of their respective vertical coordinates.

We now move to compute the $O(\delta)$ perturbation about the basic state.  First note that the dependent variable $\mathit{\Theta}_{\rm H}(X,T)$ is just the value of $\mathit{\Theta}(X,Z,T)$ evaluated at the surface $Z = H$, i.e. $\mathit{\Theta}_{\rm H}(X,T) = \mathit{\Theta}(X,H,T)$. Hence, the perturbation variables $\thb$ and $\hat{\mathit{\Theta}}({Y})$ for $\mathit{\Theta}_{\rm H}(X,T)$ and $\mathit{\Theta}(X,Z,T)$ are coupled. To them we (i) Taylor expand $\mathit{\Theta}(X,H,T)$ about the base state value $H = \Bd$ in powers of $\delta$, and (ii) equate the $O(\delta)$ terms in $\mathit{\Theta}(X,H,T)$ with those of $\mathit{\Theta}_{\rm H}(X,T)$.  We then obtain the relation:
\begin{align} \nonumber 
	\hat{\mathit{\Theta}}_{\rm H}&=\hat{\mathit{\Theta}}(\Bd)+\bar{\mathit{\Theta}}'(\Bd)\hat{H},\\ \label{eq:tHE2}
	&=\hat{\mathit{\Theta}}(\Bd)- \thb \hat{H},
\end{align}
where we have used the fact (from \eqref{eq:thetahbar}) that 
$\bar{\mathit{\Theta}}'(\Bd)= -\thb$.
Equation (\ref{eq:tHE2}) will be used to eliminate the variable $\hat{\mathit{\Theta}}_{\rm H}$ from the linear stability analysis.

To obtain the linearized equations about the basic state, we substitute the ansatz (\ref{eq:Hpert}) into the long-wavelength model given by (\ref{eq:evo})--(\ref{eq:sDN2}). Collecting the $O(\delta)$ terms, and using (\ref{eq:tHE2}) to eliminate $\hat{\mathit{\Theta}}_{\rm H}$, gives rise to equations for $\hat{H}$, $\hat{\mathit{\Theta}}(Y)$, and $\hat{\mathit{\Psi}}(Z)$:
\begin{align}
	\hat{\mathit{\Theta}}'' =0\hspace{0.5cm}&\textrm{for}\hspace{0.5cm}Y\in[0,\Bd], \label{eq:tshbzz} \\
	\hat{\mathit{\Psi}}''-\lm^2\,\hat{\mathit{\Psi}}=0\hspace{0.5cm}&\textrm{for}\hspace{0.5cm}Z\in[-1,0].\label{eq:Ths}
\end{align}
In (\ref{eq:Ths}) we have introduced
\begin{align}\label{eq:introlambda}
	\lm^2= \Omega + \beta^2,
\end{align}
which plays the role of a (complex-valued) wavenumber in the $Z$-direction for perturbations confined to the substrate domain.  Note that the sign convention assumed in (\ref{eq:introlambda}) is intentional for the subsequent stability analysis. Equations \eqref{eq:tshbzz}--\eqref{eq:Ths} are also subject to the film dispersion relation
\begin{align}
G_1\,\hat{H}
	+G_2 \, \hat{\mathit{\Theta}}(\Bd)=0\hspace{0.25cm}&\textrm{at}\hspace{0.25cm}Y=\Bd,\label{eq:disp} 
\end{align}
where
\begin{equation}
\begin{aligned}
	\begin{array}{r@{~}lr@{~}l}
		G_1 &= \Qd \Sd^2 (\lm^2-\beta^2) + \frac{1}{3} \Bd^3 \beta^4
	- \frac{1}{2} \Cd \Sd^2 \Bd^2 \,\thb  \beta^2, 
		& \quad G_2&=\frac{1}{2}\Cd \Sd^2 \Bd^2 \beta^2, \nonumber
	\end{array}
\end{aligned}
\end{equation}
and the boundary conditions are
\begin{align}
	\hat{\mathit{\Theta}}'+ \hat{\mathit{\Theta}}(\Bd) - \thb\hat{H} = 0
	\hspace{0.5cm}&\textrm{at}\hspace{0.5cm}Y=\Bd, \label{eq:hfp} \\
	\hat{\mathit{\Theta}}-\hat{\mathit{\Psi}}=0
	\hspace{0.5cm}&\textrm{at}\hspace{0.5cm}Y=Z=0, \label{eq:hfp2} \\
	\Sd\,\hat{\mathit{\Theta}}'-\hat{\mathit{\Psi}}'=0
	\hspace{0.5cm}&\textrm{at}\hspace{0.5cm}Y=Z=0, \label{eq:hfp3} \\
	\hat{\mathit{\Psi}}=0
	\hspace{0.5cm}&\textrm{at}\hspace{0.5cm}Z=-1. \label{eq:eigbc2}
\end{align}
To obtain non-zero solutions to \eqref{eq:tshbzz}--\eqref{eq:eigbc2}, we first recast the system as an eigenvalue problem for $\hat{\mathit{\Psi}}(Z)$ by eliminating the variables $\hat{H}$ and $\hat{\mathit{\Theta}}(Y)$. To first eliminate $\hat{\mathit{\Theta}}(Y)$, we solve equation (\ref{eq:tshbzz}), writing $\hat{\mathit{\Theta}}(Y)=\hat{\mathit{\Theta}}'\,Y$+$\hat{\mathit{\Theta}}(0)$ for constants $\hat{\mathit{\Theta}}'$ and $\hat{\mathit{\Theta}}(0)$.  Inserting the solution for $\hat{\mathit{\Theta}}(Y)$ into the two boundary conditions \eqref{eq:disp}--\eqref{eq:hfp} allows one to solve for the constants $\hat{\mathit{\Theta}}'$ and $\hat{\mathit{\Theta}}(0)$ in terms of $\hat{H}$ only. Writing $\hat{\mathit{\Theta}}(Y)$ in terms of $\hat{H}$, the interface conditions \eqref{eq:hfp2}--\eqref{eq:hfp3} then take the form:
\begin{align} \label{eq:interface_1}
	(1+\Bd)\,\hat{\mathit{\Psi}}'+ \Sd\,\hat{\mathit{\Psi}}
	- \Sd \,\thb\hat{H}
	&=0 
	\hspace{0.5cm}\textrm{at}\hspace{0.25cm}Z=0, \\ \label{eq:interface_2}
	G_1\,\hat{H} +G_2\, (\Bd\, \Sd^{-1} \hat{\mathit{\Psi}}'+ \hat{\mathit{\Psi}}) &=0
	\hspace{0.5cm}\textrm{at}\hspace{0.25cm}Z=0.
\end{align}
The variable $\hat{H}$ can be eliminated in the interface equations \eqref{eq:interface_1}--\eqref{eq:interface_2}, yielding a boundary condition for $\hat{\mathit{\Psi}}(Z)$ at $Z = 0$.  The resulting boundary condition at $Z = 0$, together with the ODE (\ref{eq:Ths}), and boundary condition (\ref{eq:eigbc2}) at $Z = -1$, gives rise to the following problem 
for eigenvalues $\lm^2$ and eigenfunctions $\hat{\mathit{\Psi}}(Z)$:
\begin{equation}\label{eq:eigbc1}
\left. \begin{aligned}
\hat{\mathit{\Psi}}''-\lm^2\,\hat{\mathit{\Psi}}&=0
	\hspace{0.5cm}&\textrm{for}\hspace{0.5cm}&Z\in(-1,0) \\
	a_1{\hat{\mathit{\Psi}}}+a_2\hat{\mathit{\Psi}}'+\lm^2(b_1{\hat{\mathit{\Psi}}}+b_2\hat{\mathit{\Psi}}')&=0	
	\hspace{0.5cm}&\textrm{at}\hspace{0.5cm}&Z=0 \\	
		\hat{\mathit{\Psi}} &= 0
	\hspace{0.5cm}&\textrm{at}\hspace{0.5cm}&Z=-1 
\end{aligned} \right\},
\end{equation}
with real constants
\begin{equation}\label{eq:a1a2}
\begin{aligned}
	\begin{array}{r@{~}lr@{~}l}
		a_1 &= \Sd(\frac{1}{3}\,\Bd^3 \/ \beta^4 - \Qd \/ \Sd^2 \/ \beta^2), 
		& \quad b_1&=\Qd \/ \Sd^3,  \\
		a_2&= (1+\Bd)(\frac{1}{3}\,\Bd^{3} \beta^4-\Qd \,\Sd^2\, \beta^2)
		-\frac{1}{2}\Cd \/ \Sd^2 \/ \Bd^2 \,\thb\beta^2,
		& \quad b_2&= \Qd \Sd^2 \/(1+\Bd). 
	\end{array}
\end{aligned}
\end{equation}
Note that \eqref{eq:introlambda} has been used to replace $\Omega$ in terms of $\lm$ in \eqref{eq:eigbc1}. The problem \eqref{eq:eigbc1} is irregular in the sense that the eigenvalue $\lm^2$ appears in both the boundary condition as well as the domain equation. To solve for the eigenvalues, we write the general solution for $\hat{\Psi}(Z)$ as $\hat{\mathit{\Psi}}(Z)=c_1 \lm^{-1} \sinh{(\lm{Z})}+c_2\cosh{(\lm{Z})}$, and require that it satisfies the two boundary conditions in \eqref{eq:eigbc1}. We include the extra factor of $\lm^{-1}$ in the ansatz so that $\lim_{\lm \rightarrow 0^+} \hat{\mathit{\Psi}}(Z) = c_1 Z + c_2$ solves the ODE \eqref{eq:eigbc1} when $\lm = 0$ (this will then allow for the simultaneous treatment of $\lm = 0$ and $\lm \neq 0$ in the subsequent calculations). Substitution then requires that the following determinant vanish, viz.,
\begin{equation}
	\begin{aligned}
		\begin{vmatrix}
			(a_2+\lm^2\,b_2)&a_1+\lm^2\,b_1\\\ 
			-\lm^{-1} \tanh({\lm})&1\end{vmatrix} = 0.\label{eq:disprel}
	\end{aligned}
\end{equation}
In the equation \eqref{eq:disprel}, the $\lm^{-1} \tanh{\lm}$ term has a removable singularity at $\lm = 0$ (with the limit value of $1$ when $\lm \rightarrow 0$). Equation \eqref{eq:disprel} may then be compactly written as an implicit function relating $\lm$ and $\beta$: 
\begin{align}\label{eq:ImpDisprel}
	\mF(\lm, \beta) = 0, 
\end{align}
where 
\begin{equation}
	\begin{aligned}
		&\mF(\lm, \beta) \equiv{} (a_2 + \lm^2 b_2) +  (a_1 + \lm^2 b_1) \frac{\tanh{\lm}}{\lm}.  \label{eq:dispF}
	\end{aligned}
\end{equation}
Any solution $(\lm, \beta)$ to equation \eqref{eq:disprel}, or equivalently \eqref{eq:ImpDisprel}, then determines $\Omega$ via equation \eqref{eq:introlambda}.  As a result, equation \eqref{eq:ImpDisprel} defines an (implicit) dispersion relation since it describes the values of $\lm$ (and hence $\Omega$), in terms of $\beta$, for which non-zero solutions $\hat{\Psi}(Z)$ exist.  We will therefore refer to $\mF(\lm, \beta) = 0$ as the \emph{dispersion relation}.  For values of $\lm \neq 0$ and $a_1+\lm^2b_1 \neq 0$, the dispersion relation $\mF(\lm, \beta) = 0$ can be recast into a form that is more commonly encountered in linear stability analyses of thin film models,
\begin{equation}
	\begin{aligned}
		&\Qd\, \Sd^{2} \Omega+\frac{1}{3}\Bd^{3} \beta^4  
		- \frac{\frac{1}{2}\Cd \,\Sd^{2} \Bd^{2} \,
			\thb\,\beta^2\,\sqrt{ \Omega +\beta^2 }} {\Sd\,
			\tanh{\left( \sqrt{\Omega +\beta^2 }\right)}
			+(1+\Bd)\,\sqrt{\Omega +\beta^2}
			} = 0.
			\label{eq:dispo}
	\end{aligned}
\end{equation}
In Appendix \S\ref{appB}, this form is used to readily obtain a dispersion relation for thin substrates. The third term in (\ref{eq:dispo}) describes thermocapillarity and we note that it vanishes in situations that render the free surface isothermal (this occurs if $\Cd=0$). 

To conclude the solution of the linearized system, we solve for the substrate temperature eigenfunction $\hat{\Psi}({Z})$, and the film temperature eigenfunction $\hat{\mathit{\Theta}}({Z})$ in terms of the film perturbation amplitude $\hat{H}$.  For a fixed $\beta$, take $\lm$ as a root of the dispersion relation (\ref{eq:dispF}) and fix $\Omega$ via \eqref{eq:introlambda}.  Then the ansatz \eqref{eq:Hpert} solves the linearized equations, with eigenfunction profiles given by:
\begin{align}\label{eq:Thh}
\hat{\mathit{\Theta}}_{\rm H}&=-\frac{\thb\,\lm}{\Sd\tanh{\lm}+(1+\Bd)\,\lm} \; \hat{H}, \\ \label{eq:Tsh}
\hat{\mathit{\Theta}}({Y})&=\phantom{-}\frac{\thb\left[\Sd\tanh{\lm}+\lm\,{Y}\right]}{\Sd\tanh{(\lm)}+(1+\Bd)\,\lm} \; \hat{H}, \\ \label{eq:Tfh}
\hat{\mathit{\Psi}}({Z})&=\phantom{-}\frac{\Sd\,\thb\left[\sinh{(\lm{Z})}+\tanh{(\lm)}\cosh{(\lm{Z})}\right]}{\Sd\tanh{(\lm)}+(1+\Bd)\,\lm} \; \hat{H}.
\end{align}

\section{Root structure of the dispersion relation}
\label{sec:DispRel}
In this section, we compute the growth factors $\Omega(\beta)=\lm(\beta)^2-\beta^2$ by solving the dispersion relation $\mF(\lm, \beta) = 0$ for $\lm$ in terms of $\beta$, and applying (\ref{eq:introlambda}).  The growth factors $\Omega(\beta)$ are important as they dictate the stability of the basic steady state solution, and can be used to investigate the physical regimes having (qualitatively) different linear instabilities.  

Closed form solutions for the implicit functions $\lm(\beta)$ (and hence $\Omega(\beta)$) satisfying the dispersion relation $\mF(\lm, \beta) = 0$ cannot be determined and must instead be investigated numerically.  Thus, in the work that follows, we adopt a continuation method (see \citet{Boyd14}) and use $\beta \in [0, \infty)$ as the continuation parameter.  Starting with the value $\beta = 0$ and $\lm(0)$, we will track the implicit solutions $\lm(\beta)$ to the dispersion relation (note that there are infinitely many) as continuous functions of $\beta$.  For notational purposes we will refer to the solutions $\lm(\beta)$ as \emph{roots} to the dispersion relation. In addition, we will compute the asymptotic behavior of the functions $\lm(\beta)$ for both small and large $\beta$.  Together, the asymptotic calculations and numerical continuation will provide a comprehensive picture of the values $\Omega(\beta)$, for any given set of physical parameters $(\Bd, \Sd, \Qd, \Cd)$.  This will then enable an investigation into the different physical behaviors captured by the model. 

\subsection{The continuation method}\label{sec:numerical}
We first remark on the symmetries of the dispersion relation, which will help to simplify the computation of the implicit functions $\lm(\beta)$.  Note that for any fixed $\beta$, the dispersion relation satisfies:
\begin{equation}\label{symmetries}
\begin{aligned}	
	\textrm{(Even symmetry)}  			 \quad \quad	\mF{(\lm,\beta)} &= \mF{(-\lm,\beta)}, \\
	\textrm{(Conjugation symmetry)} \quad \quad  \mF{(\overline{\lm},\beta)} &= \overline{\mF}{( \lm,\beta)}.
\end{aligned}
\end{equation}
With the above symmetries in mind, the continuation method may be restricted to the first quadrant of the $\lm$-complex plane (equivalently, to the upper half of the $\Omega$-complex plane).  Solutions $\lm(\beta)$ may then be extended to the remaining three quadrants by symmetry. 

We initialize the continuation method at $\beta = 0$, for which the roots $\lm(0)$ satisfy:
\begin{align}\label{eq:ContStart}
	f(\lm(0), 0) = 0	 \quad \Longrightarrow \quad
	\lm(0)\, \big( \tanh \lm(0) + \mS\, \lm(0) ) = 0,
\end{align}
and $\mS = \Sd^{-1}(1+\Bd) $ is a positive constant.  The initialization value $\beta = 0$ is useful as we may enumerate exactly all of the roots to equation \eqref{eq:ContStart} as follows.  

First, observe that all nonzero roots to equation \eqref{eq:ContStart} are purely imaginary. To show this, it is sufficient to write $\lm(0) = \xi + \iiii \zeta$ for real values $\xi, \zeta$ and verify that there are no solutions to \eqref{eq:ContStart} for values $\xi > 0$ and $\zeta \geq 0$ (by symmetry we may restrict to the first quadrant).  Equating the real and imaginary parts of \eqref{eq:ContStart}, the values $(\xi, \zeta)$ must satisfy the simultaneous equations:
\begin{align}\label{eq:realEq}
	\big(\sinh \xi  + \Sd\,\xi \cosh \xi \big) \cos \zeta &= \phantom{-}\mS \,\zeta \sinh \xi \sin \zeta, \\ \label{eq:imagEq}
	\big(\cosh \xi  + \Sd\,\xi \sinh \xi \big) \sin \zeta &= -\mS \,\zeta \cosh \xi \cos \zeta.
\end{align}
Note that, since $\sinh \xi  + \Sd \,\xi \cosh \xi > 0$ (similarly $\cosh \xi  + \Sd\,\xi \sinh \xi > 0$) equations \eqref{eq:realEq}--\eqref{eq:imagEq} imply that if $\sin \zeta=0 $ then $\cos \zeta = 0$ (or if $\cos \zeta =0 $ then $\sin \zeta  = 0$) --- which is not possible. Hence, $\zeta$ cannot satisfy $\sin \zeta  = 0$, or $\cos \zeta  = 0$, and we are free to divide \eqref{eq:realEq} by \eqref{eq:imagEq} to obtain the following (necessary) equation for a root:
\begin{align}\label{eq:necEq}
	\Big( 1 + \xi \,\mS \; \textrm{coth}\,\xi \Big) \Big(1 + \xi \,\mS \tanh \xi \Big) = - \mS^2 \zeta^2.
\end{align}
Equation \eqref{eq:necEq}, however, has no solutions for $\xi > 0$ since the left-hand side is (strictly) positive and the right-hand side is non-positive.  Hence, $\xi = 0$, which shows that the roots to \eqref{eq:ContStart} must have the form $\lm(0) = \iiii \zeta$.  

We can now enumerate the roots of \eqref{eq:ContStart} as $\lm_{\pm n}(0) = \pm \iiii \zeta_n$, where $0 = \zeta_0 < \zeta_1 < \ldots$, and the values of $\zeta_n$ are the non-negative solutions to the equation
\begin{equation}
	\begin{aligned}
		g(\zeta_n) = 0, \quad \textrm{where} \quad 
		&\mG(\zeta) \equiv{}\tan{\zeta}+\mS\,{\zeta} \quad \mbox{and}\quad
		\mS =(1+\Bd)/\Sd . \label{eq:dispG}
	\end{aligned}
\end{equation}
Note that in equation \eqref{eq:ContStart}, the value $\lm_{\pm 0}(0) = 0$ is a double root, and can be understood by considering $\lm_{+0}(0)$ and $\lm_{-0}(0)$ as two distinct roots. With this convention, writing $\lm_{\pm n}(0) = \pm \iiii \zeta_n$ then captures all roots of \eqref{eq:ContStart}, including multiplicity.
\begin{figure}
\begin{centering}
\includegraphics[width=40pc]{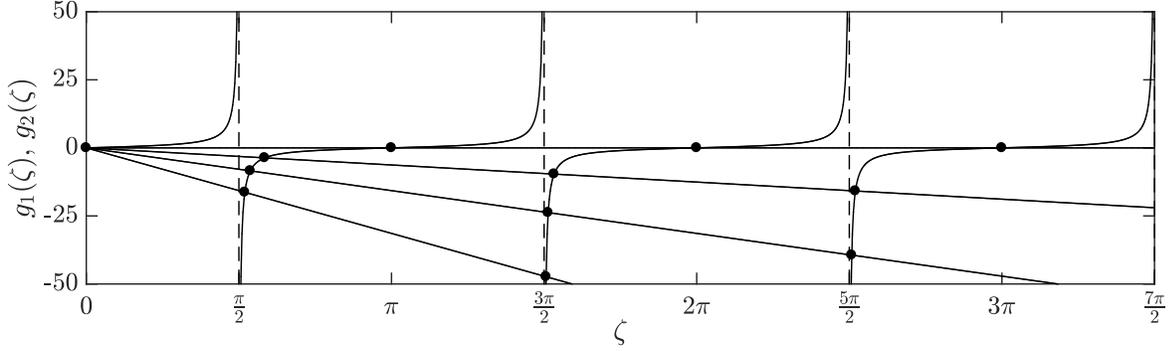}
\caption{Graphical depiction of the intersections (circles) of $g_1(\zeta) = \tan{\zeta}$ and $g_2(\zeta) = -\mS\,\zeta$ (where $g(\zeta)$ is defined in \eqref{eq:dispG})  corresponding to the first three substrate ($n=1, 2, 3$) roots $\lm=i\zeta_{n}$ (circles) that satisfy $g(\zeta_n)=0$ for $\mS=\{0,2,5,10\}$.}
\label{fig:1tanzd}
\end{centering}
\end{figure}

The roots $\zeta_n$ are presented graphically in figure \ref{fig:1tanzd} as the intersections of the functions $g_1(\zeta) = \tan{\zeta}$ and $g_2(\zeta) = -\mS\,\zeta$. In the limit of small $\mS \ll 1$ (resp. large $\mS \gg 1$), the roots $\zeta_n$ asymptotically approach the zeros of $\cos{\zeta_n}$ (resp. $\sin{\zeta_n}$). In the asymptotic limit $n \rightarrow \infty$, the roots $\zeta_n \rightarrow n \pi - \pi/2$.

We now restrict attention to the roots $\lm_{+n}(0)$, $n\geq 0$, initialized to the upper-half plane (and for brevity drop the $+$ in the subscript of $\lm_{+n}(\beta)$), since the remaining roots are negatives by symmetry of \eqref{symmetries}. With the roots $\lm(\beta)$ of (\ref{eq:dispF}) initialized to $\lm_n(0)=\iiii \zeta_n$, we continuously vary $\beta \in [0, \infty)$ and track the roots $\lm_n(\beta)$ as functions of $\beta$. In our subsequent linear stability analysis, the $\lm_0(\beta)$ root (initialized to $\lm_0(0) = 0$) will play a particularly important role.  As a result, we will refer to $\lm_0(\beta)$ as the \emph{film root}, and (from now on) write $\lmf(\beta)$. The phrase ``film root'' is motivated by the fact that the corresponding complex frequency $\Omf(\beta) = \lmf(\beta)^2 - \beta^2$ is analogous to the frequencies $\Omega(\beta)$ given by a free thin film equation (see for instance \S\ref{appB}).  The remaining roots are initialized to $\lm_{n}(0) = i\zeta_n$ for $n = 1, 2, \ldots$.

As a technical point, we stress that the roots $\lmf(\beta)$ or $\lm_n(\beta)$ are only (uniquely) identifiable by their initial values $i\zeta_n$ within an interval $0 \leq \beta \leq \betaL$ for which no collision has occurred. Once a collision occurs, i.e. two (or more) roots collide at a value $\beta = \betaL$, it is generally not possible to identify uniquely two (or more) post-collision roots $\lm(\beta)$ at values $\beta > \betaL$ with their initial values $i\zeta_n$. 

As a final remark on the numerical computations, we follow a standard continuation approach: at each step, using the value $\lm(\beta)$ as an initial guess, we use Newton's method to compute $\lm(\beta + \Delta \beta)$, where $\Delta \beta$ is the increment (chosen adaptively to ensure convergence at each step).  As a practical detail, to enable the method to find complex valued solutions, we initialize the Newton algorithm with a value that does not lie strictly on either the real or complex axis by perturbing the initial guess via $\lm(\beta) + \epsilon (1+\iiii) $, for $\epsilon \ll 1$.  This is to avoid having Newton iterates become trapped to the (invariant) real or complex axis. 

\subsection{Asymptotic behavior of the roots for small $\beta$}
The previous section demonstrates that the values $\lm_n(0) = \iiii \zeta_n$ are purely imaginary.  Hence, at $\beta = 0$, the growth rates $\Omega$ lie along the negative real axis: $\Omf(0) = 0$ (corresponding to the film root $\lmf(0) = 0$), and $\Omega_n(0) = -\zeta_n^2 < 0$ (for the roots $\lm_n(0) = \iiii \zeta_n$, $n \neq 0$); see (\ref{eq:introlambda}). The purpose of this section is to examine how the values $\Omega_n(\beta)$ and $\Omf(\beta)$ change for small values of $0 \leq \beta \ll 1$.

We first compute the small $\beta$ behavior of the film root $\lmf(\beta)$ and corresponding growth rate $\Omf(\beta)$. This can be done by expanding \eqref{eq:dispF} in powers of $\lm$ (about $\lmf(0) = 0$) to obtain:
\begin{align}\label{eq:FilmEx}
	f(\lm, \beta) = (a_1 + a_2) + (b_1 + b_2 - \frac{1}{3}a_1) \lm^2 + \mathcal{O}(\lm^4),
\end{align}
where, recall, $a_1, a_2, b_1, b_2$ depend on $\beta$ via (\ref{eq:a1a2}).  Truncating the expansion \eqref{eq:FilmEx} at $\mathcal{O}(\lm^4)$, and setting $f(\lm, \beta)$ to zero, yields an approximate solution for $\lmf(\beta)$, valid at small $\beta$:
\begin{equation}
\begin{aligned}\label{eq:filmapprox}
	\lmf(\beta) \approx \pm \sqrt{ \frac{a_1 + a_2}{b_1 + b_2 - \frac{1}{3}a_1}} = \eta_1 \beta - \eta_3 \beta^3 + \cdots .
\end{aligned}
\end{equation}
Here, the Taylor coefficients $\eta_1$ and $\eta_3$ are 
\begin{equation}
\begin{aligned}
\eta_1= \sqrt{1+\frac{\Cd\,{\Bd^2}\,\thb^2}{2\,\Qd }}, 
\quad \quad \textrm{and} \quad \quad
\eta_3=\frac{1}{6}\left[\Sd\thb\eta_1+\frac{\Bd^3}{\Qd\,\Sd^2\,\eta_1}\right],
\label{eq:zf}
\end{aligned}
\end{equation}
respectively, where $\eta_3$ has been written compactly using the definition of $\eta_1$.  The above calculation shows that the double root at $\lmf(0)=0$ splits immediately into two real nonzero roots given (approximately) by (\ref{eq:filmapprox}). As a convention, we use $\lmf(\beta)$ to denote the positive branch in \eqref{eq:filmapprox}.  Via (\ref{eq:introlambda}) we then have the small-wavenumber expansion of the corresponding growth rate $\Omf(\beta)$, viz.,
\begin{equation}
\begin{aligned}
\Omf(\beta)&=(\eta_1^2-1)\,\beta^2-2\,\eta_1\,\eta_3\,\beta^4+O(\beta^6)\\
&\approx\frac{\Cd\,\Bd^2\,\thb^2}{2\,\Qd}\beta^2-\left[\frac{\Bd^3}{3\,\Sd^2\,\Qd}+\Sd\,\thb\left(\frac{1}{3}+\frac{\Cd\,\Bd^2\,\thb^2}{6\,\Qd}\right)\right]\beta^4.\label{eq:wf}
\end{aligned}
\end{equation}
Here, the first term inside the square brackets describes capillary stabilization, and the other terms including $\thb$ pertain to thermocapillary effects.  Notably, this expression predicts that thermocapillarity acts both to destabilize small wavenumbers and stabilize large ones, in contrast to the strict thermocapillary destabilization observed for thin substrates (see Appendix \ref{appB}).  In particular, (\ref{eq:wf}) shows that the relative importance of thermocapillary stabilization will increase for large $\Sd$ (substrate thickness), large $\Cd$ (imposed temperature difference),  or small $\Qd$ (diffusive effects).  
\begin{figure}
\begin{centering}
\includegraphics[width=16pc]{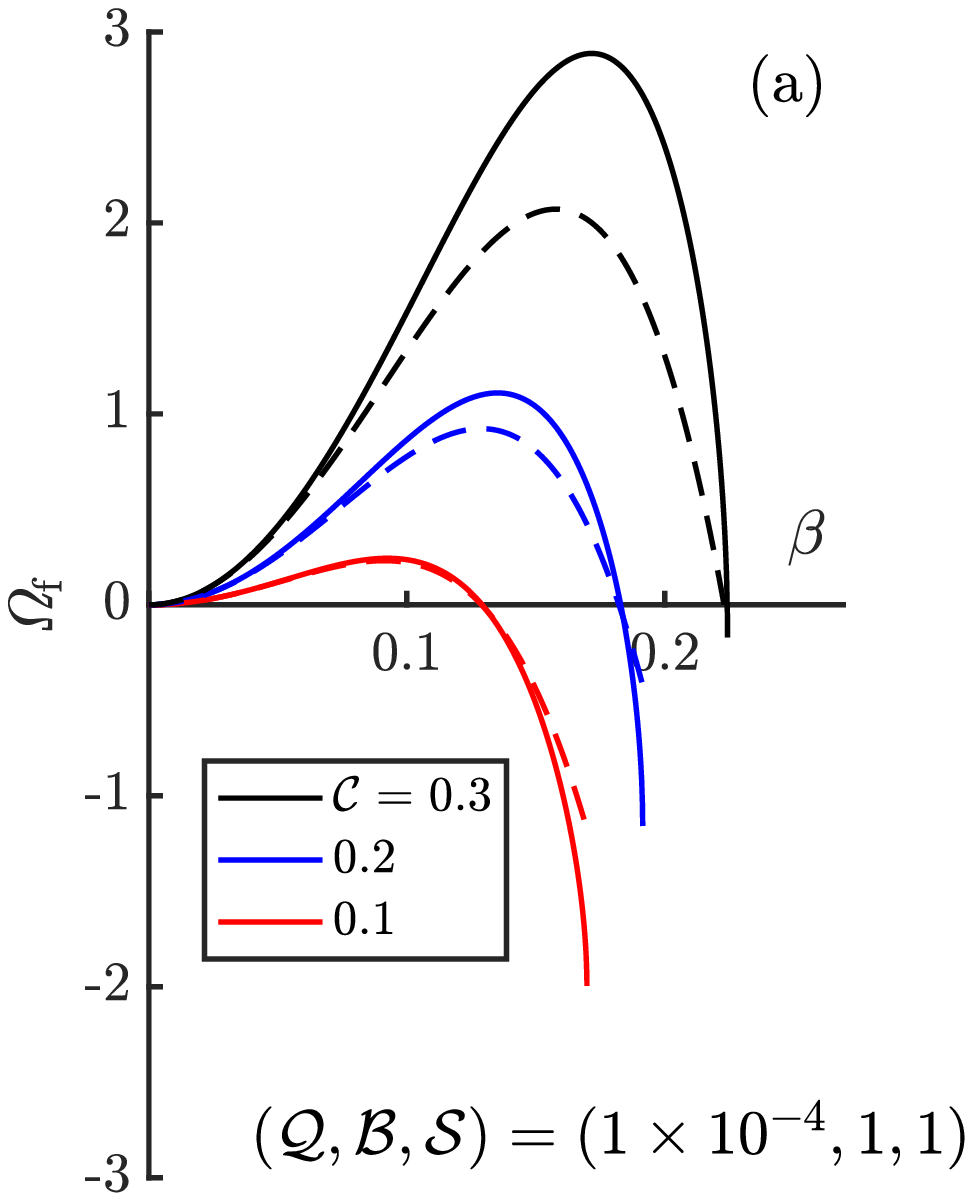}\hspace{2pc}\includegraphics[width=16pc]{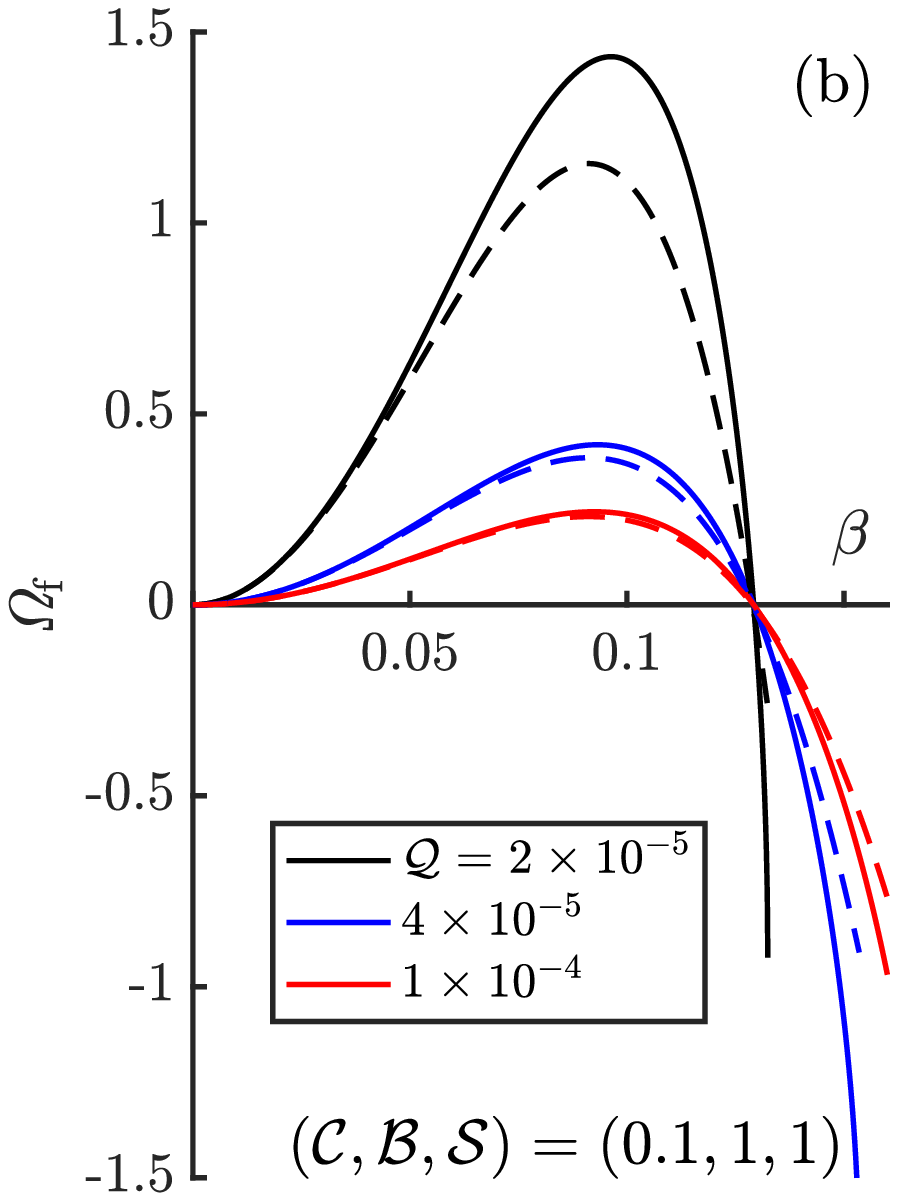}\\
\vspace{2pc}
\includegraphics[width=16pc]{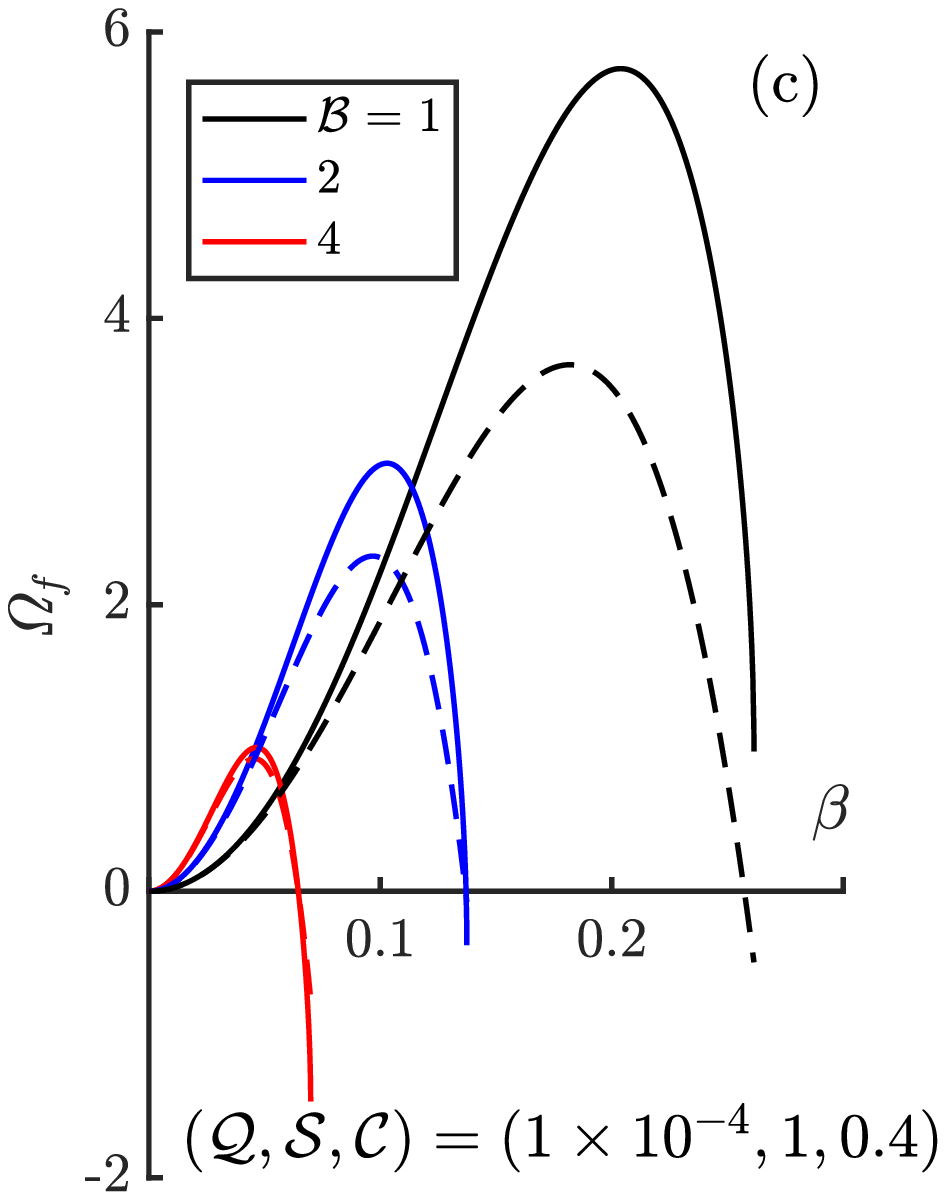}\hspace{2pc}\includegraphics[width=16pc]{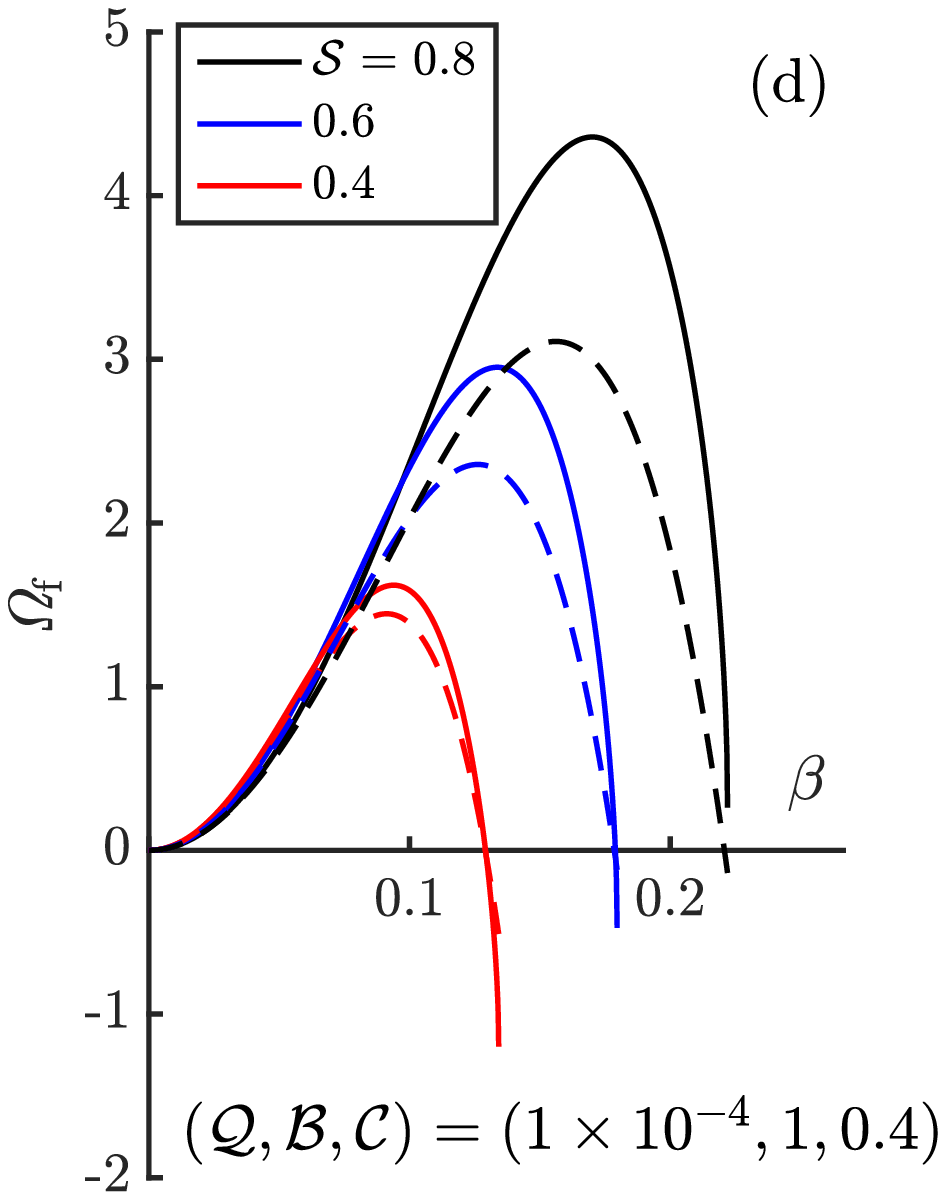}
\caption{Comparison between the numerically-calculated (solid lines, equation (\ref{eq:dispF})) and approximated (dashed lines, equation (\ref{eq:wf})) film root growth rates $\Omf(\beta)$ for varying (a) $\mathcal{C}$, (b) $\mathcal{\Qd}$, (c) $\mathcal{\Bd}$, and (d) $\mathcal{\Sd}$.}
\label{fig:2Opprox}
\end{centering}
\end{figure}

In figure \ref{fig:2Opprox}, the small-$\beta$ approximation (\ref{eq:wf}) for $\Omf(\beta)$ is compared to exact root branches calculated via numerical continuation.  In each panel, one comparison is made to demonstrate a parameter set for which the agreement is qualitatively good;  both the maximum growth rates and the cutoff wavenumbers are adequately predicted by (\ref{eq:wf}).  The surprisingly good prediction for the cutoff wavenumber is due to the original expansion in \eqref{eq:FilmEx} being based on small $\lm$: since $\lm^2=\Omega +\beta^2$, we expect the resulting approximation (\ref{eq:wf}) to be good near the origin $\beta=\Omega=0$ and at the cutoff where $\beta^2\ll 1$ still and $\Omega=0$. In the region between $\beta=0$ and the $O(1)$ cutoff values of the growthrate $\Omega$, the small-$\lm$ assumption made to obtain \eqref{eq:FilmEx} is certainly violated. This is reflected in the poor agreement observed in this region between the asymptotic approximation \eqref{eq:wf} and the numerically-calculated curves in figure \ref{fig:2Opprox}.

Each branch in figure \ref{fig:2Opprox} is presented from $\beta=0$ to a critical value of $\betaL$, beyond which the real film root ceases to exist ($\betaL$ will be defined more precisely below). Before focusing on these critical points in the next section, we note that several key physical behaviors can be inferred from figure \ref{fig:2Opprox}.  First, we see in panel (a) that larger imposed temperature differences (larger $\Cd$) increase both the unstable band of wavenumbers and their associated growth rates.  Panel (b) then shows that diffusive effects, as measured by $\Qd$, suppress the growth rates of instability without significantly changing the bandwidth of unstable wavenumbers.  The loss of agreement in panel (c) occurs because, thinner films, via small values of $\Bd$, promote the relative importance of the substrate thermal process.  

Panel (d) shows the dependence on $\Omf(\beta)$ on the parameter $\Sd$. We note that the increase in the growth rates with $\Sd$ shown in panel (d) results from having scaled time with respect to $d^2$, see (\ref{eq:scales}).  The (approximately linear) increase of $\Omf$ with $\Sd$ shown here in fact corresponds to a linear \emph{decrease} in the dimensional growth rates--due primarily to the increased thermal resistance of thicker substrates.

\subsection{The roots for $n \neq 0$}
To examine the roots $\lm_n(\beta)$ and $\Omega_n(\beta)$ with $n \neq 0$ in the small-$\beta$ limit, we substitute the power series,
\begin{equation}
\begin{aligned}
\lm_n(\beta)= \iiii (\zeta_{n} - \beta^2 \lbeta)+O(\beta^4),
\label{eq:subapproxsmall}
\end{aligned}
\end{equation}
into the dispersion relation with $\zeta_n$ as given by (\ref{eq:dispG}) (see also figure \ref{fig:1tanzd}) and unknown coefficients $\lbeta$ (which will turn out to be real).  Such an expansion in even powers of $\beta$ is justified by the symmetry relations \eqref{symmetries} and by the fact that the values $\lm_n(\beta)$ are simple near $\beta =0$. Note that this is in contrast to the roots $\pm\lmf(\beta)$ that emerge from the double root $\lmf (0)$, which does not have an even power series at $\beta = 0$ (double roots generally \emph{split} via a square root dependence on the continuation parameter).

Expanding $f(\lm_n(\beta), \beta) = 0$ (see \eqref{eq:dispF}) for small $\beta$, and setting the $O(\beta^2)$ term to zero yields
\begin{equation}
\begin{aligned}
	\lbeta= \frac{\Cd\,\Bd\,\Sd\,\thb^2}{2\,\Qd\,\zeta_{n}\,(\Sd+(1+{\Bd})^2\thb{}\,\zeta_n^2)}.
\label{eq:z20}
\end{aligned}
\end{equation}
In this expression, $\tan{\zeta_n }$ has been replaced with $-\mS \,\zeta_n$ per (\ref{eq:dispG}) (recall that $\mS = \Sd^{-1}(1+\Bd) $).
Substituting the expressions \eqref{eq:subapproxsmall}--\eqref{eq:z20} into (\ref{eq:introlambda}) yields
\begin{equation} \label{eq:ws}
	\Omega_n(\beta) = -\zeta_n^2 + \beta^2 \Big( 2 \,\zeta_n\, \lbeta - 1 \Big)
	+ O(\beta^4).
\end{equation}
At zero wavenumber, all these roots are real and negative: $\Omega_n(0)=-\zeta_n^2$. As $\beta$ increases from zero, if $2\, \zeta_n \,\lbeta > 1$, then the roots (initially) move along the negative real axis in the complex plane towards the right-half plane (RHP) (we will also use LHP to denote the left-half plane). Alternatively if $2 \,\zeta_n\,\lbeta < 1$, then the roots move to the left along the negative real axis (see figure \ref{fig:PDcp}). 

We now remark that the product $\zeta_n \,\lbeta > 0$ is (i) always positive, (ii) monotonically decreases with increasing values of $n$ (due to the fact that the values $\zeta_n$ monotonically increase with $n$), and (iii) $\lim_{n\rightarrow \infty} \zeta_n\, \lbeta \rightarrow 0$.  As a result, only a finite number of the values $\Omega_n(\beta)$ (closest to the origin) can have a positive $O(\beta^2)$ coefficient, and hence initially move towards the unstable RHP.  All other roots move (at small $\beta$) farther into the LHP.  Figure~\ref{fig:PDcp} demonstrates, via the continuation method, the motion of the roots $\{\Omf(\beta), \Omega_n(\beta)\}$, with varying $\beta$. The plot shows the film root ($\Omf(0)=0$) and $\Omega_n(0)=-\zeta_n^2$ roots at $\beta=0$. The arrows for the markers denote the numerically computed directions in which the roots move as $\beta > 0$ increases. The arrows also coincide with the small-$\beta$ approximations \eqref{eq:wf} and \eqref{eq:ws}.  Note that the figure also demonstrates that the roots become complex only after a collision, and generally move to the left (becoming more stable) with increasing $\beta$.

\section{Oscillatory instability classification}
\label{sec:charcoal}
In this section, we investigate how the frequencies $\Omf(\beta)$ and $\Omega_n(\beta)$ move in the complex plane (as functions of the continuation parameter $\beta$), and lead to oscillating in time solutions of the linearized equations \eqref{eq:disp}--\eqref{eq:eigbc2} with exponentially growing amplitudes. A value $\Omega$ (that satisfies the dispersion relation) is \emph{oscillatory unstable} if $\Omega$ lies in the strict RHP, and does not lie along the real axis. That is, $\Omega$ satisfies:
\begin{align}\label{eq:oscon}
	\textrm{(i)}  \quad \textrm{Re}(\Omega) > 0, \quad\quad\quad
	\textrm{(ii)} \quad \textrm{Im}(\Omega) \neq 0.
\end{align}
Combined with the symmetry observations from the previous section (\S\ref{sec:DispRel}), conditions (\ref{eq:oscon}) place restrictions on how exactly a root $\Omega(\beta)$ can become oscillatory unstable as the wavenumber $\beta$ increases from $0$.  

First, condition (ii) in \eqref{eq:oscon} requires that (as $\beta$ varies) two frequencies $\Omega(\beta)$ must collide at some value of $\beta$ --- that is, there is a value of $\beta$ for which two of the frequencies $\{ \Omf(\beta), \Omega_n(\beta)\}$ are equal. This is because the complex frequencies $\{\Omf(0), \Omega_n(0)\}$ are simple (at $\beta = 0$), move continuously with $\beta$, and cannot leave the real axis as long as they remain simple (due to conjugate symmetry of $\Omega(\beta)$, see \eqref{symmetries}). Hence, a necessary condition for (ii) is a collision (double frequency) at some $\beta$. 

Second, the results from section \ref{sec:DispRel} show that the frequencies $\{\Omf(0), \Omega_n(0)\}$ are on the non-positive real axis, and only a finite number of the largest roots initially move towards the RHP (as $\beta$ increases).  Hence, the two largest roots $\Omf(\beta)$ (film root) and $\Omega_1(\beta)$ are the most likely candidates to collide and satisfy \eqref{eq:oscon}.  In other words, oscillatory instabilities most likely arise from the motion and collisions of $\Omf(\beta)$ and $\Omega_1(\beta)$.

Treating the continuation parameter $\beta$ as a bifurcation parameter (holding the parameters $(\Cd,\Qd,\Bd,\Sd)$ fixed), we now classify how exactly $\Omf(\beta)$ and $\Omega_1(\beta)$ collide (bifurcate) and satisfy \eqref{eq:oscon} (lead to oscillatory instabilities).  We define the (first) collision of $\{\Omf(\beta), \Omega_1(\beta)\}$ to occur at the value $\betaL$, and denote 
\begin{align}\label{Def:OmegaCollision}
	\OmL\equiv\lim_{\beta \rightarrow \betaL^-} \Omf(\beta) = \lim_{\beta \rightarrow \betaL^-}  \Omega_1(\beta).
\end{align}
Note that the values $(\betaL, \OmL)$ are readily identifiable: in addition to the dispersion relation $\OmL = \lmL^2 - \betaL^2$, with $\mF(\lmL, \betaL) = 0$, the values $(\betaL, \OmL)$ satisfy the condition of a double root required by the implicit function theorem: $\partial_{\lm} \mF(\lmL, \betaL) = 0$.  For values of $\beta\in[0,\betaL]$, the frequencies $\Omf(\beta), \Omega_1(\beta) \in \mathbb{R}$; however, for $\beta>\betaL$ (after the collision), the roots appear as complex conjugate pairs that we denote as $\Omfone(\beta)=\Omr(\beta) + i\Omi(\beta)$ and $\Omfone^*(\beta)$, with $\Omr(\beta), \Omi(\beta) \in \mathbb{R}$.  We further denote $\betacn > \betaL$ (if it exists) as the first value at which the frequency $\Omfone(\beta)$ crosses the imaginary axis, i.e. $\Omr(\betacn) = 0$.  Summarizing the notation, we have:
\begin{equation} \label{eq:Notation}
\left. \begin{aligned}
	\begin{array}{lrrll}
		\textrm{(Pre-collision)} & 0 \leq \beta < \betaL & \; &\Longrightarrow \; & \Omega_1(\beta ) < \Omf(\beta), \textrm{ and } \Omf(\beta), \Omega_1(\beta) \in \mathbb{R}, \\
		\textrm{(At collision)} & \beta = \betaL & \; &\Longrightarrow \; & 
		\OmL \equiv \OmL(\beta) = \Omf(\beta), \textrm{ and } \OmL \in \mathbb{R},\\
		\textrm{(Post-collision)} & \beta > \betaL & \; &\Longrightarrow \; &\Omfone(\beta)=\Omr(\beta)+ i\Omi(\beta), \textrm{and }\Omfone^*(\beta), \\		
		\textrm{(Imag. axis)} & \beta = \betacn & \; &\Longrightarrow \; &\Omr(\betacn)=0,
		\textrm{   (}\betacn \geq \betaL \textrm{ may not exist)}.	
	\end{array}
\end{aligned}\right\}
\end{equation}
We now identify two characteristic ways for the roots $\Omf(\beta), \Omega_1(\beta)$ to satisfy conditions \eqref{eq:oscon} and give rise to oscillatory instabilities.
\begin{figure}
\begin{centering}
\includegraphics[width=15pc]{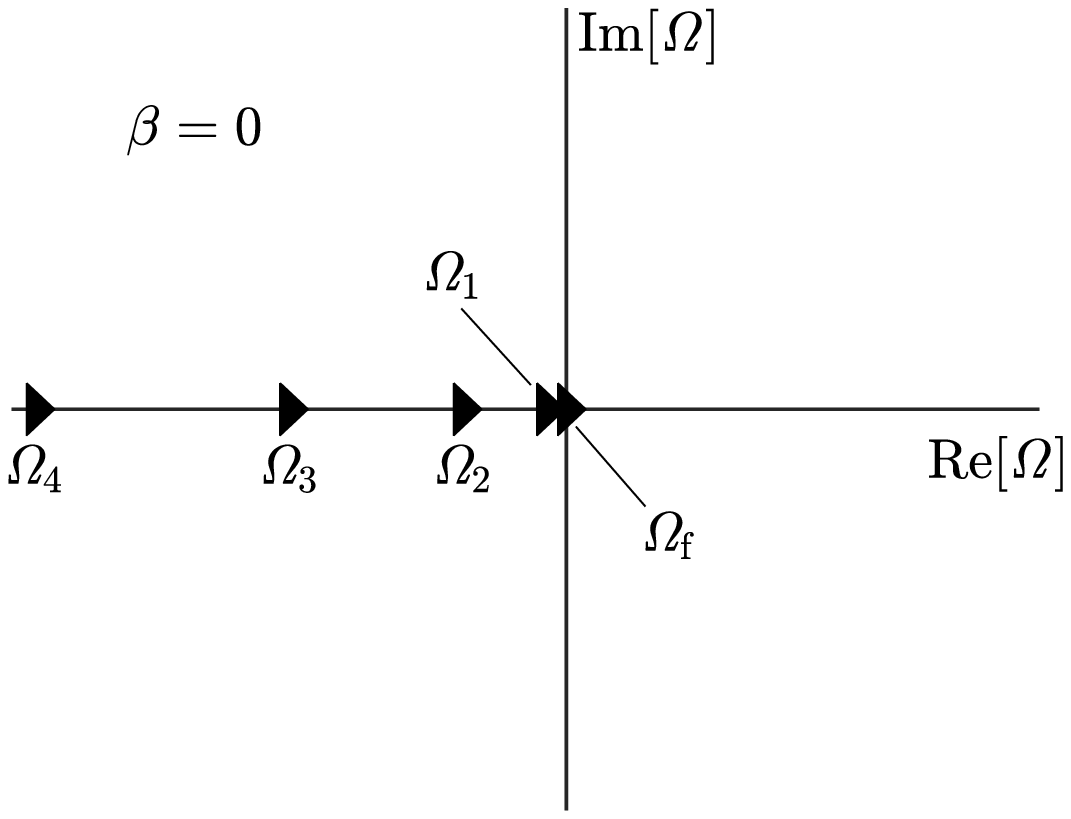}\includegraphics[width=15pc]{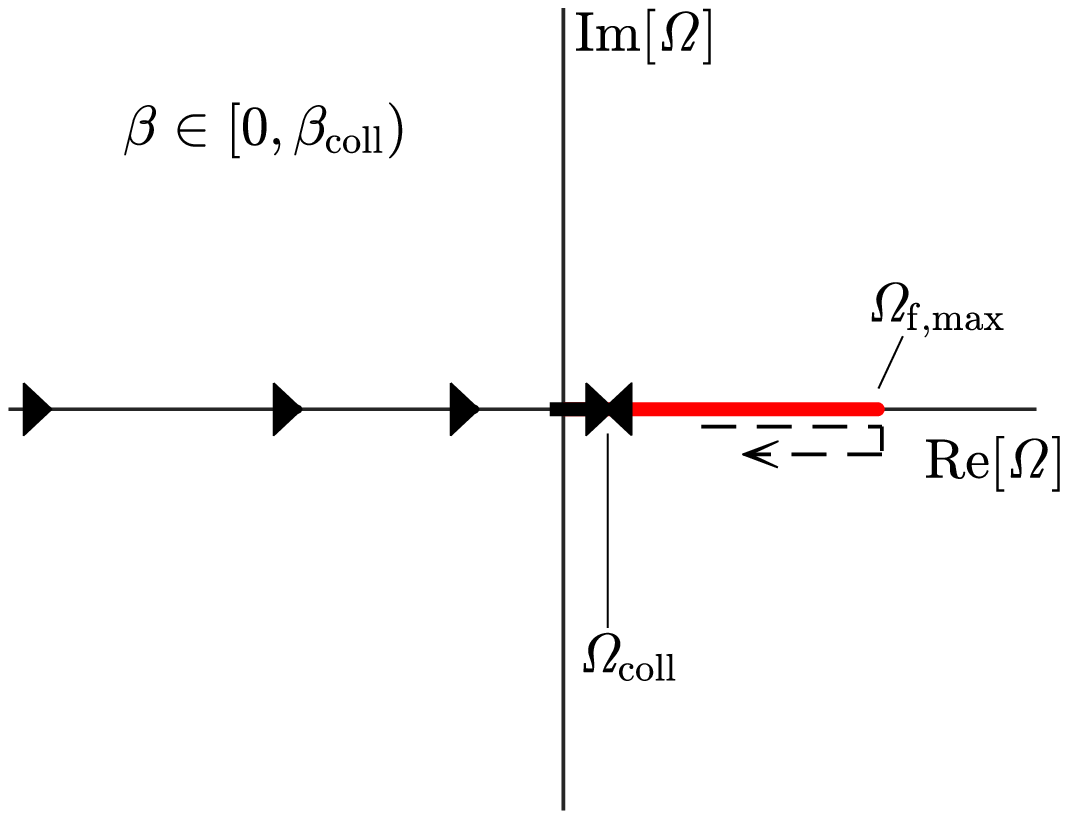}\\
\includegraphics[width=15pc]{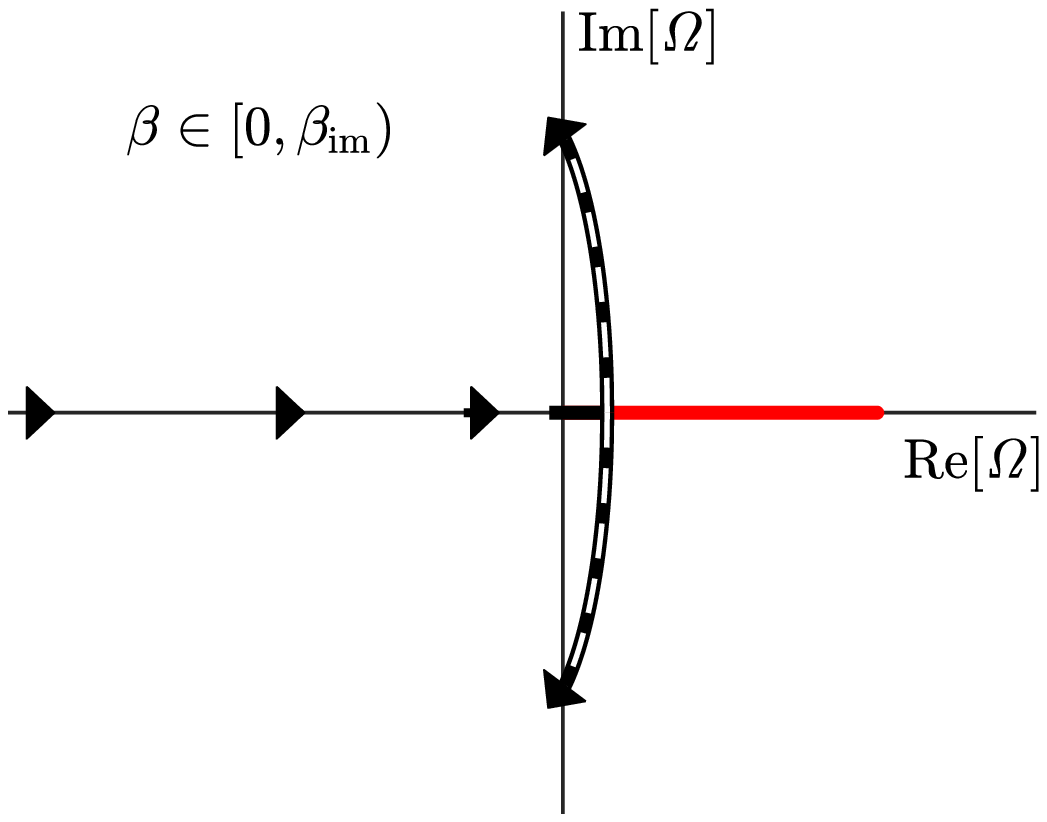}\includegraphics[width=15pc]{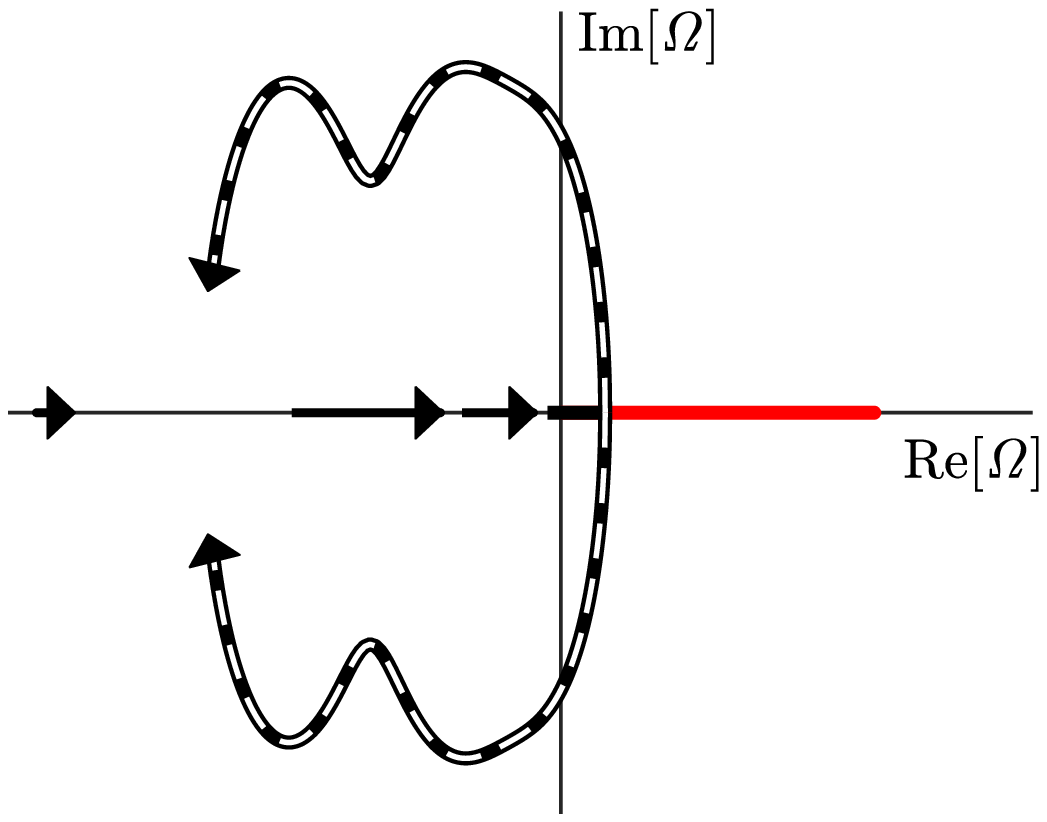}
\caption{Shows the motion of the roots for a Type I oscillatory instability at increasing values of $\beta$: $\beta = 0$ (top left), $\beta = \betacn$ (top right), $\beta = \betacn$ (bottom left) and $\beta > \betacn$ (bottom right). The arrows denote the instantaneous location and direction of motion of the roots; the dark curves trace out the motion of the roots (with the white inset dashed line showing the motion after collisions); the red curve traces out the motion of the film root $\Omf$ prior to collision.}
\label{fig:PDcp}
\end{centering}
\end{figure}
\begin{figure}
\begin{centering}
\includegraphics[width=32pc]{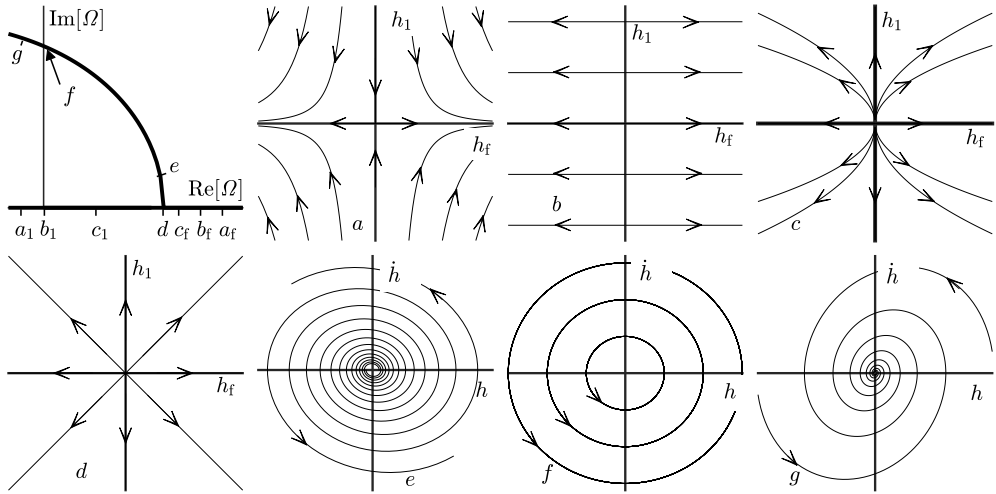}
\caption{Type I oscillatory instabilities. Top-left panel (boxed region in figure~\ref{fig:PDcp} 
shows the motion of $\Omega_1(\beta)$, 
$\Omf(\beta)$ and $\Omfone(\beta)$ (in bold) as $\beta \geq 0$ varies to 
satisfy conditions \eqref{eq:oscon}. 
Subfigures (a)--(g) are the linear phase plane for the time evolution of a 
perturbed solution from \eqref{eq:Hpert}, made to the film height. Perturbations are excited at frequencies
$\Omega_1(\beta), \Omf(\beta)$ or $\Omfone(\beta)$ that coincide with 
locations a--g for different $\beta$ values in the top-left panel. }
\label{fig:PD}
\end{centering}
\end{figure}

Type I: $\OmL > 0$.  The top-left subfigure~\ref{fig:PD} highlights how the frequencies $\Omf(\beta), \Omega_1(\beta)$ move from locations $a$--$c$ at wavenumbers $\beta < \betaL$; collide at location $d$ ($\beta = \betaL$ with $\OmL > 0$); are unstable and satisfy \eqref{eq:oscon} at any point $e$ between $d$ and $f$; cross over into the LHP at $f$; and are stable at points $g$ in the LHP.	The behavior of the perturbations \eqref{eq:Hpert}qualitatively changes at different wavenumbers $\beta$ through a series of bifurcations. To provide a visual characterization of the linearized dynamics of the perturbations \eqref{eq:Hpert} at different points $a$--$g$, we plot the phase plane trajectories in the eigenmodes with frequencies $\Omf(\beta),\Omega_1(\beta)$  (when $0 \leq \beta \leq \betaL$) or frequencies $\Omfone(\beta), \Omfone^*(\beta)$ (for $\beta > \betaL$). For wavenumbers $\beta \leq \betaL$, a perturbation in $H(X,T)$ with amplitude $\delta \hat{H}_{\rm f}$ in frequency $\Omf(\beta)$, and amplitude $\delta \hat{H}_1$ in frequency $\Omega_1(\beta)$,	evolves as:	

\begin{align}\label{Eq:PertHReal}
		H(X,T) = \Bd  + \delta\; \cos{(\beta{X})} 
		\Big[ 
			\underbrace{\hat{H}_{\rm f} \exp{\big( \Omf(\beta) {T} \big)}}_{h_{\rm f}(T)}+ 
			\underbrace{\hat{H}_1 \exp{\big( \Omega_1(\beta){T} \big)}}_{h_1(T)} \Big] 
			+ \mathcal{O}(\delta^2).
\end{align}
Figure~\ref{fig:PD}$(a-d)$ plots the phase plane dynamics in the $h_{1}(T)$--$h_{\rm f}(T)$ plane.  For wavenumbers $\beta > \betaL$, the frequencies are complex $\Omfone(\beta)$,$\Omfone(\beta)^*$ and we write the perturbation as
\begin{align} \nonumber
		H(X,T) &= \Bd  + \delta\; \cos{(\beta{X})} 
		\Big[ 
			\underbrace{\hat{H}_{\rm R} \textrm{Re[}\exp{\big(\Omfone(\beta) T\big)}\textrm{]}}_{h(T)} +
			\underbrace{\hat{H}_{\rm I} \textrm{Re[} \Omfone(\beta) \exp{\big(\Omfone(\beta) T\big)}\textrm{]}}_{dh(T)} 
		\Big] \\ \label{Eq:PertHComplex}
			&+ \mathcal{O}(\delta^2),
\end{align}
where $\hat{H}_{\rm R}$ and $\hat{H}_{\rm I}$ are the two amplitudes of the perturbation. The panels $e$--$g$ in figure~\ref{fig:PD}) plot the phase plane trajectories in the $h(T)$--$dh(T)$ plane. Here we use the short-form notation $dh(T)$ for the second linearly independent term in \eqref{Eq:PertHComplex} since it is proportional to $\frac{dh}{dt}$. The subfigures show the qualitatively different phase plane behavior, and emergence of oscillatory instabilities, as the bifurcation parameter $\beta$ varies.  The boundary in the parameter space $(\Cd,\Qd,\Bd,\Sd)$, for Type I behavior to occur, must satisfy (as a necessary condition) $\OmL = 0$. 	
\begin{figure}
\begin{centering}
\includegraphics[width=15pc]{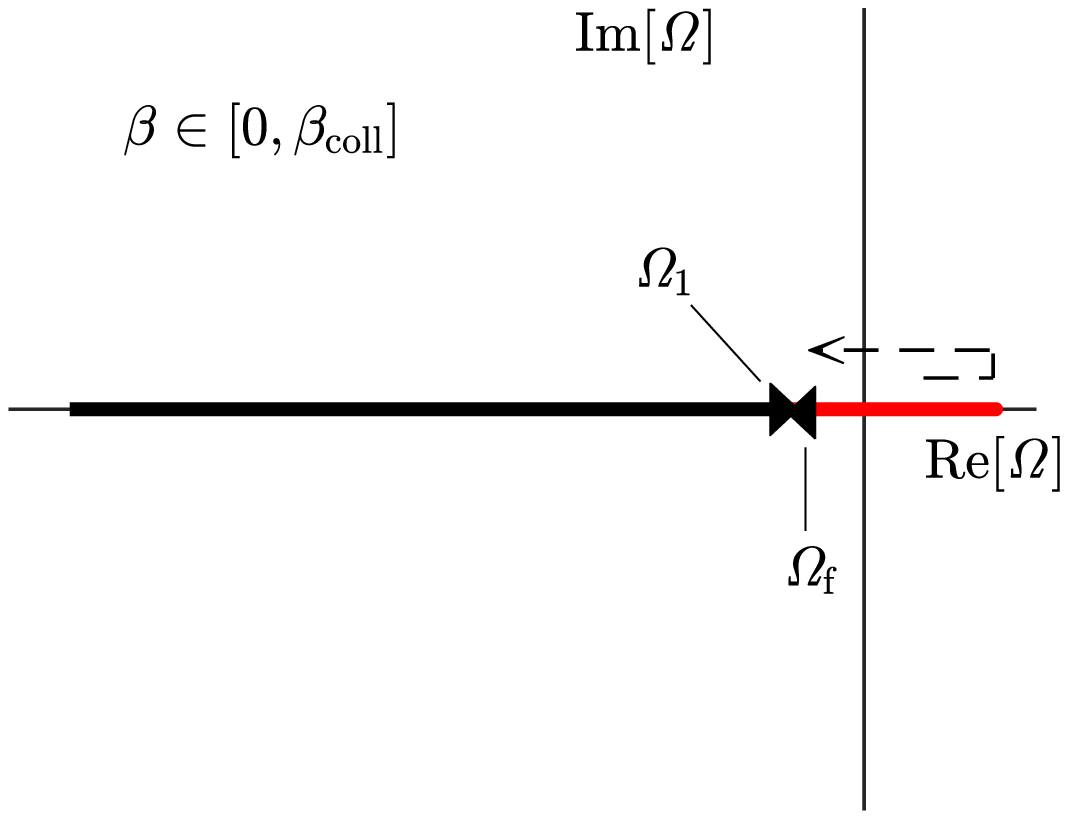}\includegraphics[width=15pc]{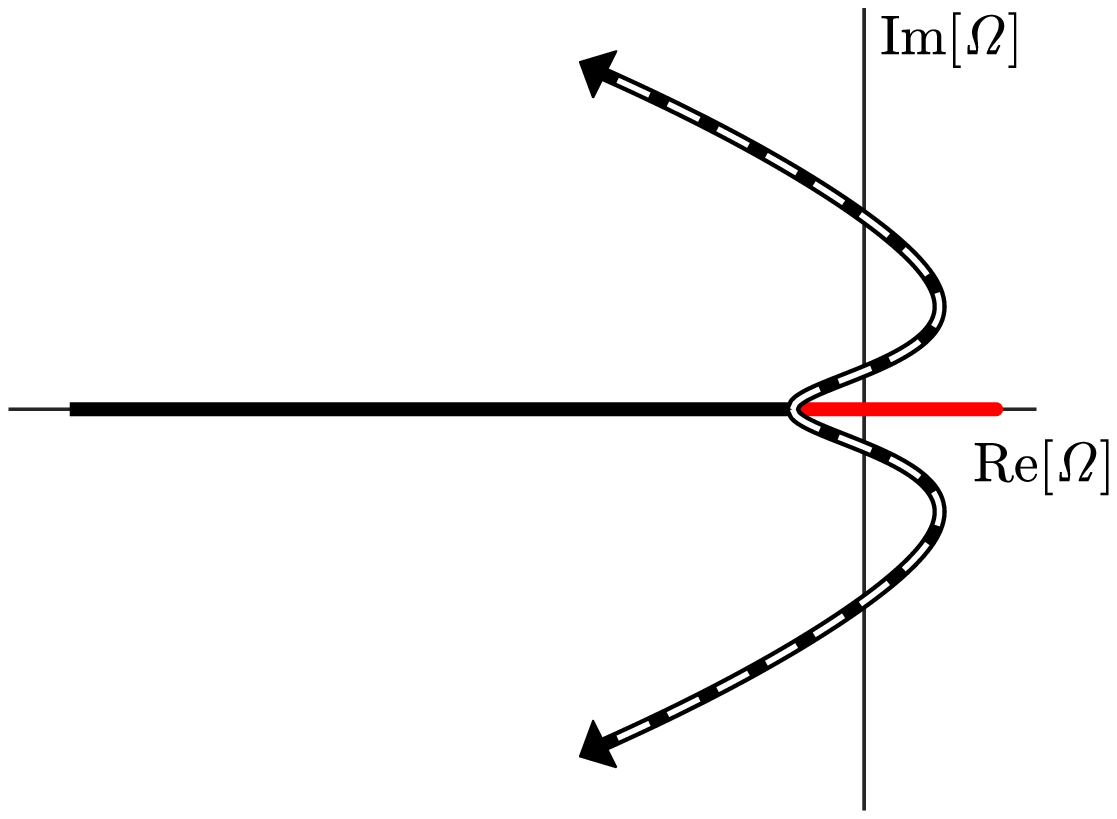}
\caption{Shows the motion of the roots for a Type II oscillatory instability at increasing values of $\beta$: $\beta = \betacn$ (left), $\beta > \betacn$ (right). The arrows denote the instantaneous location and direction of motion of the roots; the dark curves trace out the motion of the roots (with the white inset dashed line showing the motion after collisions); the red curve traces out the motion of the film root $\Omf$ prior to collision. The collision occurs on the negative real axis ($\OmL<0$) and before entering into the RHP.}
\label{fig:PDcp2}
\end{centering}
\end{figure}
\begin{figure}
\begin{centering}
\includegraphics[width=32pc]{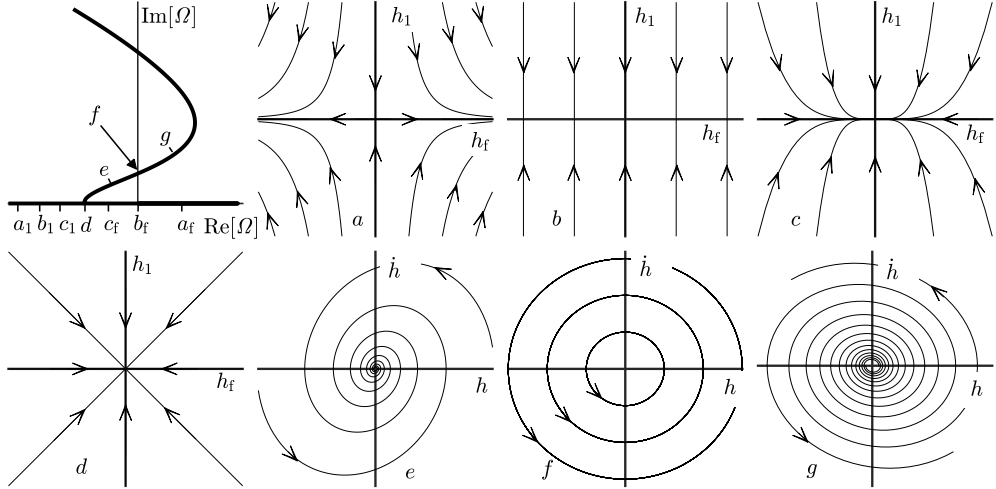}
\caption{Type II oscillatory instabilities. Top-left panel (boxed region in left panel of figure \ref{fig:PDcp2}) shows the motion of $\Omega_1(\beta)$, $\Omf(\beta)$ and $\Omfone(\beta)$ (in bold) as $\beta \geq 0$ varies to satisfy conditions \eqref{eq:oscon}.  Subfigures (a)--(g) are the linear phase plane portraits for the time evolution of a perturbation made to the film height excited at frequencies $\Omega_1(\beta), \Omf(\beta)$ or $\Omfone(\beta)$ and different $\beta$'s. }
\label{fig:PD2}
\end{centering}
\end{figure}

Type II: $\OmL \leq 0$ and $\Omr(\beta) > 0$ for some $\beta > \betaL$. Figure~\ref{fig:PDcp2} highlights (via arrows) the motion of the roots $\Omf(\beta)$ (red), $\Omega_1(\beta)$ (black), and post collision roots $\Omfone(\beta)$ (black with white dashed line) with varying $\beta$.  The top-left panel in Figure~\ref{fig:PD2} shows again the motion of the frequencies $\Omf(\beta)$, $\Omega_1(\beta)$, and $\Omfone(\beta)$. The frequencies $\Omf(\beta)$, $\Omega_1(\beta)$ move from locations $a$--$c$ and collide at $d$ with $\OmL \leq 0$. The values $\Omfone(\beta)$ are complex at location $e$, and then move into the right-half plane $\Omr(\beta) \geq 0$ at $f$--$g$; thereby satisfying conditions \eqref{eq:oscon}. The subfigures also show the phase plane trajectories for perturbations to the film height (in a fashion completely analogous to figure~\ref{fig:PD}) given by equations \eqref{Eq:PertHReal}--\eqref{Eq:PertHComplex}. The boundary in the parameter space $(\Cd,\Qd,\Bd,\Sd)$, for Type II behavior, requires (as a necessary condition) that $\max_{\beta \geq \betaL} \Omr(\beta) = 0$. 
	
Having criteria for the boundaries in the parameter space $(\Cd,\Qd,\Bd,\Sd)$ of Types I or II oscillatory instabilities will become useful in the following section.  Specifically, we will use these conditions to help plot phase diagrams and identify model parameters and experimental conditions that may yield oscillatory instabilities. 

The underlying distinction between Types I and II instabilities occurs from viewing $\beta$ as a bifurcation parameter.  Each root $\Omega$ for any value of $\beta$ is associated with a linear dynamical system for the variables in equations \eqref{eq:Hpert}.  Type I oscillatory instabilities occur from one bifurcation when the roots collide at $\beta=\betaL$. Meanwhile, Type II oscillatory instabilities occur through two bifurcations: the first at $\beta = \betaL$ when the roots collide, and the second at the value $\beta = \betacn$, when the roots enter into the RHP (e.g., see, \citet{Strogatz}, chapter 8).  Categorizing oscillatory instabilities as Type I or Type II through different bifurcations provides criteria that we will use in \S\ref{sec:emerge} to systematically determine which parameter values  $(\Cd,\Qd,\Bd,\Sd)$ give rise to oscillatory instabilities. Type I and Type II instabilities are physically distinguishable through their bands of unstable wavenumbers: Type I does not contain a band of wavenumbers with values $\beta < \betaL$ where $\Omf(\beta)$ is stable (figure~\ref{fig:PD} at points $a$, $b$, $c$ and $d$ are unstable); in contrast, Type II does contain an interval of wavenumbers with values $\beta < \betaL$ where $\Omf(\beta)$ is stable (figure~\ref{fig:PD2} at points $c$ and $d$ are unstable).

In addition to classifying Types I and II oscillatory instabilities, we further distinguish whether a set of parameters $(\Cd,\Qd,\Bd,\Sd)$ gives rise to \emph{global oscillatory instabilities}.  A set of parameters $(\Cd,\Qd,\Bd,\Sd)$ is said to be globally oscillatory unstable if there is a $\betacm$ and $\Omegacm=\Omega(\betacm)$ satisfying the dispersion relation \eqref{eq:ImpDisprel} that is oscillatory unstable, and has the largest growth rate $\textrm{Re}(\Omegacm) \geq \textrm{Re}(\Omega(\beta))$ for all $\beta$, $\Omega(\beta)$ (satisfying the dispersion relation  \eqref{eq:ImpDisprel}). Global oscillatory unstable parameter values are physically significant because they represent experimental situations where the most unstable perturbation to the linearized system \eqref{eq:Hpert} is oscillatory unstable (and hence the most likely to be observed).  

The bottom panel of figure~\ref{fig:PDcp} provides additional information for the motion (as functions of $\beta$) of the frequencies $\Omf(\beta), \Omega_n(\beta)$. This highlights the fact that the complete motion of the roots is quite complicated. In the figure, bold lines show 
the trajectories of the roots in the complex plane over an interval $0 \leq \beta \leq \beta_*$ 
($\beta_* >0$ is chosen somewhat arbitrarily). The bold curves trace out several collisions of the roots (including collisions made by roots $\Omega_n(\beta)$ for $n > 1$), and highlight that they always appear in complex conjugate pairs. The arrows and red circles show roots at the value $\beta = \beta_*$, with the arrows indicating the direction of motion for the roots at $\beta = \beta_*$. Although several of the roots depicted at $\beta=\beta_*$ are moving in the positive direction along the real axis, we note that each of these roots ultimately reverses direction and moves into the left-half plane for sufficiently large $\beta$. 

In the numerical continuation of the roots $\{\Omf(\beta), \Omega_n(\beta)\}$, we have always observed that oscillatory instabilities arise as Type I or Type II, as described in this section.  It may be possible that oscillatory instabilities occur from the collision of other roots (for instance, a collision including $\Omega_2(\beta)$); however we did not observe this in any of our investigations.  If the only possible mechanism to obtain oscillatory instabilities is through Type I or Type II, and the largest growth rate $\Omegacm$ occurs at a value $\Omf(\beta)$ (which we also numerically observe to be the case in our studies), then we may simplify the condition for global oscillatory instabilities to growth rates computed in terms of $\Omf(\beta)$ and $\Omfone(\beta)$ by defining:
\begin{align}\label{eq:GOI_def}
	\Omfmax \equiv \max_{0 \leq \beta \leq \betaL}\Omf(\beta),\quad \quad
	\Omrmax \equiv \max_{\beta \geq \betaL}\Omr(\beta).	
\end{align}
The condition for global oscillatory instabilities is then
\begin{align}\label{eq:GOI_cond}
	&\textrm{(Global oscillatory instabilities)} \quad \quad \Omfmax \leq \Omrmax.
\end{align}
To characterize oscillatory instabilities, we will also make use of the \emph{most unstable wavenumber}, $\beta_{\rm max}$ defined as:
\begin{align}
	\Omega_{\rm r,max} = \Omega_{r}(\beta_{\rm max}),
\end{align}
or alternatively written as the argument of the maximum $\beta_{\rm max} = \textrm{argmax}_{\beta \geq \betaL} \Omega_{\rm r}(\beta)$. We will also denote the imaginary frequency of the most unstable wave number as $\Omega_{\rm i,max} = \Omega_{\rm i}(\beta_{\rm max})$.

In practice, the maximization $\max_{\beta\geq \betaL} \Omr(\beta)$ in ondition~\eqref{eq:GOI_cond} implies that we maximize the real value of the root $\Omfone(\beta)$ over a suitably large range of $\beta$ values (by taking $\beta$ large enough $\Omr(\beta)$ will eventually become negative).  Replacing the inequality $(\leq)$ in \eqref{eq:GOI_cond} with an equality $(=)$ then provides a condition for the boundary of the global oscillatory instability region.

\section{Emergence of oscillatory instabilities}
\label{sec:emerge}
The purpose of this section is to explore the model parameter space $(\Cd,\Qd,\Bd,\Sd)$ and characterize which parameter regions give rise to oscillatory instabilities.  This will provide a guide for experimental scenarios in which one may likely see oscillatory instabilities. To compute these regions we use the formulas for the boundaries of the parameter regions, satisfied by Type I and Type II instabilities, developed in \S\ref{sec:charcoal}. As a general guide, we also introduce a heuristic value 
\[
	\partial_{\beta} \OmL \equiv \lim_{\beta \rightarrow \betaL^+} \frac{d}{d\beta}\Omr(\beta),
\] 
as the rate of change of the real value of the roots $\Omfone(\beta) = \Omr(\beta)+i \Omi(\beta)$ immediately after the collision $\beta \rightarrow \betaL^+$. A positive value $\partial_{\beta} \Omr > 0$ (resp. $<$) implies the roots $\Omfone(\beta)$ move towards the right (resp. left) in the complex plane as $\beta$ increases past $\betaL$.  Knowing whether the roots $\Omfone(\beta)$ move towards the left (more stable, viz. figures \ref{fig:PDcp} and \ref{fig:PD}) or right (more unstable, viz. figures \ref{fig:PDcp2} and \ref{fig:PD2}) in the complex plane after the collision is a useful heuristic when identifying regions of global oscillatory instabilities. Specifically, numerical evidence shows that parameter values $(\Cd,\Qd,\Bd,\Sd)$ that have $\partial_{\beta} \Omr < 0$ (solutions initially become more stable after collisions) do not exhibit global oscillatory instabilities.

This section is organized as follows: in \S\ref{sec:CdltCdg} we plot and detail the behavior of the phase diagram for oscillatory instabilities, with a focus on the parameters $(\Cd, \Bd, \Sd)$.  In \S\ref{sec:deponQ} we examine the effect of the parameter $\Qd$ (material property dependent) on the behavior of the phase diagram. The results from \S\ref{sec:deponQ} will help guide realistic choices of material properties and experimental conditions for observing oscillatory instabilities. Guided by the results in \S\ref{sec:CdltCdg}--\S\ref{sec:deponQ}, in \S\ref{sec:varC} we discuss materials that give rise to reasonable model parameter values for observing oscillatory instabilities.

\subsection{Material phase diagrams and oscillatory instabilities}
\label{sec:CdltCdg}
In this section we plot phase diagrams that show for which model parameters oscillatory instabilities occur. Our approach for plotting the diagrams is motivated by experimental considerations.  The parameter $\Qd$ depends on the material properties, and is the most difficult to change in experiments (it requires changing the substrate or fluid materials in the experiment). The values of $\Bd, \Sd$ can by varied by modifying the thickness of the film ($\Bd$) and substrate ($\Sd$), while $\Cd$ may be varied easily by modifying the temperature difference across the film and substrate.  Since we have four parameters $(\Cd,\Qd,\Bd,\Sd)$ we adopt the following approach to visualize the phase diagrams: we fix a value of $\Qd$, and then plot the phase diagram in the $\Bd\,$--$\,\Sd$ plane for different values of $\Cd$.  This is equivalent to plotting cross-sections of the three dimensional phase diagram $(\Cd, \Bd, \Sd)$ (holding $\Qd$ constant). For the purpose of developing better intuition, one may think of $\Bd$ and $\Sd$ being the thicknesses of the film and substrate (respectively) and $\Cd$ the imposed temperature difference.
\begin{figure}
\begin{centering}
\includegraphics[width=36pc]{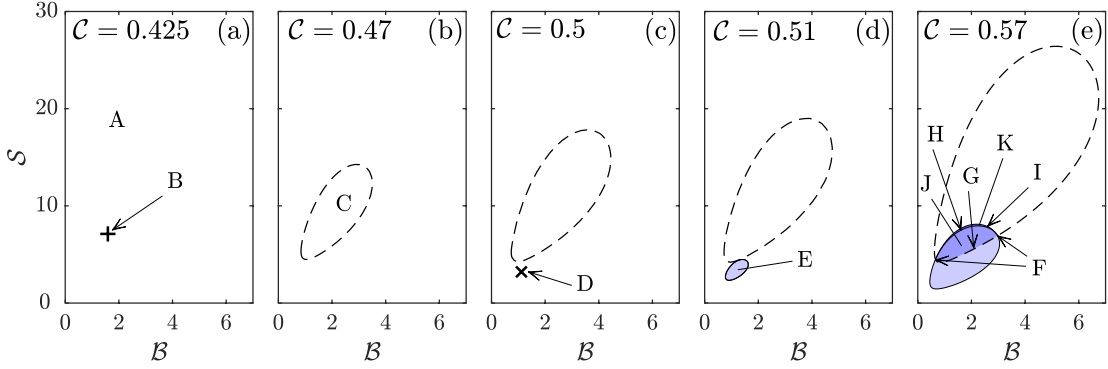}
\caption{Phase diagrams for $(\Bd,\Sd)$, at different $\Cd$, showing oscillatory instabilities (shaded regions) with Type II instabilities (dark purple, region K) and Type I instabilities (lighter shades of purple, regions E and J).  Oscillatory instabilities grow from point D, $(\Bd_{\rm o},\Sd_{\rm o})=(1.12,3.21)$, as $\Cd$ increases.  Stability types A-K corresponding to regions/lines/points are labelled as they appear in (a)-(e) and depicted characteristically in  figure \ref{fig:charcoal1}; they are distinguished by the signs of $\OmL$ and the heuristic $\partial_\beta\OmL$ (the heuristic zero level set is given by the dashed contour).}
\label{fig:drw0}
\end{centering}
\end{figure}

\begin{figure}
\begin{centering}
\includegraphics[width=40pc]{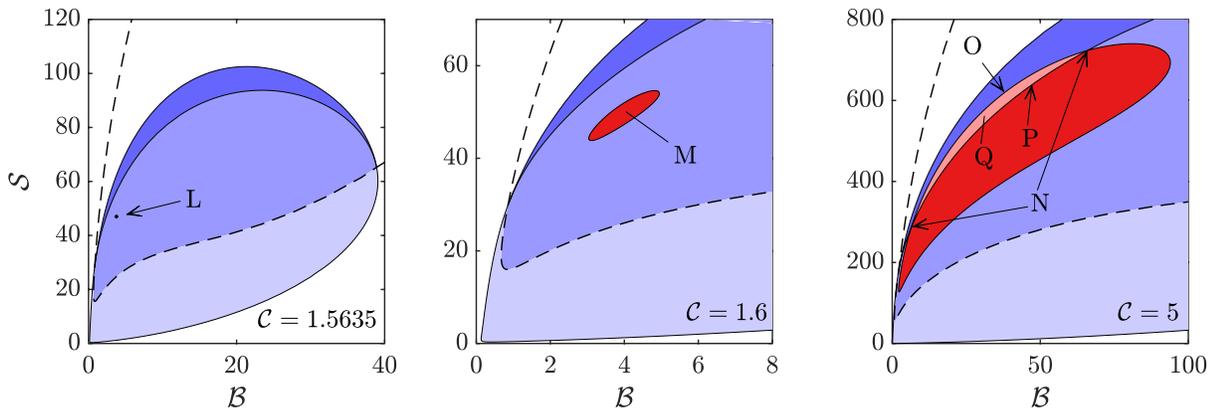}
\caption{Phase diagrams for $(\Bd,\Sd)$, at different $\Cd$, showing oscillatory instabilities (shaded regions) with Type II instabilities (dark purple) and Type I instabilities (two light shades of purple/blue bounded bounded by the solid line).  Global oscillatory instabilities (pink which are Type II and red which are Type I) grow from the point L, $(\Bd_{\rm o},\Sd_{\rm o})=(3.766,46.991)$, as $\Cd$ increases.  The heuristic contour $\partial_{\beta} \OmL = 0$ (dashed line) is plotted as a numerical guide to bound the red/pink region.}
\label{fig:drw02}
\end{centering}
\end{figure}

Figure~\ref{fig:drw0} plots the $\Bd\,$--$\,\Sd$ phase diagram for values of $0.425 \leq \Cd \leq 0.57$, holding $\Qd=5.114\times{10}^{-4}$ fixed. The subfigures in \ref{fig:drw0} reveal an onset parameter value $\Cd_{\rm o} = 0.5$ such that for $\Cd < \Cd_{\rm o}$ there is no region of oscillatory instability in the $\Bd\,$--$\,\Sd$ phase diagram; while values $\Cd > \Cd_{\rm o}$ give one connected region (shown in color) of oscillatory instability ($\Qd$ is chosen to three decimals so that $\Cd_{\rm o}$ is a single decimal). At $\Cd=0.5$, the region of oscillatory instability emerges from a single point $(\Bd_{\rm o},\Sd_{\rm o})=(1.12,3.21)$ (labeled D in the figure).  The different shadings in figure~\ref{fig:drw0} provide details on how the roots $\Omf(\beta), \Omega_1(\beta)$ become unstable (i.e. Type I or II), as well as the sign of the heuristic quantity $\partial_{\beta}\OmL$. The region inside the dashed curves indicates where $\partial_{\beta}\OmL>0$. When investigating the four dimensional parameter space $(\Cd,\Qd,\Bd,\Sd)$, the heuristic $\partial_{\beta}\OmL>0$ is helpful in identifying regions that have global oscillatory instabilities, as they numerically appear inside the heuristic, see {\it e.g.} figure \ref{fig:drw02} (note that figure \ref{fig:drw0} shows no regions of global oscillatory instability). The boundary of the heuristic is easy to compute and can then be used to restrict the region where a refined search for global oscillatory instabilities can be done. In addition to $\Cd_{\rm o}$, we introduce $\Cd_{\rm g}$ as the critical parameter value for which global oscillatory instabilities occur, i.e. global oscillatory instabilities occur when $\Cd > \Cd_{\rm g}$, while for values of $\Cd < \Cd_{\rm g}$ all oscillatory instabilities are non-global (such as those in figure \ref{fig:drw0}). A key observation from the figure is that oscillatory instabilities do not occur at low temperatures ($\Cd < \Cd_{\rm o}$). 
\begin{figure}
\begin{centering}
\includegraphics[width=10pc]{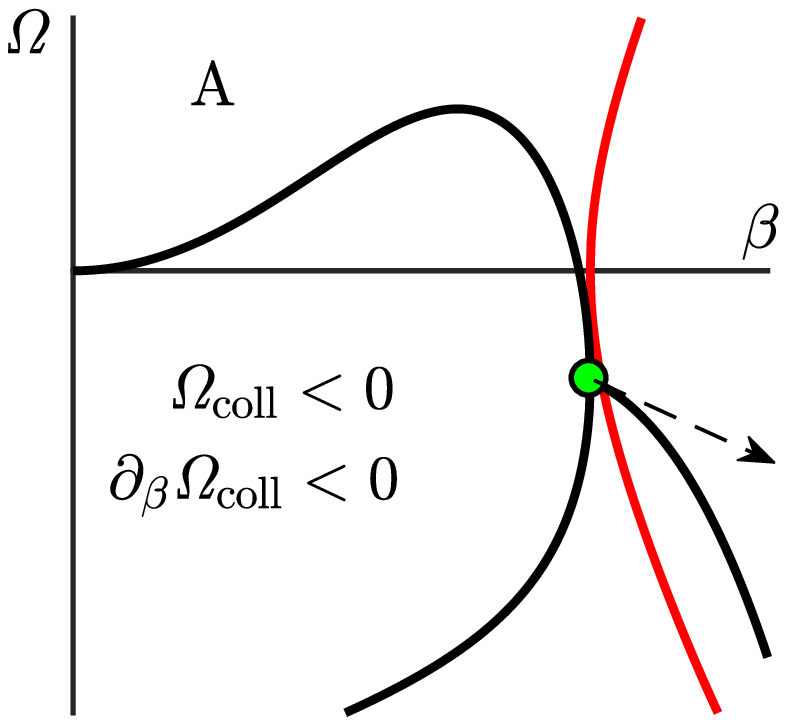}\includegraphics[width=10pc]{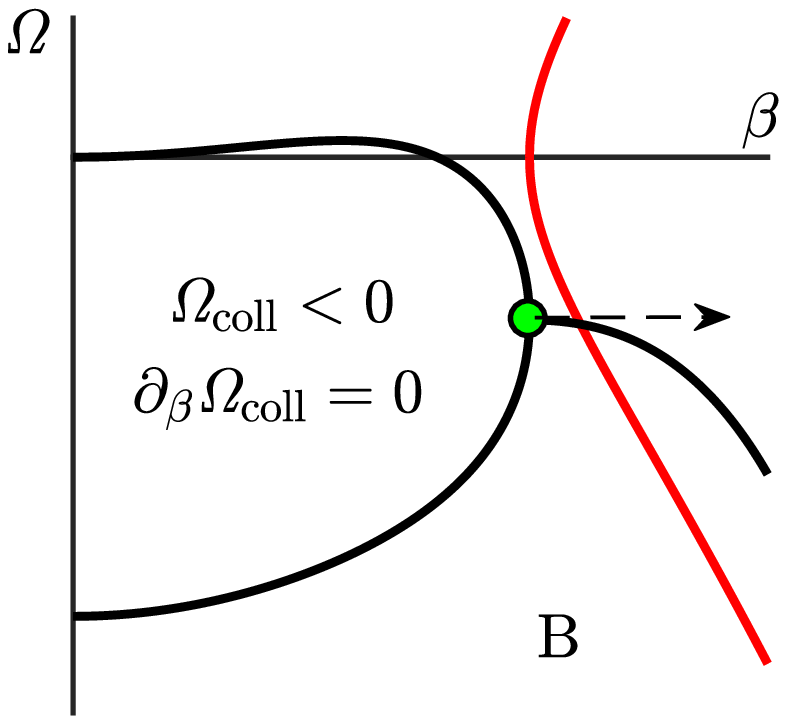}\includegraphics[width=10pc]{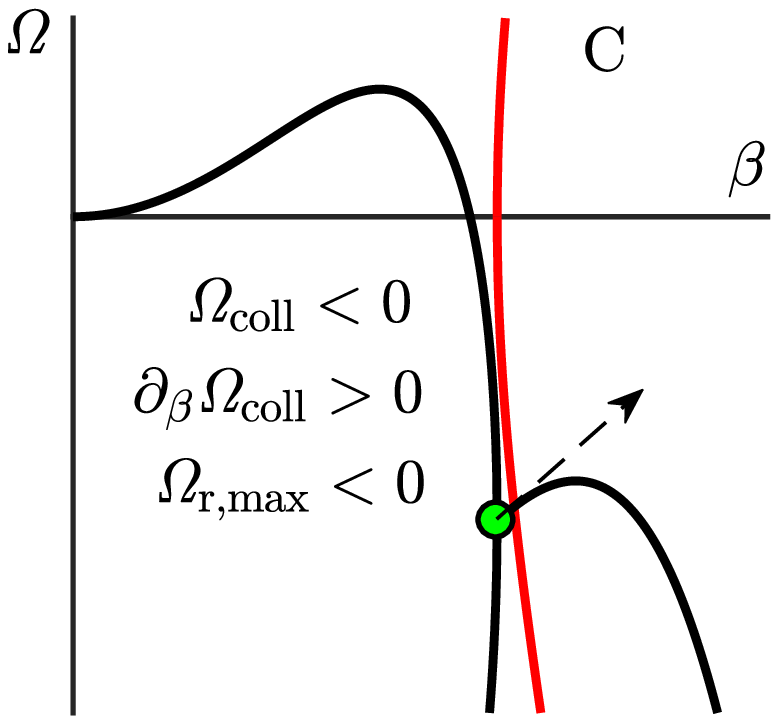}\includegraphics[width=10pc]{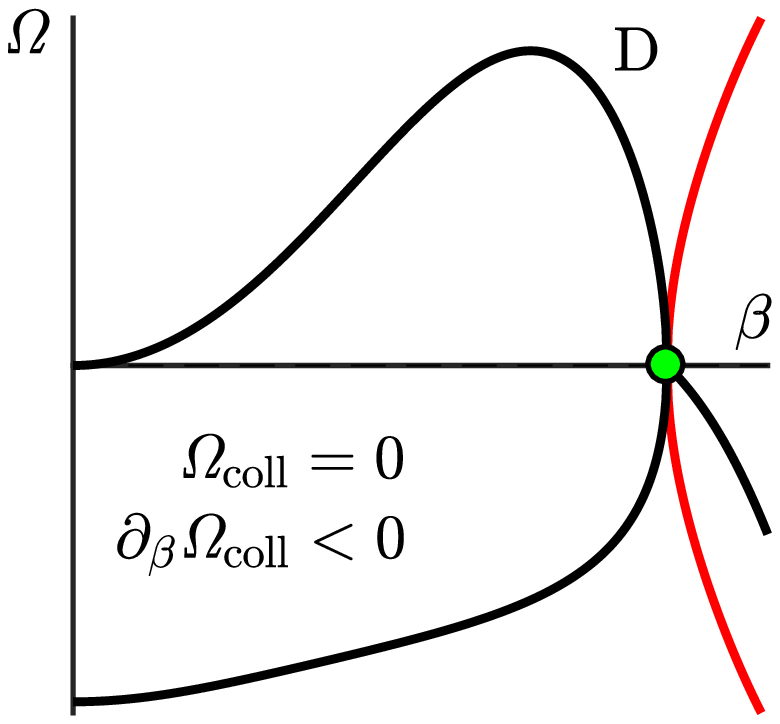} \\
\includegraphics[width=10pc]{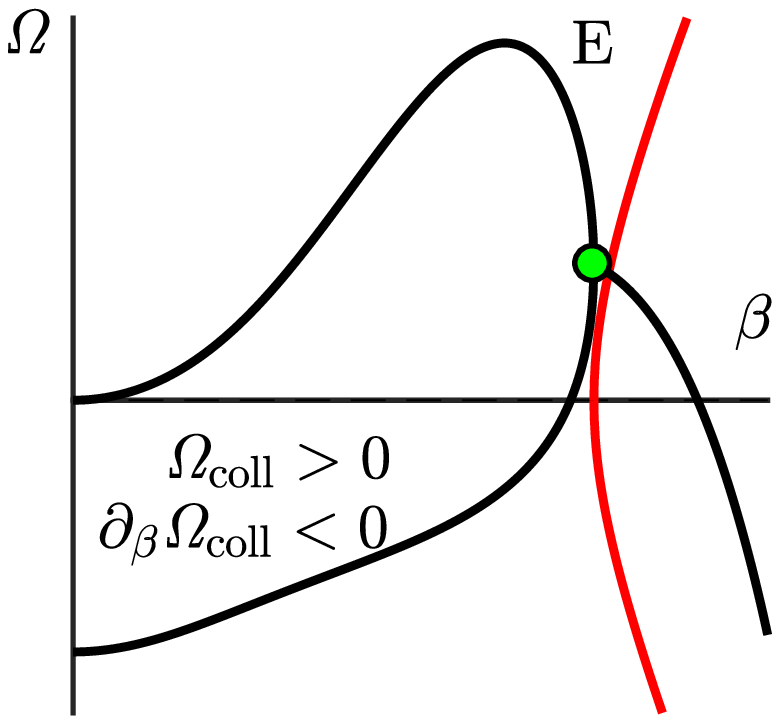}\includegraphics[width=10pc]{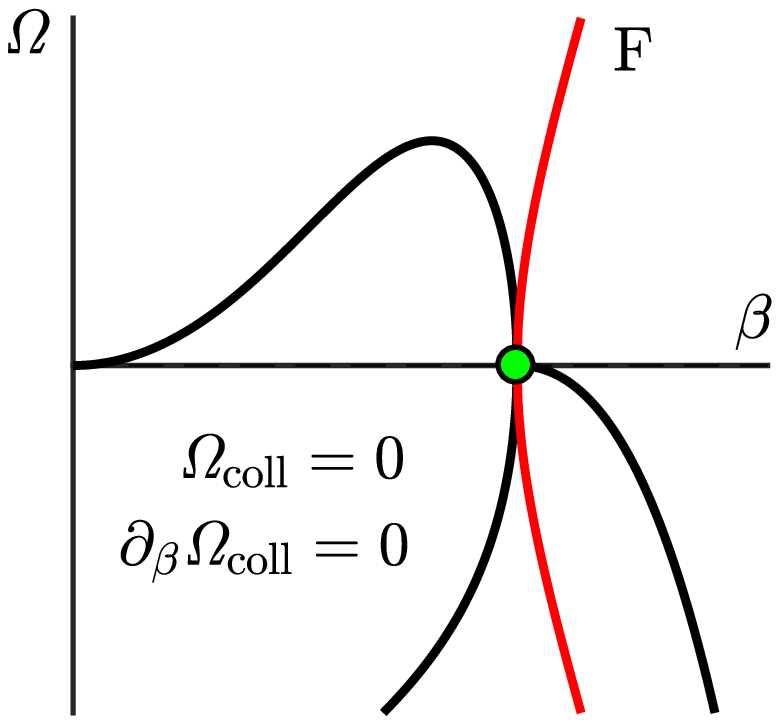}\includegraphics[width=10pc]{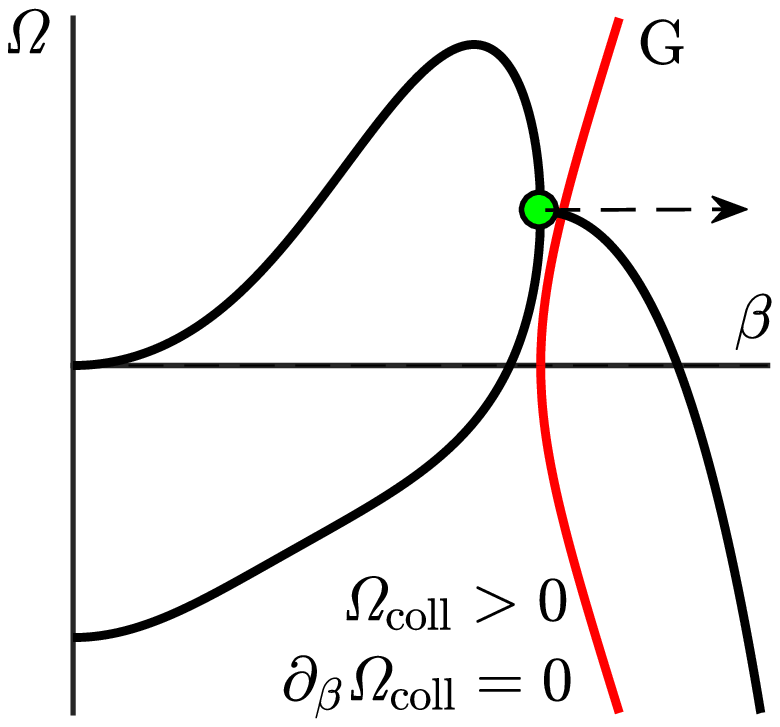}\includegraphics[width=10pc]{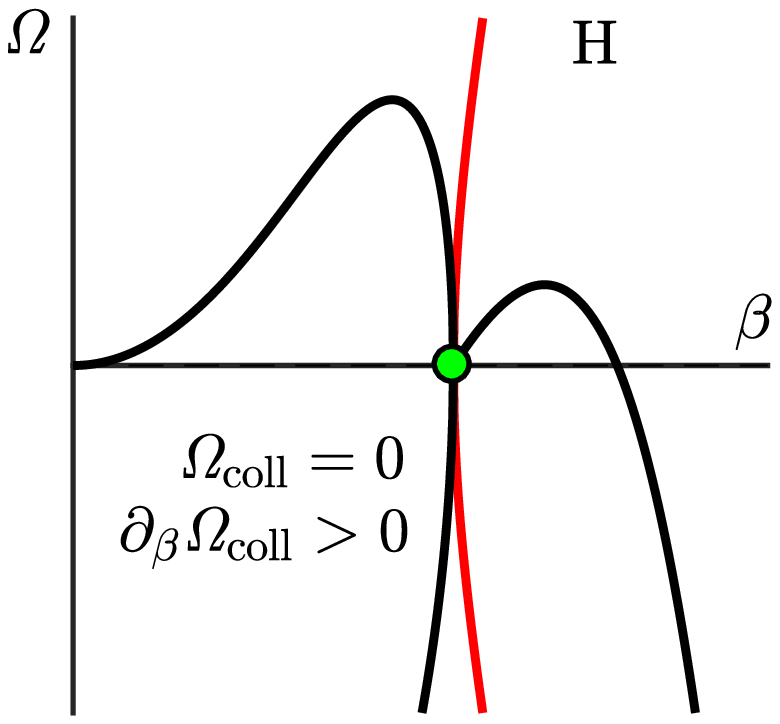} \\
\includegraphics[width=10pc]{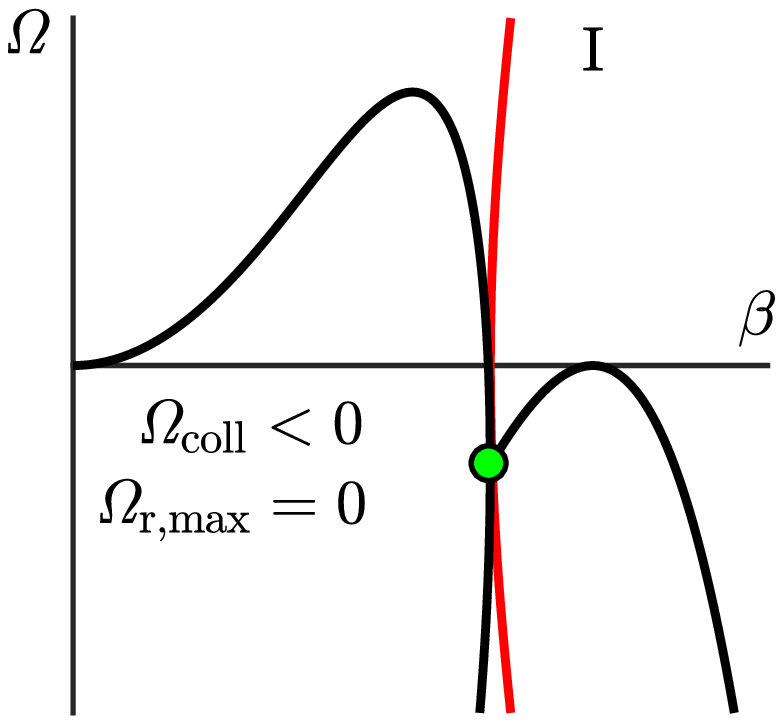}\includegraphics[width=10pc]{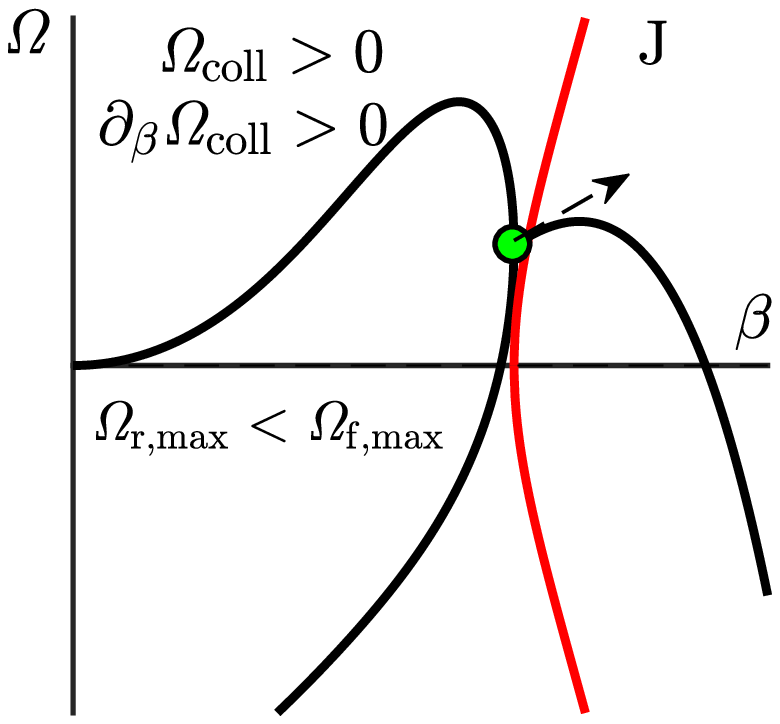}\includegraphics[width=10pc]{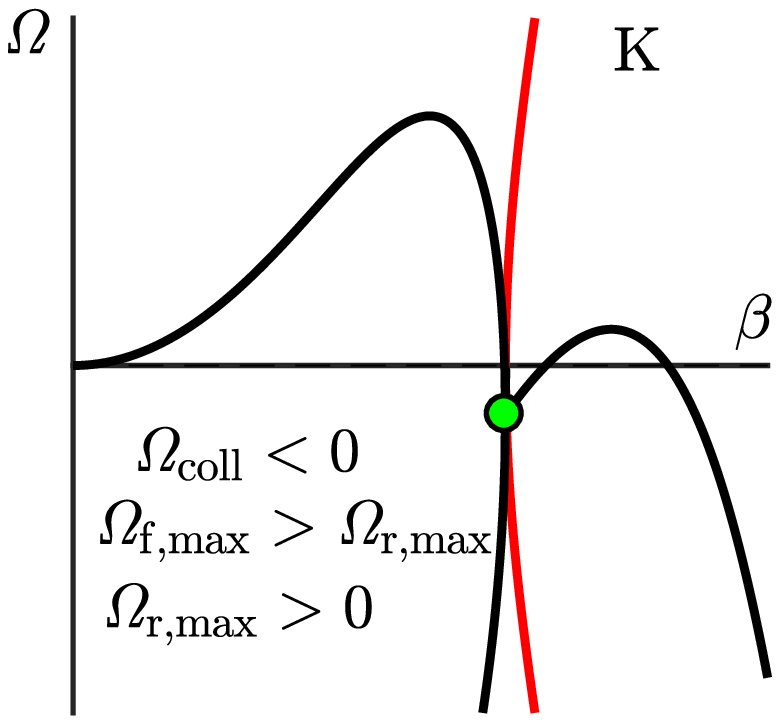}
\caption{Plots of the real Re[$\Omega(\beta)$] (black lines), and imaginary Im[$\Omega(\beta)$] (red) parts of the frequencies $\Omf(\beta)$, $\Omega_1(\beta)$ and $\Omfone(\beta)$ for varying $\beta$. The frequencies $\Omf(\beta)$ and $\Omega_1(\beta)$ collide at $(\betaL,\OmL)$ (shown in green). All diagrams are for values $\Cd<\Cd_{\rm g}$ that do not support global oscillatory instabilities (i.e. $\Omrmax < \Omfmax$), and correspond to the points with parameters labelled A--K in figure~\ref{fig:drw02}.}
\label{fig:charcoal1}
\end{centering}
\end{figure}

The qualitative differences in how the roots become unstable (as $\beta$ varies) are shown in figure~\ref{fig:charcoal1}.  Each part of figure \ref{fig:charcoal1} plots $\Omf(\beta)$, $\Omega_1(\beta)$, and the real and imaginary values of $\Omfone(\beta)$ (defined in \eqref{eq:Notation}) for parameter values that capture the behavior at locations A--K in figure~\ref{fig:drw0}.  Specifically, in figure~\ref{fig:charcoal1}, panels A--C first show the change in sign of the heuristic quantity $\partial_{\beta}\OmL$; panels E, G, and J correspond to Type I instabilities; panels H and K to Type II instabilities.  Meanwhile, panels D, F, H and I provide a comprehensive survey of parameter values that lie on the boundaries of Type I or II instabilities. Type I oscillatory instabilities have one continuous band of unstable wavenumbers $0 \leq \beta < \betacn$, of which $[0, \betaL]$ is monotonically unstable and $(\betaL, \betacn)$ is oscillatory unstable.  Type II oscillatory instabilities have two continuous bands of unstable wavenumbers, separated by a gap of stable wavenumbers that includes the interval $(\betaL, \betacn]$. Together, the panels in figure~\ref{fig:charcoal1} characterize all the different possibilities for (possibly oscillatory) instability development. Note that figure~\ref{fig:charcoal1} does not admit global oscillatory instabilities: all the subfigures are plotted for values of $\Cd < \Cd_{\rm g}$, where $\Cd_{\rm g}$ is the critical parameter such that global oscillatory instability can only occur for $\Cd > \Cd_{\rm g}$ (oscillatory instabilities for $\Cd < \Cd_{\rm g}$ are non-global).
\begin{figure}
\begin{centering}
\includegraphics[width=10pc]{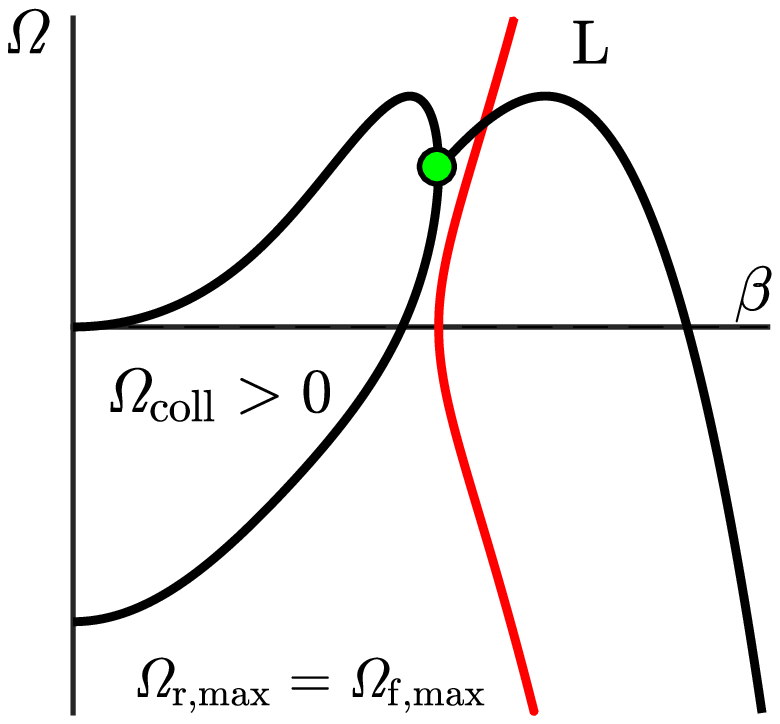}\includegraphics[width=10pc]{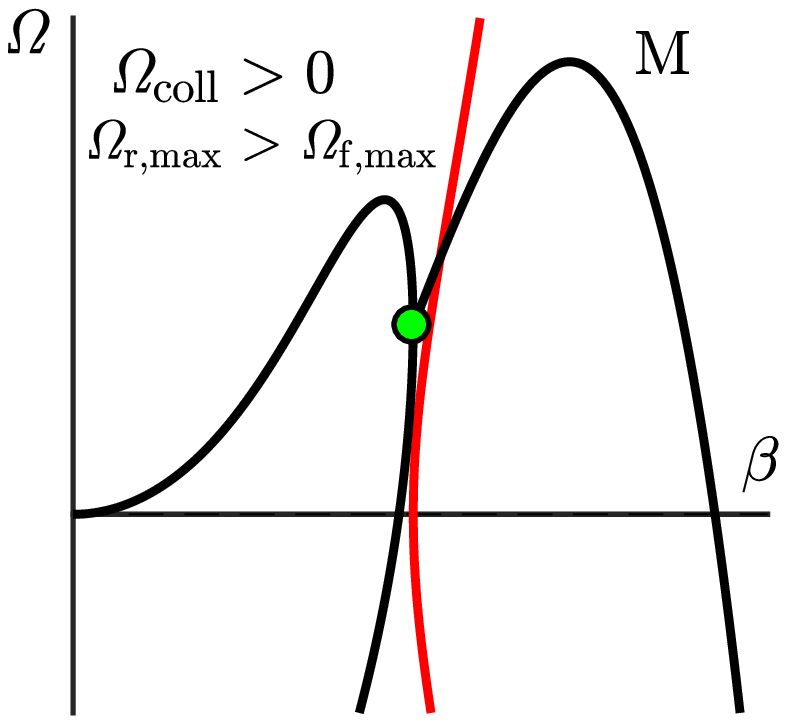}\includegraphics[width=10pc]{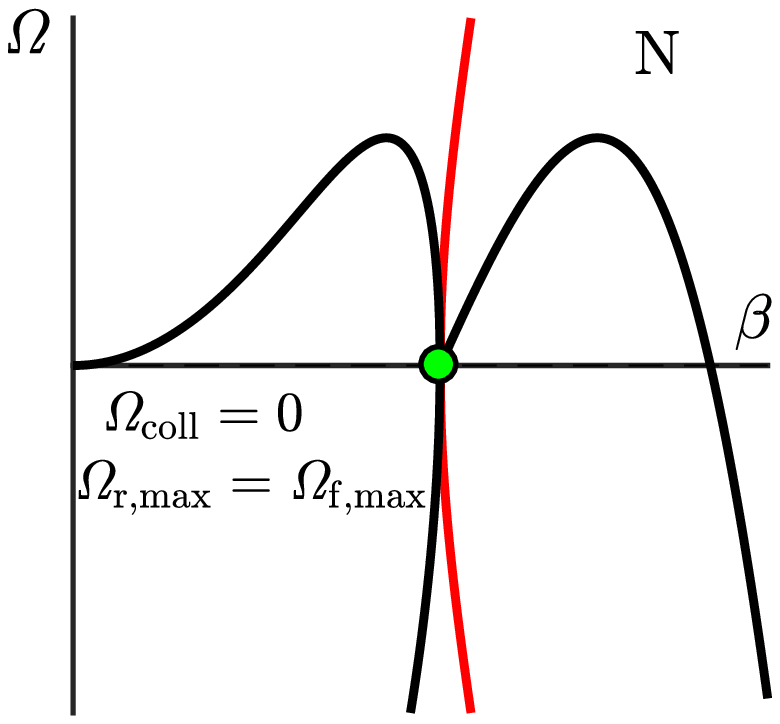}\includegraphics[width=10pc]{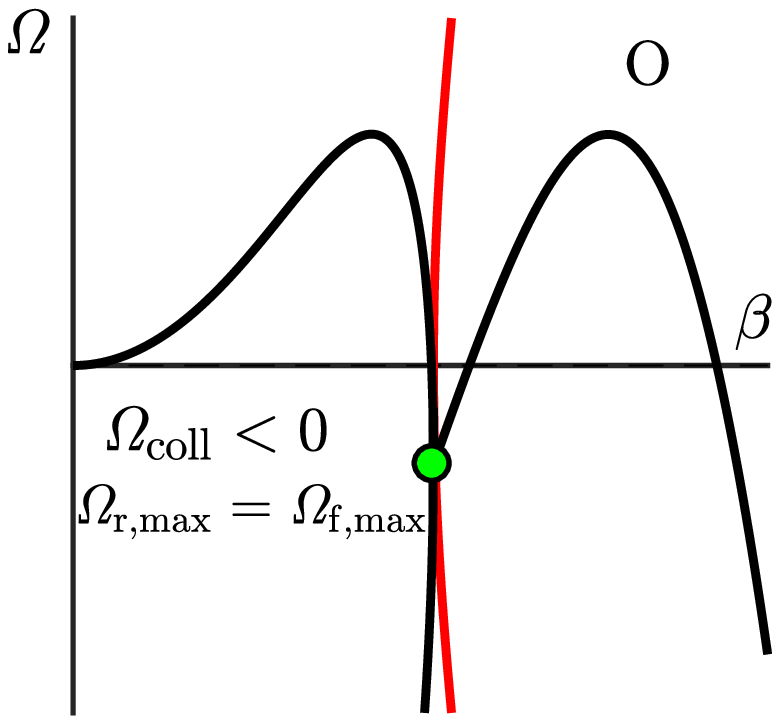} \\\includegraphics[width=10pc]{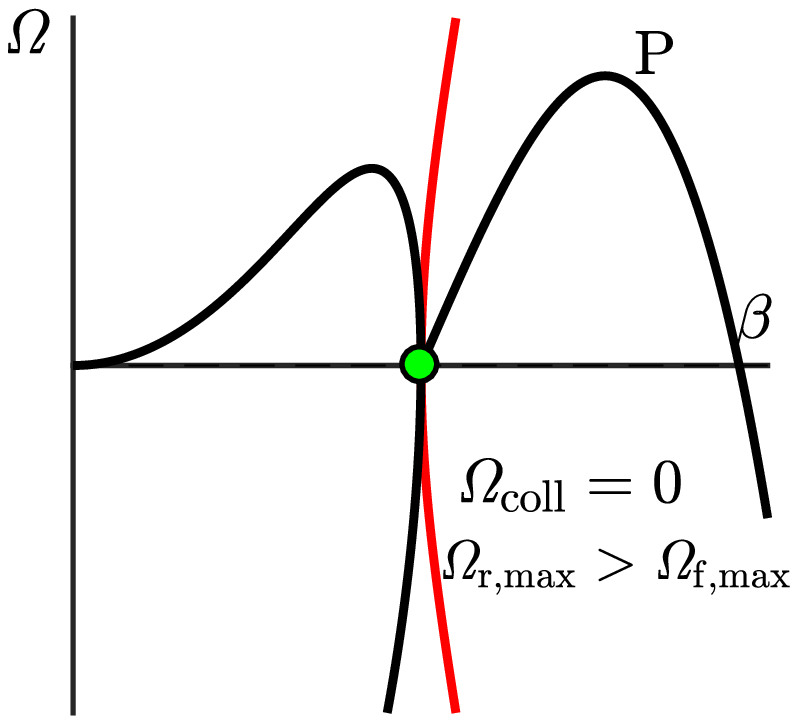} \includegraphics[width=10pc]{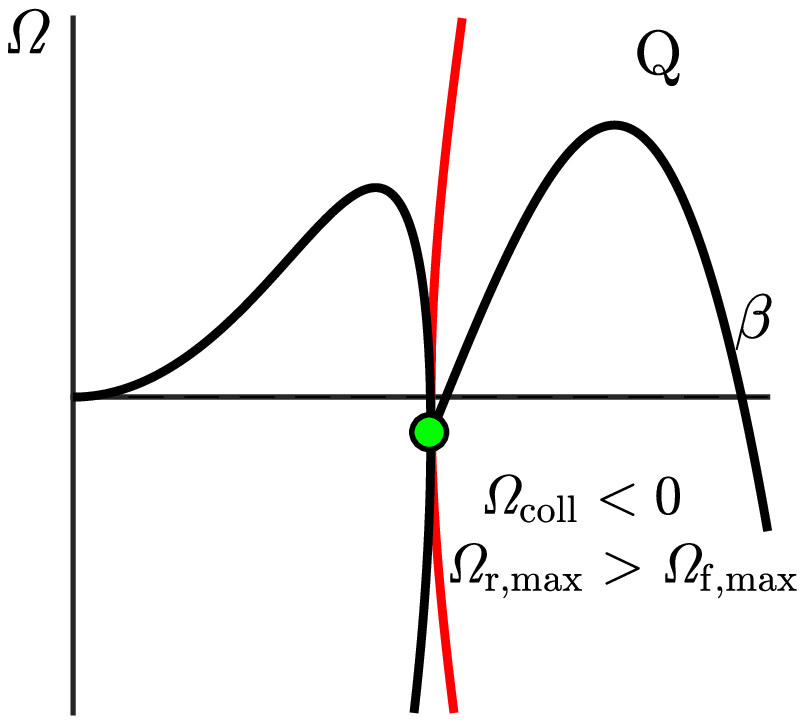}
\caption{Plots of the real Re[$\Omega(\beta)$] (black lines), and imaginary Im[$\Omega(\beta)$] (red) parts of the frequencies $\Omf(\beta)$, $\Omega_1(\beta)$ and $\Omfone(\beta)$ for varying $\beta$. The frequencies $\Omf(\beta)$ and $\Omega_1(\beta)$ collide at $(\betaL,\OmL)$ (shown in green). In contrast to figure~\ref{fig:charcoal1}, the plots are for values $\Cd\geq\Cd_{\rm g}$ at which global oscillatory instabilities may occur. Panels M, P, and Q show that the maximum growth rate ($\Omrmax$) for the real frequency $\Omf(\beta)$ is smaller than the maximum real growth rate ($\Omfmax$) for frequencies $\Omfone(\beta)$ with a non-zero imaginary part. The panels correspond to the points with parameters labelled L--Q in figure~\ref{fig:drw02}.}
\label{fig:charcoal2}
\end{centering}
\end{figure}

Figure \ref{fig:drw02} continues the phase diagrams in figure \ref{fig:drw0} to the values $1.5635 \leq \Cd \leq 5$ (again with $\Qd{}=5.114\times{10}^{-4}$).  The value of $\Cd_{\rm g} = 1.5635$ in the first panel of figure \ref{fig:drw02} is significant: it is the onset value for global oscillatory instabilities, which emerge at the point labeled L. For $\Cd > \Cd_{\rm g}$, there are regions of parameter values $(\Bd, \Sd)$ (red shading, figure \ref{fig:drw02}) that are globally oscillatory unstable.  The value $\Cd = 5$ in figure \ref{fig:drw02} highlights that global oscillatory unstable modes can be either Type I or Type II.  Figure~\ref{fig:charcoal2} plots the roots $\Omf(\beta), \Omega_1(\beta)$ and $\Omfone(\beta)$ for behavior indicative of parameter values L--Q in figure~\ref{fig:drw02}.  Subfigures~\ref{fig:charcoal2} L, N and O correspond to parameter values on the boundary of the global oscillatory instability region (L being Type I, O being Type II and N being on the boundary of Type I).  The remaining panels M, P and Q show points for parameter values in the interior of the global oscillatory instability region.
\begin{figure}
\begin{centering}
\includegraphics[width=40pc]{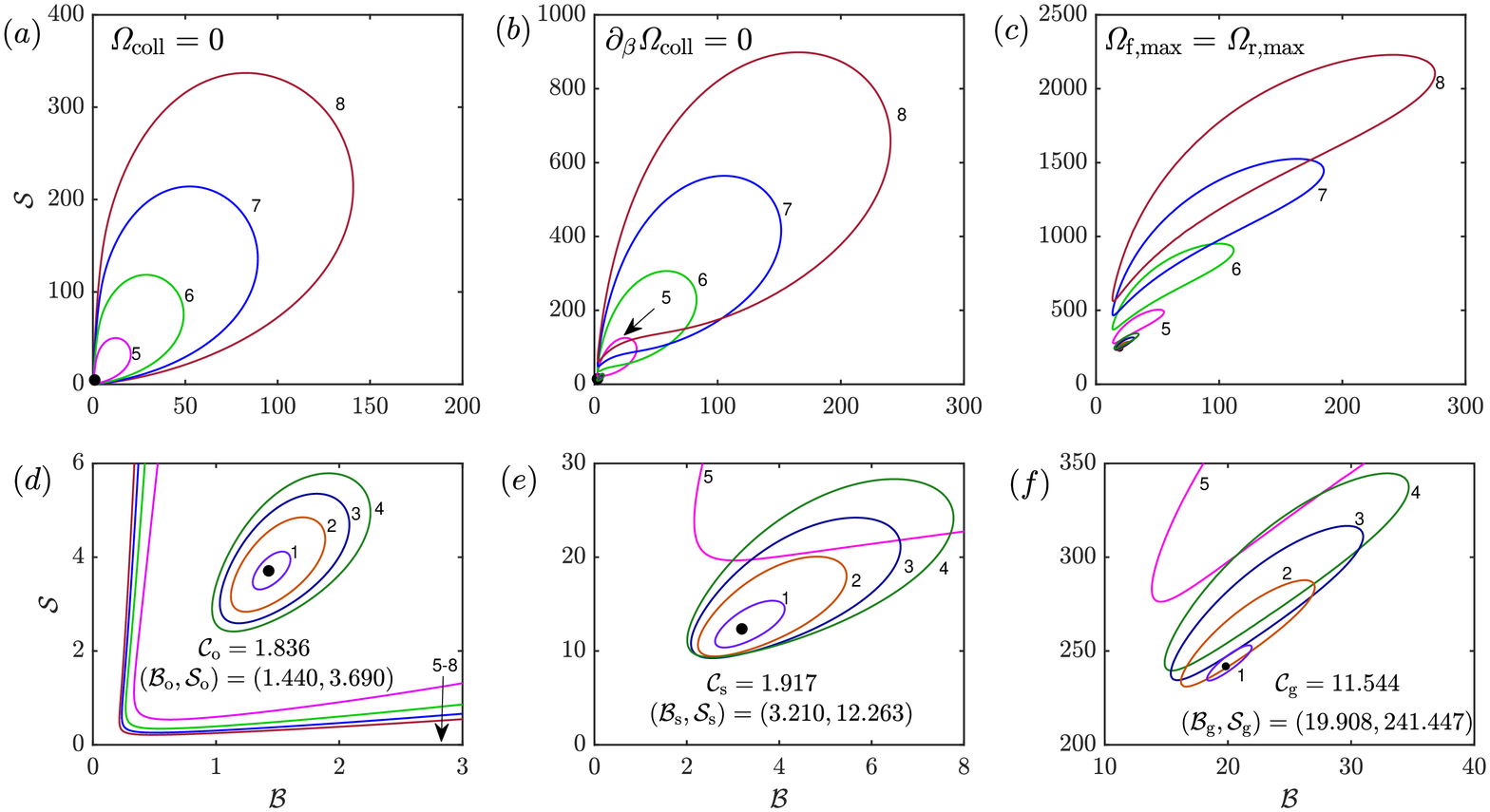}
\caption{Left panels (a), (d): show the contours (given by $\OmL = 0$) that enclose regions of Type I oscillatory instabilities. The contours (enumerated $1$--$8$) correspond to values $\Cd=(1.84, 1.86, 1.88, 1.90, 4, 6, 8, 10)$. Middle panels (b), (e): show the contours for the heuristic ($\partial_\beta\OmL=0$) used to help identify regions of global oscillatory instabilities. The curves (enumerated $1$--$8$) are for $\Cd=(1.95, 2.05, 2.15, 2.25, 4, 6, 8, 10)$. Right panels (c), (f): contours (enumerated $1$--$8$) enclose regions of global oscillatory instabilities (given by $\Omfmax = \Omrmax$) and are for values $\Cd=(11.6,12,12.4,12.8, 15,20,25,30)$. All plots are for $\Qd=0.006$.}
\label{fig:drw0evo1}
\end{centering}
\end{figure}

Lastly, we characterize the phase diagrams as $\Cd$ (which one may think of as the temperature difference) becomes large.  Figure~\ref{fig:drw0evo1} fixes  $\Qd=0.006$ and plots contours (corresponding to the curve $\OmL = 0$) in subpanels (a), (d) that enclose Type I instabilities; contours in subpanels (b), (e) that define the heuristic (corresponding to the curve $\partial_\beta\OmL=0$); and contours in subpanels (c), (f) that enclose regions of global oscillatory instabilities (corresponding to the curve $\Omfmax = \Omrmax$). In the subpanels the contours are enumerated 1--8 for convenience, and correspond to increasing values of $\Cd$, with the smallest value of $\Cd$ labelled $1$ and the largest labelled $8$ (the numerical values of $\Cd$ are stated in the caption). We define $\Cd_{\rm s}$ to be the (smallest) onset value of $\Cd$ for which the zero contour of the heuristic $\partial_{\beta}\OmL$ emerges in the $\Bd\,$--$\,\Sd$ plane; and $(\Bd_{\rm s},\Sd_{\rm s})$ as the point at which the heuristic $\partial_{\beta}\OmL = 0$ first emerges. The bottom panels (d)--(f) of figure \ref{fig:drw0evo1} show the onset coordinate  values $(\Bd_{\rm o}, \Sd_{\rm o})$ for oscillatory instabilities (subpanel (d));  $(\Bd_{\rm o}, \Sd_{\rm o})$ for the heuristic (subpanel (e)); and $(\Bd_{\rm g}, \Sd_{\rm g})$ for global oscillatory instabilities (subpanel (f)). 

Subfigures (a), (d) show that the contours 1--8 are nested: given two contours $1\leq j, k \leq 8$ with $\Cd_j > \Cd_k$, then contour $\Cd_j$ encloses contour $\Cd_k$.  This fact implies that the regions of oscillatory instability in the phase diagrams $\Bd\,$--$\,\Sd$ become large as $\Cd$ increases. The regions of global oscillatory instabilities in subfigures (c), (f) are not nested, but still grow in size as $\Cd$ increases. How the oscillatory unstable regions change as $\Cd$ varies is significant. Generally speaking,  contours 5--8 (subpanel (a)) show that these contours expand in all directions with increasing $\Cd$, and eventually cover the entire $\Bd\,$--$\,\Sd$ plane. This suggests that (for this value of $\Qd$), any pair $(\Bd, \Sd)$ will lead to oscillatory instabilities for a sufficiently large $\Cd$ (temperature difference) value. In contrast, the existence of global oscillatory unstable regions (subpanels (c), (f)) in the $\Bd\,$--$\,\Sd$ plane depends on $\Cd$:
\begin{itemize}
	\item There are values of $(\Bd,\Sd)$ (for instance, if the ratio $\Sd/\Bd$ is sufficiently small) that will not be globally oscillatory unstable for any value of $\Cd$.  
	\item An arbitrary set of parameters $(\Bd,\Sd)$ (material thicknesses) will only be 	subject to a global oscillatory instability for a range of $\Cd$ (temperature differences) values (between a minimum value $\Cd_{\rm g}$ and some maximum value).
\end{itemize} 
Similar conclusions can be drawn that describe the change in the zero contours of the heuristic $\partial_\beta\OmL$ with $\Cd$ (subpanels (b), (e)).  These observations provide a valuable guide for choosing materials that yield realistic experimental setups for observing Type II and global oscillatory instabilities.

\subsection{The effect of $\Qd$ on the behavior of the phase diagrams}\label{sec:deponQ}
\begin{figure}
\begin{centering}
\includegraphics[width=40pc]{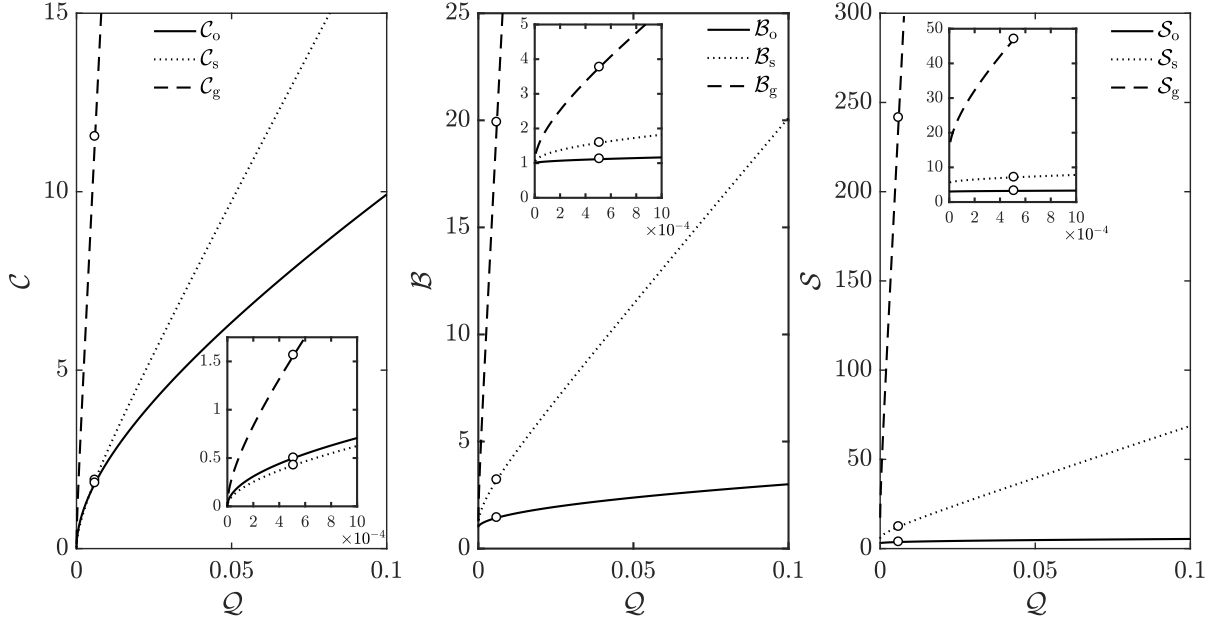}
\caption{Left panel: plots the onset values for which oscillatory instabilities occur ($\Cd_{\rm o}$); the heuristic changes sign ($\Cd_{\rm s}$); and global oscillatory instabilities occur ($\Cd_{\rm g}$). The middle and right panels plot the same three onset values for $\Bd$ and $\Sd$ respectively. Note that all onset values are a function of $\Qd$ only -- once a value of $\Qd$ is fixed, the onset values are uniquely determined. The insets depict the behavior in the vicinity of $\Qd=0$ and the circles correspond to values at $\Qd=5.114\times{10}^{-4}$ used in figures \ref{fig:drw0} and \ref{fig:drw02}; the circles in the primary panels are for $\Qd=0.006$ used in figure \ref{fig:drw0evo1}.}
\label{fig:neutdiag}
\end{centering}
\end{figure}
As previously stated, the value of $\Qd$ depends on the material properties in the experiment and cannot be modified by the experimental setup (i.e. experimental geometry or temperature).  Therefore, identifying values of $\Qd$ that give rise to oscillatory instabilities (and in particular, global oscillatory instabilities) will guide the choice of experimental materials. 

Numerical experiments show that modifying the value of $\Qd$ does not change the qualitative behavior of the phase diagrams in \S\ref{sec:CdltCdg}. Changes in the value of $\Qd$ can, however, result in a (potentially significant) quantitative change in the onset values $\Cd_{\rm o}$, $\Cd_{\rm g}$ (temperature difference), as well as the locations in the $\Bd\,$--$\,\Sd$ plane (i.e. thicknesses of the film and substrate) for which the regions of oscillatory and global oscillatory instabilities emerge: $(\Bd_{\rm o}, \Sd_{\rm o})$ and $(\Bd_{\rm g}, \Sd_{\rm g})$.  

Figure \ref{fig:neutdiag} plots the onset values $\Cd_{\rm o}, \Bd_{\rm o}, \Sd_{\rm o}$ (solid line) for oscillatory instabilities, as well as the onset values $\Cd_{\rm g}, \Bd_{\rm g}, \Sd_{\rm g}$ (dashed line) for global oscillatory instabilities, as functions of $\Qd$.  Onset values $\Cd_{\rm s}, \Bd_{\rm s}, \Sd_{\rm s}$ of the emergence of the zero of the heuristic 
(dotted lines) are also given. Figure \ref{fig:neutdiag} restricts the range of $0 \leq \Qd \leq 0.1$ to an experimentally feasible range. The plots are significant since $\Cd_{\rm o}$ and $\Cd_{\rm g}$ are the minimum values for which oscillatory and global oscillatory instabilities occur. In addition, the values of $\Bd_{\rm g}$ and $\Sd_{\rm g}$ provide information on choosing $\Bd$ and $\Sd$. Guided by the qualitative behavior in figure~\ref{fig:drw0}, the diagrams show that choosing $(\Bd, \Sd)$ close to $(\Bd_{\rm g}, \Sd_{\rm g})$ will likely yield global oscillatory instabilities for some range of $\Cd > \Cd_{\rm g}$.

As a computational remark, the values $\Cd_{\rm g}$ and $\Cd_{\rm o}$ are calculated by minimizing the value of $\Cd$ in the region of $(\Cd, \Bd, \Sd)$ parameter space that satisfies the Type I instability criterion (i.e. satisfy the condition $\OmL=0$) or the global oscillatory instability criterion (condition \eqref{eq:GOI_cond}). 

\subsection{Experimental considerations for oscillatory instabilities}
\label{sec:varC}
\begin{table}
  \begin{center}
\def~{\hphantom{0}}
  \begin{tabular}{lcc}
   & & silicone oil \\
      $\mu$ & (kg/m$\cdot{}$s)  & 4.94$\times{10^{-4}}$ \\
      $\kappa_{\rm f}$ & (W/m$\cdot{}$K) & 0.1 \\
      $\sigma_0$ & (kg/s$^{2}$) & 1.59$\times{10}^{-2}$ \\
      $\gamma$ & (kg/s$^2\cdot$K) & 6.4$\times{10}^{-5}$ \\
      $\theta_0$ & (K) & 293 
  \end{tabular}
  \quad
  \begin{tabular}{lccc}
       & & copper & PMMA \\
      $\kappa_{\rm s}$ & (W/m$\cdot{}$K)  & 400 & 0.19 \\
      $\chi_{\rm s}$ & (m$^2$/s) & 1.16$\times{10}^{-4}$ &  1.15$\times{10}^{-7}$ \\
\\ $q$ & (W/m$\cdot{}$K) & 5 & 5
            \\      $\Qd$ & (dimensionless) & $2.88\times{10}^3$ & $6.45\times{10^{-7}}$ 
  \end{tabular}  
  \caption{Physical properties for silicone oil, copper and PMMA (resp. \citet{Hintz01}, \citet{Araki92}, and \citet{Assael05}).  The two values of $\Qd$ are obtained by pairing each substrate with the silicone oil and estimating the heat transfer coefficient.  }
  \label{tab:filmprop}
  \end{center}
\end{table}
\begin{figure}
\begin{centering}
\includegraphics[width=40pc]{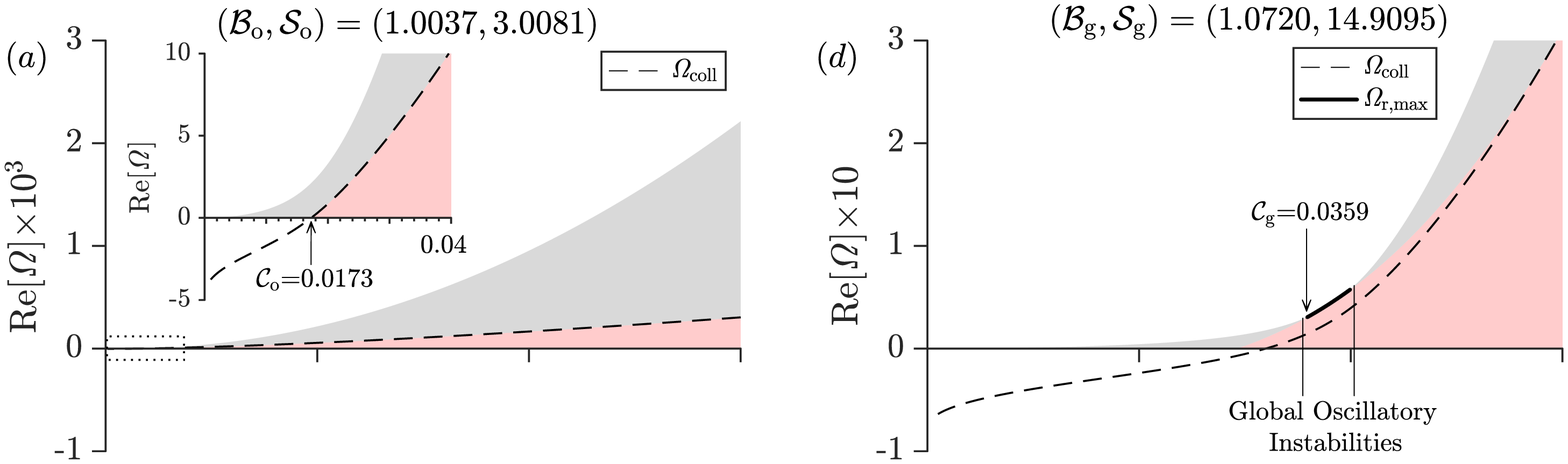}\\
\includegraphics[width=40pc]{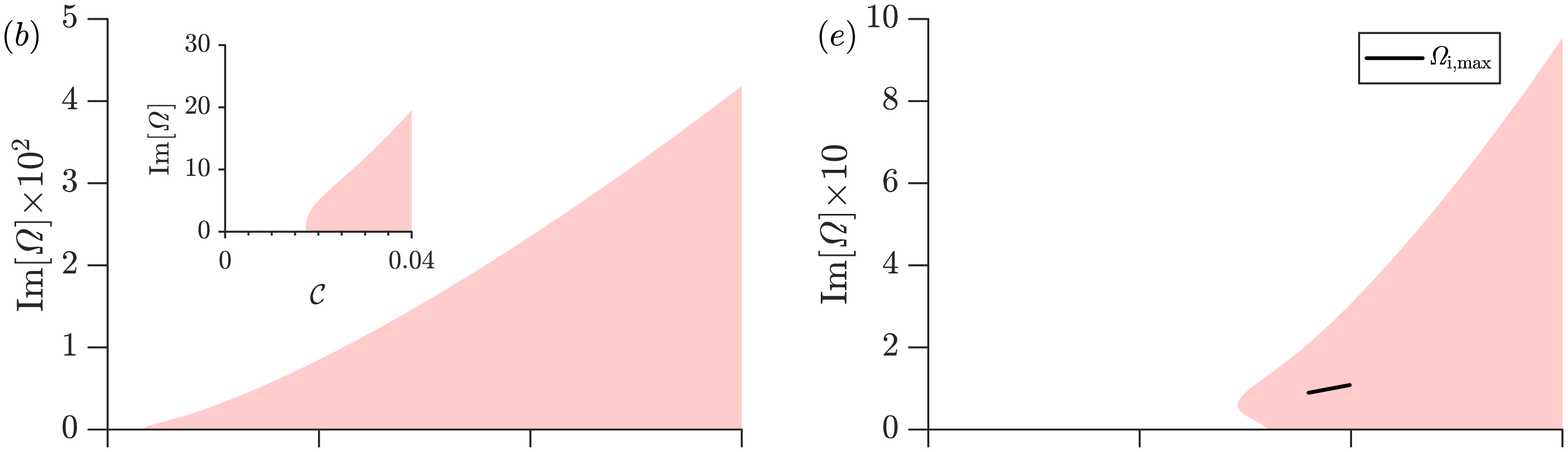}\\
\includegraphics[width=40pc]{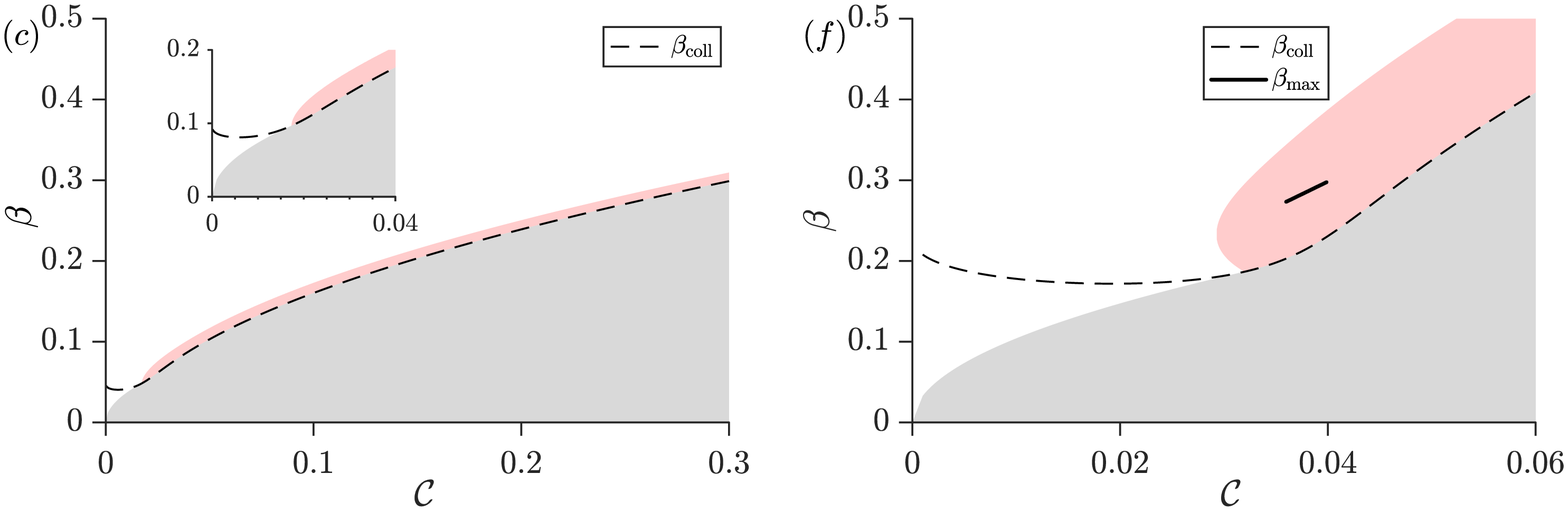}
\caption{The figure plots the unstable frequencies $\Omega$ (with real parts in subfigures (a) and (d), and imaginary parts in (b) and (e)) and wavenumbers $\beta$ (insubfigures (c) and (f)) versus $\Cd$ for two different sets of $\Sd$, $\Bd$ values holding $\Qd=6.449\times{10}^{-7}$ fixed. The shaded regions in grey show values of $\Omega$ and $\beta$ that are unstable but not oscillatory unstable; while pink regions contain oscillatory instabilities (and may also contain non-oscillatory instabilities as well). The solid black lines denote the frequencies and wavenumbers for global oscillatory instabilities with the largest growth rate (see equation \eqref{eq:GOI_def}). This figure demonstrates that global oscillatory instabilities may occur only in an interval range of $\Cd$ values; and that increasing the temperature difference (proportional to $\Cd$) between the film and substrate promotes the range of frequency and wavenumber values for which oscillatory instabilities may occur.}
\label{fig:varC}
\end{centering}
\end{figure}
In this section, we discuss oscillatory instabilities in the context of an experimental setting. This will shed light on the physical mechanism for oscillatory instabilities. In particular, we will contrast two cases: a substrate that is a conductor (having a large thermal conductivity); and a substrate that is an insulator (having a small thermal conductivity).  We conclude that oscillatory instabilities are far more likely to be observed for films heated by substrates that are insulators.

To draw this conclusion, we estimate the rate of heat transfer between the surface and the gas $q=5$ W/m$\cdot{}$K (see table~\ref{tab:filmprop}) and consider low-viscosity silicone oil films for all the cases presented in this section.  Physical properties of the silicone oil, copper (a conductor), and PMMA (poly(methyl methacrylate), an insulator) are given in Table \ref{tab:filmprop}. We first consider films of silicone oil heated by copper substrates. The results from the previous section will demonstrate that oscillatory instabilities for silicone oil-copper systems are not likely in experimentally feasible conditions.

Together, the silicone oil--copper system yields a parameter value $\Qd = 2.88\times{10}^3$, which is four orders of magnitude larger than the range plotted in figure \ref{fig:neutdiag}. The magnitude of $\Qd$ is large primarily because the thermal conductivity ratio $\kappa_{\rm s}/\kappa_{\rm f}$ is large.  This value of $\Qd$ yields onset parameters:  $(\Cd_{\rm o},\Bd_{\rm o},\Sd_{\rm o})=(1.187\times{}10^5,1.03\times{}10^4,8.65\times{}10^3)$. Using these onset values as a rough guide to estimate experimental conditions yields the parameter values:  $(\Delta,\bar{h},d)=(1.85 \textrm{ K}, 2.05\times{}10^2 \textrm{ m}, 6.90\times{}10^5 \textrm{ m})$.  Although the temperature difference ($\Delta$) is feasible, the thicknesses are clearly not.  

We now shift our focus to a substrate material that does lead to oscillatory instabilities under experimentally feasible experimental conditions.  PMMA is a readily-available insulating material with low thermal conductivity and diffusivity.   Again using the properties from Table \ref{tab:filmprop}, the silicone oil-PMMA system has a value of $\Qd=6.45\times{10}^{-7}$.  Substituting this value of $\Qd$ into figure \ref{fig:neutdiag} yields the onset parameter values, which can then be used to estimate experimental conditions:
\begin{equation*} 
\begin{aligned}
	\begin{array}{lcl}
	(\Cd_{\rm o},\Bd_{\rm o},\Sd_{\rm o})=(0.0173,1.004,3.008) &\quad \Longrightarrow \quad& 
	(\Delta,\bar{h},d)=(1.19\textrm{ K},0.020\textrm{ m}, 0.114 \textrm{ m}), \\
	(\Cd_{\rm g},\Bd_{\rm g},\Sd_{\rm g})=(0.0359,1.072,14.910) &\quad \Longrightarrow \quad&
	(\Delta,\bar{h},d)=(2.47\textrm{ K}, 0.0214\textrm{ m}, 0.5\textrm{ m}).
	\end{array}
\end{aligned}
\end{equation*}
The above dimensional variables (temperature and thicknesses) provide a guide for predicting the range of experimental values for which oscillatory instabilities occur. Figure \ref{fig:varC} presents results for the silicone oil-PMMA system with different experimental parameters $\Delta, \bar{h}, d$.  The figure varies the parameter $0 \leq \Cd \leq 0.3$ which corresponds to a dimensional temperature (difference) range $0 \leq \Delta \leq 26.5$ K.  In particular, the top row of figure \ref{fig:varC} plots the real values of several important unstable frequencies versus $\Cd$ at different $\Bd, \Sd$ values: $\Omfmax$ (maximum real growth rate), $\Omegacm$ (maximum real growth rate of oscillatory instabilities) and $\OmL$. Note that these values are defined in \S\ref{sec:charcoal}.  The middle row in figure \ref{fig:varC} plots the range of imaginary values $\Omi$ that are oscillatory unstable. Having information on the possible imaginary values of the complex frequencies that are oscillatory unstable is useful, since these frequencies may be excited via parametric resonance by external forcing.  Lastly, the bottom row plots the band of wavenumbers that are unstable, and oscillatory unstable. Plotting the unstable wavenumbers provides information on the length scales (thereby influencing which experimental domain sizes one can utilize) that lead to instability. 

Figure~\ref{fig:varC} demonstrates the effect of temperature difference ($\Cd$) on oscillatory instabilities. Specifically, the figure plots frequency values $\Omega$ (real parts are in subfigures (a) and (d)) and imaginary parts in subfigures (b) and (e)) and wavenumbers $\beta$ versus $\Cd$ (in subfigures (c) and (f)).  The grey regions correspond to values for which only monotonic (non-oscillatory) instabilities occur, while the pink regions correspond to values at which oscillatory instabilities may occur. Note that since there are an infinite number of roots $\Omega_{n}(\beta)$, the pink regions -- which always have oscillatory instabilities -- may also contain monotonic instabilities as well as oscillatory instabilities. Figure~\ref{fig:varC} contrasts the stability behavior of $\Omega$ and $\beta$ versus $\Cd$ for two different sets of $\Bd$ and $\Sd$ values. Here the values of $\Bd = \Bd_{\rm o}$ and $\Sd = \Sd_{\rm o}$ were chosen to ensure oscillatory instabilities in the subfigures (a--c), while $\Bd = \Bd_{\rm g}$ and $\Sd = \Sd_{\rm g}$ were chosen to ensure that global oscillatory instabilities occur for a range of $\Cd$ values in subfigures (d--f). The dashed lines in the subfigures correspond to $\beta_{\rm coll}$ (in (a), (d)) and $\Omega_{\rm coll}$ (in (c) and (f)) and are defined in equation \eqref{eq:Notation}. The solid lines plot the most unstable wavenumber and frequencies for which global oscillatory instabilities occur; and correspond to the variables $\beta_{\rm max}$, $\Omega_{\rm r,max}$ and $\Omega_{\rm i,max}$ as defined in equation \eqref{eq:GOI_def}. Figure~\ref{fig:varC} shows that the range of unstable wavenumbers and frequencies increase with $\Cd$, and that in general, large $\Cd$ values tend to drive oscillatory instabilities.

\subsection{Summary}
We now recapitulate the most important results of this section.  First, this section classifies oscillatory instabilities, for any $(\Qd,\Cd,\Bd,\Sd)$, as Type I, Type II, and subsequently determines whether they are globally oscillatory unstable.  Several important conclusions can be reached:
\begin{itemize}
	\item The choice of materials (i.e. the fluid and the substrate) dictate the	parameter $\Qd$. The value of $\Qd$ then guides which experimental conditions, such as the film and substrate thicknesses, as well as temperature difference,  lead to oscillatory instability. Generally speaking, the onset values $\Sd_{\rm g}$ and $\Bd_{\rm g}$ provide guides (i.e. order-of-magnitude estimates) that can be used as minimum material thicknesses. 
	\item Temperature drives instability.  It is well-known that temperature gradients can drive instabilities in fluids (i.e. Rayleigh-Benard convection). This result is also true in the current setting: oscillatory instabilities are more likely when there is a larger temperature difference across the film and substrate. Crucially, we find that oscillatory instabilities, or global oscillatory instabilities arise for $\Cd$ values (i.e. temperature values) that exceed predefined thresholds $\Cd > \Cd_{\rm o}$, or $\Cd > \Cd_{\rm g}$, respectively.  
	\item Insulating substrates are more likely to give rise to oscillatory instabilities than conducting substrates.  The physical reason is that substrates that  are thermally conducting transfer heat, and consequently equilibrate their temperatures,  over time scales much faster than the characteristic time scales in the thin film. Oscillatory instabilities require thermal coupling between the substrate and the film, and can occur when the natural time scales of the film are on the same order as the time scale governing thermal diffusion in the substrate.
\end{itemize}

\section{Discussion and conclusion}
In this work we derived a nonlinear model that couples the thermocapillary dynamics of a liquid film heated by a thermally conductive and diffusive substrate.  This was done by assuming a large substrate-to-film thermal conductivity ratio and a substrate thickness that is asymptotically larger than both the mean film thickness and the characteristic lateral disturbance.  In order to highlight parameter regimes that are subject to oscillatory instabilities, a scaling was incorporated that grouped the effects of the substrate thermal diffusivity, the imposed temperature difference, the film thickness, and the substrate thickness via four separate dimensional parameters:    $(\Qd,\Cd,\Bd,\Sd)$.

For any set of model parameters, linear stability of the model can be described by the wavenumber-dependent interaction between a perturbation associated with the governing film evolution equation and an infinite number of perturbations associated with the substrate heat equation.  The film root coalesces with the root $\Omega_{1}(\beta)$ at certain wavenumbers; at these points the participating roots bifurcate into the complex plane and become oscillatory unstable. To investigate the emergence of these instabilities, complex numerical continuation and optimization algorithms were used in section \ref{sec:emerge} to describe the emergence of oscillatory instabilities.   Notably, we showed that parameter sets subject to a global oscillatory instability  occur only for substrate-to-film thickness ratios that are sufficiently large.   

We have not provided quantitative predictions of the exact experimental conditions at which one will be able to observe oscillatory thermocapillary instability.  This is due to the difficulties associated with prescribing the heat transfer coefficient $q$ that describes the rate of heat transfer between the film and the bounding cold gas layer.  In many cases it is probably not possible to determine this parameter prior to running an experiment.  One way to handle this issue could be to first conduct experiments on thin substrates, and then use the observed instability wavelength to determine $q$.  Still, several workers have highlighted the weakness of Newton's Law of Cooling, in particular, due to transport effects in the gas layer, see \citet{VanHook97}, or in applications where heat transfer rates are large, see \citet{Besson12}.  Therefore, we suggest that the results of our work serve as a foundation upon which more elaborate models can be developed.

To conclude we note that, although most of the oscillatory instabilities we discussed in this work were not global, the coupled model may also serve as a foundation upon which time-periodic excitation could be investigated as a means of driving instability.  


\appendix
\section{Thin substrate limit}\label{appB}
For sufficiently thin substrates, lateral heat conduction and the thermal diffusivity can be neglected and a single nonlinear PDE can be derived for the evolution of the local film thickness.  Instead of re-deriving the long wave-model starting with these these assumptions, we equivalently obtain its dispersion relation by taking the limit of (\ref{eq:dispo}) as $\sqrt{\Omega+\beta^2}\rightarrow{0}$, viz.,
\begin{align}
\Qd\,\Sd^2\,\Omega+\frac{1}{3}\,\Bd\,\beta^4-\frac{1}{2}\,\Cd\,\Bd\,\Sd^2\,\thb^2\,\beta^2=0.
\label{eq:dispD}
\end{align}
Effectively we have restricted consideration to film and substrate temperature profiles depending only on the vertical coordinate, as described by the basic state solutions (\ref{eq:thetahbar}).  In doing so, the full dispersion relation reduces to an explicit expression for strictly real values of ${\Omega}$ in terms of the model parameters.  

The dimensional equivalent to (\ref{eq:dispD}) is obtained by making substitutions (\ref{eq:dpar}) and (\ref{eq:scales}) and solving for the dimensional growth rate $\omega$ as a function of the wavenumber $k$, viz.,

\begin{align}
\mu\,\omega=-\frac{\sigma_0\,\hb^3k^4}{3}+\frac{\gamma\,\Delta\,\hb\,k^2}{2}\dfrac{q\,\hb}{\kappa_{\rm f}}
\left[1+\dfrac{q\,\hb}{\kappa_{\rm f}}+\dfrac{q\,d}{\kappa_{\rm s}}\right]^{-2}.
\label{eq:dispdim}
\end{align}
Aside from $\chi_{\rm s}$ (negligible for thin substrates), this expression for $\omega(k)$ describes the influence of material properties and dimensions on film stability.  It is clear that viscosity modifies only the growth rate. Solving for the cutoff wavenumber $k_{\rm c}$ at which $\omega=0$ we obtain
\begin{align}
k_{\rm c}^2=\frac{3\,\gamma\,\Delta}{2\,\sigma_0\,\hb^2}\dfrac{q\,\hb}{\kappa_{\rm f}}\left[1+\dfrac{q\,\hb}{\kappa_{\rm f}}+\dfrac{q\,d}{\kappa_{\rm s}}\right]^{-2}.
\label{eq:kct}
\end{align}
This wavenumber divides the continuous bands of unstable ($0<k<k_{\rm c}$) and stable ($k>k_{\rm c}$) wavenumbers for a given set of system parameters.  Instability described by (\ref{eq:kct}) is clearly driven by increasing values of the coefficient $\gamma\,\Delta/\sigma_0$ and decreasing film thickness $\hb$.  

It is also evident that $\lim_{q\to{0}}k_{\rm c}=\lim_{q\to\infty}k_{\rm c}=0.$  For $q\rightarrow{0}$, the resistance to heat transfer at the film-gas interface becomes infinite and, as a result, the perturbed free surface is uniformly equal to the blackbody temperature $\psi_{\rm b}$.  In the absence of variations in the free surface temperature, no thermocapillary stresses arise and perturbations of all wavelengths are stable.  Likewise, all values of $k$ are stabilized in the limit $q\rightarrow{\infty}$, which uniformly sets the free surface temperature to the gas temperature $\theta_{\rm g}$.  For finite values of $q$, interfacial resistance to heat transfer introduces variations in the free surface temperature that depend locally on the perturbed film thickness.  Specifically, with the rate at which heat is removed from the film fixed by $q$, local hot and cold spots form at troughs and crests, respectively, due to their relative proximities to the heating source.

Finally, inspecting the limits 
\begin{gather}
\lim_{d\to{0}} k_{\rm c}^2= \lim_{\kappa_{\rm s}\to{\infty}} k_{\rm c}^2= \frac{3\,\gamma\,\Delta}{2\,\sigma_0\,\hb^2}\dfrac{q\,\hb}{\kappa_{\rm f}}\left[1+\dfrac{q\,\hb}{\kappa_{\rm f}}\right]^{-2},
\label{eq:nosublim}
\end{gather}
we see that $k_{\rm c}$ is maximized for situations that effectively transfer the isothermal blackbody temperature directly to the film-substrate interface ($\Sd\rightarrow{0}$).  We conclude by stating that placing a substrate between a film and the blackbody necessarily stabilizes films for finite values of $d$ and $\kappa_{\rm s}$ relative to the case of heating a film directly without a substrate.

\section*{Acknowledgments}
The research was supported by a fellowship from the New Jersey Institute of Technology Department of Mathematical Sciences (Batson); by NSF CBET--1604351 (Batson, Kondic); by NSF DMS--1815613 (Cummings, Kondic); and by NSF DMS--1719693 (Shirokoff). 
D. Shirokoff was supported by a grant from the Simons Foundation ($\#359610$).

\bibliographystyle{plainnat}
\setlength{\bibsep}{0pt}
\bibliography{2018-06-28-v8-wrb}

\begin{thebibliography}{30}
\providecommand{\natexlab}[1]{#1}
\providecommand{\url}[1]{\texttt{#1}}
\expandafter\ifx\csname urlstyle\endcsname\relax
  \providecommand{\doi}[1]{doi: #1}\else
  \providecommand{\doi}{doi: \begingroup \urlstyle{rm}\Url}\fi

\bibitem[Anderson and Worster(1996)]{Anderson96}
D.~M. Anderson and M.~G. Worster.
\newblock A new oscillatory instability in a mushy layer during the
  solidification of binary layers.
\newblock \emph{J. Fluid Mech}, 307:\penalty0 245--267, 1996.

\bibitem[Araki et~al.(1992)Araki, Makino, and Mihara]{Araki92}
N.~Araki, A.~Makino, and J.~Mihara.
\newblock Measurement and evaluation of the thermal diffusivity of two-layered
  materials.
\newblock \emph{Int. J. Thermophys.}, 13:\penalty0 331--349, 1992.

\bibitem[Assael et~al.(2005)Assael, Botsios, Gialou, and Metaxa]{Assael05}
M.~J. Assael, S.~Botsios, K.~Gialou, and I.~N. Metaxa.
\newblock Thermal conductivity of polymethyl methacrylate ({PMMA}) and
  borosilicate crown glass {BK7}.
\newblock \emph{Int. J. Thermophys.}, 26:\penalty0 1595--1605, 2005.

\bibitem[Atena and Khenner(2009)]{Atena09}
A.~Atena and M.~Khenner.
\newblock Thermocapillary effects in driven dewetting and self assembly of
  pulsed-laser-irradiated metallic films.
\newblock \emph{Phys. Rev. B}, 80:\penalty0 075402, 2009.

\bibitem[Beerman and Brush(2007)]{Beerman07}
M.~Beerman and L.~N. Brush.
\newblock Oscillatory instability and rupture in a thin melt film on its
  crystal subject to freezing and melting.
\newblock \emph{J. Fluid Mech.}, 586:\penalty0 423--448, 2007.

\bibitem[Besson(2012)]{Besson12}
U.~Besson.
\newblock The history of the cooling law: When the search for simplicity can be
  an obstacle.
\newblock \emph{Science \& Education}, 21:\penalty0 1085--1110, 2012.

\bibitem[Bestehorn and Borcia(2010)]{Bestehorn10}
M.~Bestehorn and I.~D. Borcia.
\newblock Thin film lubrication dynamics of a binary mixture: Example of an
  oscillatory instability.
\newblock \emph{Phys. Fluids}, 22:\penalty0 104102, 2010.

\bibitem[Boyd(2014)]{Boyd14}
J.~P. Boyd.
\newblock \emph{Solving Transcendental Equations}.
\newblock SIAM, 2014.

\bibitem[Craster and Matar(2009)]{Craster09}
R.~V. Craster and O.~K. Matar.
\newblock Dynamics and stability of thin liquid films.
\newblock \emph{Rev. Mod. Phys.}, 81:\penalty0 1131--1198, 2009.

\bibitem[Dietzel and Troian(2009)]{Dietzel09}
M.~Dietzel and S.~M. Troian.
\newblock Formation of nanopillar arrays in ultrathin viscous films: The
  critical role of thermocapillary stresses.
\newblock \emph{Phys. Rev. Lett.}, 103:\penalty0 074501, 2009.

\bibitem[Dong and Kondic(2016)]{Dong16}
N.~Dong and L.~Kondic.
\newblock Instability of nanometric fluid films on a thermally conductive
  substrate.
\newblock \emph{Phys. Rev. Fluids}, 1:\penalty0 063901, 2016.

\bibitem[Hintz et~al.(2001)Hintz, Schwabe, and Wilke]{Hintz01}
P.~Hintz, D.~Schwabe, and H.~Wilke.
\newblock Convection in a czochralski crucible--part 1: non-rotating crystal.
\newblock \emph{J. Cryst. Growth}, 222:\penalty0 343--355, 2001.

\bibitem[Morozov et~al.(2014)Morozov, Oron, and Nepomnyashchy]{Morozov14}
M.~Morozov, A.~Oron, and A.~A. Nepomnyashchy.
\newblock Long-wave {M}arangoni convection in a layer of surfactant solution.
\newblock \emph{Phys. Fluids}, 26:\penalty0 112101, 2014.

\bibitem[Nepomnyashchy et~al.(2001)Nepomnyashchy, Velarde, and
  Colinet]{Nepomnyashchy01}
A.~Nepomnyashchy, M.~Velarde, and P.~Colinet.
\newblock \emph{Interfacial phenomena and convection}.
\newblock CRC Press, 2001.

\bibitem[Nepomnyashchy and Simanovskii(2007)]{Nepom07}
A.~A. Nepomnyashchy and I.~B. Simanovskii.
\newblock {M}arangoni instability in ultrathin two-layer films.
\newblock \emph{Phys. Fluids}, 19:\penalty0 122103, 2007.

\bibitem[Oron et~al.(1997)Oron, Davis, and Bankoff]{Oron97}
A.~Oron, S.~H. Davis, and S.~G. Bankoff.
\newblock Long-scale evolution of thin liquid films.
\newblock \emph{Rev. Mod. Phys.}, 69:\penalty0 931--980, 1997.

\bibitem[Podolny et~al.(2005)Podolny, Oron, and Nepomnyashchy]{Podolny05}
A~Podolny, A~Oron, and AA~Nepomnyashchy.
\newblock Long-wave {M}arangoni instability in a binary-liquid layer with
  deformable interface in the presence of {S}oret effect: Linear theory.
\newblock \emph{Phys. Fluids}, 17:\penalty0 104104, 2005.

\bibitem[Pototsky et~al.(2005)Pototsky, Bestehorn, Merkt, and
  Thiele]{Pototsky05}
A.~Pototsky, M.~Bestehorn, D.~Merkt, and U.~Thiele.
\newblock Morphology changes in the evolution of liquid two-layer films.
\newblock \emph{J. Chem. Phys.}, 122:\penalty0 224711, 2005.

\bibitem[Rednikov et~al.(1998)Rednikov, Colinet, Velarde, and
  Legros]{Rednikov98}
A.~Ye. Rednikov, P.~Colinet, M.~G. Velarde, and J.~C. Legros.
\newblock Two-layer {B}\'enard-{M}arangoni instability and the limit of
  transverse and longitudinal waves.
\newblock \emph{Phys. Rev. E}, 57:\penalty0 2872--2884, 1998.

\bibitem[Saeki et~al.(2011)Saeki, Fukui, and Matsuoka]{Saeki11}
F.~Saeki, S.~Fukui, and H.~Matsuoka.
\newblock Optical interference effect on pattern formation in thin liquid films
  on solid substrates induced by irradiative heating.
\newblock \emph{Phys. Fluids}, 23:\penalty0 112102, 2011.

\bibitem[Saeki et~al.(2013)Saeki, Fukui, and Matsuoka]{Saeki13}
F.~Saeki, S.~Fukui, and H.~Matsuoka.
\newblock Thermocapillary instability of irradiated transparent liquid films on
  absorbing solid substrates.
\newblock \emph{Phys. Fluids}, 25:\penalty0 062107, 2013.

\bibitem[Scriven and Sternling(1964)]{Scriven64}
L.~E. Scriven and C.~V. Sternling.
\newblock On cellular convection driven by surface-tension gradients: effects
  of mean surface tension and surface viscosity.
\newblock \emph{J.~Fluid Mech.}, 19:\penalty0 321--340, 1964.

\bibitem[Seric et~al.(2018)Seric, Afkhami, and Kondic]{Seric18}
I.~Seric, S.~Afkhami, and L.~Kondic.
\newblock Influence of thermal effects on stability of nanoscale films and
  filaments on thermally conductive substrates.
\newblock \emph{Phys. Fluids}, 30:\penalty0 012109, 2018.

\bibitem[Shklyaev et~al.(2012)Shklyaev, Alabuzhev, and Khenner]{Shklyaev12}
S.~Shklyaev, A.~A. Alabuzhev, and M.~Khenner.
\newblock Long-wave {M}arangoni convection in a thin film heated from below.
\newblock \emph{Phys. Rev. E}, 85:\penalty0 016328, 2012.

\bibitem[Singer(2017)]{Singer17}
J.~P. Singer.
\newblock Thermocapillary approaches to the deliberate patterning of polymers.
\newblock \emph{J. Polym. Sci. B}, 55:\penalty0 1649--1668, 2017.

\bibitem[Sternling and Scriven(1959)]{Sternling59}
C.~V. Sternling and L.~E. Scriven.
\newblock Interfacial turbulence: Hydrodynamic instability and the {M}arangoni
  effect.
\newblock \emph{AIChE J.}, 5:\penalty0 514--523, 1959.

\bibitem[Strogatz(2015)]{Strogatz}
S.~H. Strogatz.
\newblock \emph{Nonlinear dynamics and chaos: with applications to physics,
  biology, chemistry and engineering}.
\newblock Westview Press, second edition, 2015.

\bibitem[Takashima(1981)]{Takashima81}
M.~Takashima.
\newblock Surface tension driven instability in a horizontal liquid layer with
  a deformable free surface. ii. overstability.
\newblock \emph{J. Phys. Soc. Jap.}, 50:\penalty0 2751--2756, 1981.

\bibitem[Trice et~al.(2007)Trice, Thomas, Favazza, Sureshkumar, and
  Kalyanaraman]{Trice07}
J.~Trice, D.~Thomas, C.~Favazza, R.~Sureshkumar, and R.~Kalyanaraman.
\newblock Pulsed-laser-induced dewetting in nanoscopic metal films: Theory and
  experiments.
\newblock \emph{Phys. Rev. B}, 75:\penalty0 235439, 2007.

\bibitem[VanHook et~al.(1997)VanHook, Schatz, Swift, McCormick, and
  Swinney]{VanHook97}
S.~J. VanHook, M.~F. Schatz, J.~B. Swift, W.~D. McCormick, and H.~L. Swinney.
\newblock Long-wavelength surface-tension-driven {B}\'enard convection:
  experiment and theory.
\newblock \emph{J.~Fluid Mech.}, 345:\penalty0 45--78, 1997.

\end{thebibliography}

\end{document}